\documentclass{aa}  

\usepackage{amssymb,amsmath}
\usepackage{graphicx,booktabs}
\usepackage[T1]{fontenc}
\usepackage{ae,aecompl}
\usepackage[normalem]{ulem}
\usepackage{xcolor}
\usepackage{multicol}
\usepackage{multirow}
\usepackage[varg]{txfonts}
\usepackage{AAS_macro}
\usepackage{blindtext}

\usepackage[final]{hyperref}
\hypersetup{
    colorlinks=true,
    linkcolor=blue,
    filecolor=blue,      
    urlcolor=blue,
    citecolor = blue
    }
\usepackage{cleveref}

\newcommand{\isotope}[2]{$\rm ^{#1}{#2}$}

\graphicspath{{fig/}}

\begin{document} 

   \title{On the origin of the spectral features observed in the cosmic ray spectrum}
   \subtitle{}

   \author{S. Recchia
          \inst{1, 2}\fnmsep\thanks{sarah.recchia@unito.it}
          \and
          S. Gabici\inst{3}
          }
   \institute{
   INAF-Osservatorio Astronomico di Brera, Via Bianchi 46, I-23807 Merate, Italy 
   \and
   Dipartimento di Fisica,           Universit\'a di Torino, via P. Giuria 1, 10125 Torino, Italy
         \and
    Istituto Nazionale di Fisica Nucleare, Sezione di Torino, Via P. Giuria 1, 10125 Torino, Italy
         \and 
    Universit\'e Paris Cit\'e, CNRS, Astroparticule et Cosmologie, F-75013 Paris, France
    }

   \date{Received September 15, 1996; accepted March 16, 1997}

% \abstract{}{}{}{}{} 
% 5 {} token are mandatory
 
  \abstract
{Recent measurements revealed the presence of several features in the cosmic ray spectrum.
In particular, the proton and helium spectra
exhibit a spectral hardening at $\approx 300 \, \mathrm{GV}$ and a spectral steeping at $\approx 15 \, \mathrm{TV}$, followed by the well known \textit{knee}-like feature at $\approx 3 \mathrm{TV}$.
The spectra of heavier nuclei also harden at $\approx 300 \, \mathrm{GV}$, while no claim can be currently done about the presence of the $\approx 15 \, \mathrm{TV}$ softening, due to low statistics.
In addition, the B/C ratio flattens at  $\approx 1 \mathrm{TeV/n}$. 
}
{We present a novel scenario for cosmic ray sources and transport in the Galaxy that may explain all of the observed spectral features.}
{The proposed scenario is based mainly on two assumptions.
First, in the Galactic disk, where magnetic field lines are mainly oriented along the Galactic plane, particle scattering is assumed to be very inefficient. Therefore, the transport of cosmic rays from the disk to the halo is set by the magnetic field line random walk induced by large scale turbulence.
Second,  we propose that the  spectral steepening at
$\approx 15 \, \mathrm{TV}$
is related to the typical maximum rigidity
reached in the acceleration of cosmic rays by the majority of supernova remnants, while we assume that only a fraction of sources, contributing to $\approx 10-20\% $ of the
cosmic ray  population, can accelerate particles up to $\sim$ PV.}
{We show that, within this framework, it is possible to reproduce 
the proton and helium spectra from GV to multi-PV, and the $p$/He ratio,  the spectra of cosmic ray  from lithium to iron, 
the  $\Bar{p}$ flux and the $\Bar{p}/p$ ratio and the abundance ratios B/C, B/O, C/O, Be/C, Be/O, Be/B. We also discuss the \isotope{10}{Be}/\isotope{9}{Be} ratio in view of the recent AMS02 preliminary measurements.
}
{}

\keywords{ISM:cosmic rays --
Diffusion -- Galaxy: disk -- Galaxy: halo}

   \maketitle
%
%-------------------------------------------------------------------

%%%%%%%%%%%%%%%%%%%%%%%%%%%%%%%%%%%%%%%%%%%%%%%%%%

%%%%%%%%%%%%%%%%% BODY OF PAPER %%%%%%%%%%%%%%%%%%

\section{Introduction}
\label{sec:intro}
The standard paradigm of the origin of cosmic rays (CRs) is based on three main pillars: \textit{i)} the bulk of primary CRs is accelerated  out of the interstellar medium (ISM), presumably up to the energy of the \textit{knee} ($ E \approx 3$ PeV) and beyond, %for protons,  
at the shocks of supernova remnants (SNRs). The acceleration mechanism, diffusive shock acceleration (DSA), is rigidity-dependent ($R$) and produces power-law CR spectra $\propto R^{-\alpha}$ ($\alpha \approx 2.0-2.4$);  \textit{ii)} CRs are diffusively confined in a magnetized  Galactic halo (GH) of height $H \approx $ few kpc. The CR residence time in such halo decreases with rigidity as $\tau_{\rm res}(R) = H^2/D(R) \propto R^{-\delta}$ ($\delta \approx 0.3-0.6$). 
The equilibrium spectrum is then $\propto R^{-\alpha -\delta}$ with $\alpha + \delta \sim 2.7$, which roughly describes the CR observed primary spectra; 
\textit{iii)} secondary CRs (such as Li, Be, B, $\Bar{p}$) are produced in CR interaction with matter during the repeated crossings of the Galactic disk (GD, height $h \sim 150$ pc, density $n_d \sim 1\, \rm cm^{-3}$) that CR perform before leaving the Galaxy.  The relative abundance of secondary to primary elements at high enough $R$ are expected to be proportional to the grammage (average amount of matter) traversed by CRs during the Galactic  propagation, namely, $X(R) = \Bar{n}\mu v \tau_{\rm res}(R)$, where $\Bar{n} = n_{d} h/H$ is the average ISM density in the propagation volume,  $\mu\sim 1.4\,m_p$ the mean mass of ISM nuclei,  $v\sim c$ the CR particle velocity. Note also that $\tau_{\rm d} \sim \left(\frac{h}{H}\right)\tau_{\rm res}$ corresponds to the residence time in the disk, which can be then derived directly from the measured grammage $X = n_d\mu v \tau_{\rm d}$. 
The total residence time in the Galaxy, $\tau_{\rm res}$, can be determined by the measure of the grammage combined with that of the short-lived radioisotope \isotope{10}{Be}, which can be used as a CR clock \citep{Ptuskin-Soutoul1998-review-CR-clocks}. 

By doing so, one gets $\tau_{\rm res}(R) \approx \tau_0 (R/{10\, \rm GV})^{-\delta}$, with $\tau_0 \approx 30\, H_{\rm kpc}\, \rm Mry$ and $\delta \approx 0.3-0.6$ \citep{Blasi-2013-review, Gabici-2019-review}.

In recent times, an impressive amount of high-precision CR data and refined theoretical models and simulations became available. This has challenged the standard picture in several ways \citep[see e.g.][for a critical review of the SNR paradigm]{Gabici-2019-review}. Some of the main issues that emerged are briefly illustrated in the remaining of this Section.
%%%%%%%%%%%%%%%%%%%
\subsection*{Cosmic rays up to $\approx 1$ TV:  propagation and grammage}
The PAMELA \citep{PAMELA-2011-Hard} and AMS02 \citep{AMS02-2015-hard} experiments reported on a spectral hardening at $R\sim 300$ GV in the proton and helium spectra. Such feature is explained by a change in the rigidity dependence of the diffusion coefficient \citep{Tomassetti-2012, Aloisio-Balsi-2013-self, Genolini-2017-hard-secondary}, due, for instance, to a change from scattering on  self-generated turbulence to pre-existing turbulence. This hypothesis as been recently confirmed by the measurement of
an  hardening in the  spectra of  secondary CRs (e.g. Li, Be, B) which is  twice  more pronounced than that observed  in the proton and  helium  spectra \citep{AMS02-5yr-LiBeB}. 
Alternative explanations invoking the presence of a nearby source \citep{Thoudam-2012} or non-linear effects in particle acceleration  \citep{Ptuskin-2013-hard-nonlin-acc, Recchia-2018-pamela} seem now to be disfavored.

More recent data revealed the presence of another spectral structure, i.e., a flattening of the B/C ratio at $\approx$ 1 TeV/n \citep{AMS02-7yr, CALET-2020-C-O, DAMPE-2022-BC-BO}.  
The measured value is $\rm B/C ({TeV/n}) \approx 0.04$ (see Fig.~\ref{fig:B-C-O-ratio})
which can be converted into a grammage equal to \citep{Evoli-2019-ams02}:
\begin{equation}\label{eq:grammage-TeV-BC}
    X({\rm TeV/n}) \approx 0.8 \, \rm g/cm^2,
\end{equation}
which in turn corresponds to a residence time in the disk
\begin{equation}\label{eq:restime-TeV-BC}
    \tau_d({\rm TeV/n}) \approx \frac{4\times 10^{5}}{n_d}\, \rm yr.
\end{equation}

The $\Bar{p}/p$ ratio also appears rather flat above $\approx 30$ GV (\citealt{AMS02-2016-pbar}, see also Fig.~\ref{fig:pbar})  and  a similar behavior is found in the $e^{+}/p$ ratio. The latter is related to the issue of the rising positron fraction \citep{AMS02-2013-pos-frac}, which however is typically explained with the contribution to positrons form middle-aged pulsars (see e.g. \citealt{Hooper-2009-pulsar}). 

The flat scaling with energy of these abundance ratios remains unexplained even considering a slower rigidity dependence of the diffusion coefficient above $R\sim 300$ GV, and posed the question on whether the grammage is uniquely accumulated during the large-scale propagation of CRs in the Galaxy or if it may include a relevant contribution accumulated inside or in the vicinity of CR sources.

Having in mind SNRs as the main accelerators of CRs, the source or near-source contribution to the grammage may come from: \textit{i)} secondaries produced at the shock that either leave rapidly the acceleration site or  get reaccelerated before escaping. In the former case the secondary spectrum is similar to the primary spectrum, while in the latter case the secondary spectrum is harder (see e.g. \citealt{Tomassetti-Donato-2012-sec-src, Bresci-2019-src-grammage, Mertsch-2021-sec-src-old-SNR} and references therein); \textit{ii)} 
seed secondary CRs (produced by spallation in the ISM) that are re-accelerated by shocks, a process that produces secondary spectra similar to primaries \citep{Tomassetti-Donato-2012-sec-src, Bresci-2019-src-grammage}; \textit{(iii)} production of secondaries by CRs which are confined in the vicinity of SNRs. In this case, the spectra of secondaries are typically steeper than those of primaries, depending on the rigidity-dependent residence time, as due e.g. a to self-confinement processes \citep{Dangelo-2016-grammage, Recchia-2022-grammage}.

The first two effects are unavoidable and potentially significant above $R \approx$ TV, as shown in recent works \citep{Bresci-2019-src-grammage, Mertsch-2021-sec-src-old-SNR}, but their actual relevance is a matter of debate. Indeed, their impact on secondary production ultimately depends on the very poorly understood time evolution of the maximum rigidity $R_{\rm max}(t)$ at SNR shocks, on the duration of effective acceleration up to and above TV, on the density of the medium in which SNRs explode (for the first process) and on the volume spanned by SNRs during efficient acceleration  up to TeV energies (relevant for re-acceleration) \citep{Tomassetti-Donato-2012-sec-src}.
As for the third process, its relevance depends on the effectiveness of confinement in the source region, the density of the background and its level of ionization and turbulence, and on the energy dependence of the residence time (see e.g. the discussion by \citealt{Recchia-2022-grammage} and references therein).

In this context \citet{Cowsik-2010-cocoon} proposed a radically different point of view (see also \citealt{Cowsik-2016-cocoon}), where CRs up to $\sim $ TV are confined for long and $R$-dependent times in dense cocoons around accelerators, and escape from the Galaxy in a $R$-independent way. The residence time in the Galaxy is assumed to be rather small, so that positrons would not suffer much from energy losses, an hypothesis also discussed in detail by \citealt{Lipari-2017-prop-new}. In this scenario the authors try to explain the flat $\Bar{p}/p $ and $e^{+}/p$.  
Such class of models are surely very interesting, but the formation of cocoons and the possibility of long confinement times in such regions remains an open issue and may be plagued by the same difficulties of models of CR self-confinement around sources \citep{Recchia-2022-grammage}. Moreover it requires a number of conditions that needs to be justified theoretically,
such as a  steep injection spectrum of protons, little energy losses for leptons and at the same time  different source spectra for nuclei and leptons.

\subsection*{Cosmic rays in the  multi TV - PV range: sources and maximum energy}
While there is a wide consensus on the fact that SNRs may accelerate the bulk of CRs, the  maximum $R$ that can be reached in the acceleration process is matter of intense debate. The most popular hypothesis is that the spectral steepening at $\sim 3$ PeV (the \textit{knee}) is the result of  SNRs  accelerating protons up to such energy, and heavier nuclei up to energies $Z$ (ion charge)  times larger. 
The detection of gamma-rays from SNRs of GeVs to multi-TeV confirmed that efficient particle acceleration is indeed taking place there, but so far there has been
no clear evidence of PeV acceleration (see e.g. \citealt{Gabici-2016-SNR-PeVatrons, Cristofari-2021-review-PeV}). 

Theoretical models show that particle acceleration to PeV energies requires a substantial amplification of  magnetic field at the shock. Remarkably, X-ray measurements at SNR shocks detected values of $B$ much larger than the typical $\rm \mu G $ scale of the ISM, even at the level of hundreds $\rm \mu G $ (see e.g.   \citealt{Vink-2012-SNR-X-review}).  However, current theories show that in SNRs a PeVatron phase may be present only in a subclass of peculiar objects, which explode in a dense environment and have large shock speeds, and that such phase may last for short times, of the order of $\rm \approx~100$s yrs \citep{Bell-2013-acc-escape, Cristofari-2020-low-rate-PeV, Cristofari-2021-review-PeV}. 

Alternative scenarios propose 
powerful stellar winds and supernovae in compact clusters of massive stars and OB associations (see e.g. \citealt{Bykov-2020-review-acc-star-forming} for a review) as potential PeVatrons and as a possible explanation for the $\rm ^{22}Ne/ ^{20}Ne$ anomalous isotopic ratio, with a overall contribution to the CR population  
of $\approx 5-10\%$ \citep{Tatischeff-2021-composition, Vieu-2023-PeV-massive-star-cluster}. 

It is also possible that only a subset of SNRs accelerates particles up to PeV energies and beyond.
In this case, 
a crucial issue would be to determine the maximum energy reached in "typical" SNRs,  the overall CR luminosity of such typical sources and PeV sources, and the possible spectral features emerging from the presence of (at least) these two classes of sources (typical SNRs and PeV SNRs). Interestingly, recently  DAMPE \citep{DAMPE-2019-H, DAMPE-2021-He} found an additional spectral steepening in the proton and helium spectra at  $\rm \approx 15~TV$, between 
the spectral hardening at $\rm \approx 300~GV$ and the \textit{knee}. 
Similar to the case of the $\rm \approx 300~GV$ break, possible explanations that have been put forward involve  a nearby source \citep{Fornieri-2021PhRvD-DAMPE-10TeV-loca-src, Malkov-2022-DAMPE-bump-10TeV, Liu-2022-DAMPE-10TeV-local-src, Lagutin-2023-DAMPE-break-pHe-local-src}, effects related to the acceleration process \citep{Ohira-2022PhRvD-perp-shocks-10TeV} or a change in the propagation regime \citep{Chernyshov-2023-NLLD}.

\subsection*{Cosmic ray propagation: the halo and the disk}
The CR diffusive transport in the GH is thought to be due to the particle scattering on magnetic inhomogeneities 
whose wavelength is
comparable to the particle’s Larmor radius, $r_L$, a process called resonant scattering  \citep{Skilling-1975-effect-waves}. The
rigidity dependence of the diffusion coefficient reflects the spectral
properties of the turbulence at the resonant scale (see e.g. \citealt{Fornieri-2021-MHD}). 

As for the GD, and in general for near-disk region, it is not clear whether CRs can be well scattered, due e.g. to a very effective  damping of the turbulence by ion-neutral friction (\citealt{Recchia-2022-grammage}, see also the discussion on the formation of the  diffusive halo  by \citealt{Evoli-2018-halo-formation}). In general, the GD is either regarded as an infinitesimally thin (compared to the GH) passive target for secondary production, where CR crossing takes place at the speed of light, or the propagation there is assumed to be diffusive with the same diffusion coefficient of the GH. 

However, while CR data at $R\lesssim \,\rm TV $ can be well reproduced with such assumptions, hints that the particle transport in the GD and in the GH may be substantially different emerged in a variety of observations and theoretical investigations.
For instance, the extended $\gamma-$ray halos detected around the Geminga and Monogem pulsars \citep{HAWC-2017-Geminga} and likely produced my runaway electrons and positrons of $\approx 20-200\, \rm TeV$, suggest that the CR diffusion coefficient in the disk for such high energy particles  may be $\approx 100-1000$ times smaller than that inferred from B/C and extrapolated to high energies. Moreover, \citet{Mertsch-2023-Be10-low-diff} recently suggested that 
a diffusion coefficient in GD a few times smaller than in the GH may help in reproducing the preliminary $\rm ^{10}Be/ ^{9}Be$ ratio measurements reported by AMS02.

Finally, the particle propagation 
in a magnetized environment is generally anisotropic, with substantial differences between the particle motion along and across the average magnetic field \citep{Shalchi-2020-perp-transp-review, Mertsch-2020-test-particle-transp-review}.  If the intensity of the turbulent
field is significantly smaller than the mean large-scale field, particle diffusion primarily takes place along field lines. The particle propagation perpendicular to the average field is a  combination of the random  motion of field lines (field line random walk, FLRW), induced by large scale fluctuations and usually dominant, and of perpendicular (cross-field) diffusion.
Interestingly, the large scale Galactic magnetic field in the GD is mainly parallel to the Galactic plane (GP) while in the GH is preferentially perpendicular to the plane. The transition between these two configurations also depends on the galactocentric distance and has been invoked in the literature as a possible solution to the CR gradient problem \citep{Cerri-2017-gradient-problem-anisotropic}. 

\subsection*{Alternative framework for the interpretation of the cosmic ray data up to the \textit{knee}}
Starting from the issues discussed above, in this paper we propose a novel interpretation of available CR data based on the following assumptions:
\begin{enumerate}
    \item the spectral steepening in the proton and helium spectra at $\sim 15$ TV, reported by DAMPE,
    is related to the typical $R_{\rm max}$ reached in the acceleration of CRs by the majority of SNRs. Only a fraction of sources, contributing to $\approx 10-20\%$ of the CR population can accelerate particles up to PeV energies;\\
    \item  
    the propagation of CRs in the GD is assumed to be
    characterized by inefficient scattering along field lines, assumed to be mostly parallel to the GP,  and by a motion perpendicular to the mean field (and disk) dominated by the FLRW induced by the large-scale turbulence in that region. As we will show in  detail in Section~\ref{sec:prop-disk}, if the scattering mean free path is larger than a characteristic length $L_{RR}$ (Eq.~\ref{eq:LRR}, typically a few times the field coherence length), beyond which a particle trajectory is no more correlated with its initial field line, the motion perpendicular to the mean field (and so to the GP) is found to be diffusive,  roughly rigidity independent and   with an effective diffusion coefficient substantially smaller than that along field lines;
\\
    \item  the CR transport in the kpc-scale GH is the typical diffusive motion, with the typical rigidity-dependent diffusion coefficient commonly adopted in a large variety of propagation model.
\end{enumerate}

We show that, with these assumptions, it is possible to reproduce quite well: \textit{i)}
the proton and helium spectra from GV to multi-PV rigidities and the $p$/He ratio; \textit{ii)} the available spectral data for nuclei ranging from lithium to iron; \textit{iii)} 
the  $\Bar{p}$ flux and the $\Bar{p}/p$ ratio; \textit{iv)} the  abundance ratios B/C, B/O, C/O, Be/C, Be/O, Be/B.
Moreover, we will also discuss the \isotope{10}{Be}/\isotope{9}{Be} ratio in view of the recent AMS02 preliminary measurements.

Such results are obtained with a relatively small number of free parameters, 
without introducing spectral breaks in the injection spectra or in the GH diffusion coefficient. However, similar to previous works (see e.g. \citealt{Evoli-2019-ams02}),   the  required injection spectra of protons, helium and  heavier nuclei are found to be slightly different, with protons systematically steeper than nuclei, and helium systematically flatter than  protons and nuclei (see Table~\ref{tab:param}).\\

The paper is organized as follows. In Section~\ref{sec:prop-disk}
we discuss  the microphysics of the Galactic propagation of CRs and describe of model for the CR transport in the GD. In Section~\ref{sec:prop-setup} we report  the  details of 
the proposed CR propagation setup and discuss the analytic 
solution of the transport equation,  which is derived in Appendix~\ref{sec:appendix-stable-nuclei} and \ref{sec:appendix-unstable-nuclei}. 
In Section~\ref{sec:results} we discuss  our results and compare our predicted CR fluxes with available data. We draw conclusions in Section~\ref{sec:conclusion}.

\section{Model of cosmic ray propagation in the disk}
\label{sec:prop-disk}

The diffusive propagation of CRs in the ISM is dictated by their interaction with the turbulent Galactic magnetic field. The CR random motion is  thought to be due to scattering on plasma waves, most notably  magneto-hydrodynamic (MHD) incompressible Alfvénic and compressible (fast and slow) magnetosonic fluctuations \citep{Mertsch-2020-test-particle-transp-review, Lazarian-2023-review-MHD-transport}, at the resonant scale $\lambda \sim r_L$ (despite non-resonant interaction may also be relevant, \citealt{Lazarian-2023-review-MHD-transport}). 

MHD turbulence is ubiquitous in the interstellar space and is produced by astrophysical sources on scales $\approx 10-100$ pc. In order to be able to effectively scatter CRs, such turbulence has to cascade to resonant scales ($\ll$ pc for sub-PV particles) without being damped or becoming inefficient for scattering. 
In particular, a variety of studies showed that the scattering associated to Alfvénic turbulence is highly
suppressed due to the anisotropy of the cascade, while  magnetosonic modes can play a dominant role in CR diffusion \citep{Fornieri-2021-MHD, Lazarian-Xu-2021-mirror-diffusion, Lazarian-2023-review-MHD-transport}.
Alternatively, also CRs can produce  plasma waves through e.g.  the excitation of Alfvén waves at the resonant scale \citep{Skilling-1971}. At the same time, turbulence is also subject to a variety of damping processes, such ion-neutral damping in a partially ionized medium and the non-linear Landau damping \citep{Recchia-2022-grammage}. Such effects are typically very important in the GD while their impact is expected to be less dramatic in the GH (where ion-neutral damping does not operate). 

The actual CR transport in a magnetized and turbulent environment is typically way more complex than a purely isotropic diffusive motion and may be anisotropic with respect to the average magnetic field \citep{Mertsch-2020-test-particle-transp-review, Shalchi-2020-perp-transp-review}.
Indeed, let us assume that a region is embedded with a mean magnetic field $B_0$, on top of which there is a turbulent field $\delta B(k)$, with a given spectrum in wave-number $k$ characterized by a maximum (injection) scale $L_{\rm max} = 1/k_{\rm min}$ and a coherence length parallel ($L_{\parallel}$) and perpendicular ($L_{\perp}$) to $B_0$. In the following we assume for simplicity that the turbulence is such that $L_{\parallel} = L_{\perp} \equiv L_c$.
Moreover, let us assume that such perturbations are small, namely $\delta B/B_0 \ll L_{\perp}/L_{\parallel}$, a condition likely to be met in the ISM and necessary in order to apply the so-called quasi-linear theory of particle transport \citep{Kadomtsev-Pogutse-1979}. 
The perturbations at the resonant scale, $\delta B_{\rm res}$, produce a pitch angle scattering of the particle, which eventually result in a spatial diffusion \textit{along} field 
lines, characterized by the parallel diffusion coefficient $D_{\parallel}$. Such process is typically rigidity dependent, as a result of the spectral properties of the turbulence \citep{Mertsch-2020-test-particle-transp-review}. 
In the absence of scattering particles would simply gyrate in the local field.

The particle transport perpendicular to $B_0$  is a combination of the FLRW, namely the stochastic motion of field lines, which is due to fluctuations on large scale (much larger than the resonant scale) and of the diffusion across field lines due to particle jumping from one line to another \citep{Rechester-Rosenbluth-1978, Kirk-Duffy-Gallant-1996, Shalchi-2020-perp-transp-review} . When   $\delta B/B_0 \ll L_{\perp}/L_{\parallel}$ the latter process is much suppressed and the former can be approximated as a diffusion process characterized by a field line diffusion coefficient $D_m$.

Both incompressible  Alfvénic turbulence and compressive modes
can contribution to FLRW \citep{Shalchi-2021-FLRW-MHD}, depending on their specific properties and amplitude compared to the mean field.
If the power in  Alfvén modes is large, while that in compressive modes  is substantially  smaller, CRs would experience at the same time a weak scattering along field lines and an effective FLRW. 

Assuming that $B_0$ is along the z-axis and the x-y plane is perpendicular to $B_0$, the meaning of such quantity is that two field lines  that at  $z = 0$  pass in  the vicinity of a point ($x_0$, $y_0$),  are spread over a larger region at position $z$, whose size is characterized by a Gaussian distribution such that 
\begin{equation}\label{eq:FLRW-Dm}
    \langle (x-x_0)^2 \rangle = \langle (y-y_0)^2 \rangle = 2 D_m z.
\end{equation}
For a broad-band spectrum of turbulence an estimate of $D_m $ is given by \citep{Kadomtsev-Pogutse-1979, Kirk-Duffy-Gallant-1996} 
\begin{equation}\label{eq:Dm-estimate}
    D_m = \left(\frac{\delta B}{B_0}\right)^2 \frac{L_c}{4} = 0.25 \, \left(\frac{b^2}{0.1}\right) \left( \frac{L_c}{10\, \rm pc}\right) \, \rm pc,
\end{equation}
where $b^2 \equiv (\delta B/B_0)^2$.

In the absence of  scattering, the CR gyromotion leads to a motion along $B_0$  with a speed $v\approx c/2$ that only depends on the particle pitch angle and not on rigidity. In this limit, in a time $t$, a particle travels along $B_0$ by $z(t) \sim v\,t$. At the same time, the FLRW produces a perpendicular displacement 
\begin{equation}
    \langle (\Delta x)^2 \rangle = \langle (\Delta y)^2 \rangle = 2 D_m\, v\, t,
\end{equation}
namely a diffusive motion perpendicular to $B_0$ characterized by an effective diffusion coefficient
\begin{equation}\label{eq:D-FLRW-perp-gyro}
D^{\rm gyromotion}_{\rm FLRW, \perp} = D_m\, v.
\end{equation}

Instead, in the case of particle diffusion along field lines with parallel diffusion coefficient $D_{\parallel}$, $\, z(t) \propto \sqrt{D_{\parallel} t}$ and the perpendicular displacement reads
\begin{equation}
    \langle (\Delta x)^2 \rangle = \langle (\Delta y)^2 \rangle = 2 D_m\, \sqrt{D_{\parallel} t},
\end{equation}
with an effective perpendicular diffusion coefficient
\begin{equation}\label{eq:D-FLRW-perp-subdiff}
D^{\rm sub-diffusion}_{\rm FLRW, \perp} =  D_m\, \sqrt{\frac{D_{\parallel}}{t}}. 
\end{equation}
This is the compound diffusion regime illustrated by e.g. \citet{Rechester-Rosenbluth-1978}.
If a particle follows the same field line indefinitely, then the effective diffusion coefficient  perpendicular to the average field,
$D^{\rm sub-diffusion}_{\rm FLRW, \perp} \propto t^{-1/2}$ vanishes as t increases.  

However, small motions  perpendicular to
field lines can restore the diffusive behavior of particles because of the exponential divergence of neighboring field
lines \citep{Jokipii-Parker-1969, Rechester-Rosenbluth-1978}. The separation $d$ between two  neighboring lines increases on average with the distance along  the field line as
\begin{equation}
    d(l) \sim d(0)e^{l/L_k},
\end{equation}
where $L_k $ is the Lyapunov length.
Estimates of $L_k$ gives $L_k \approx L_c$ (see \citealt{Chandran-2000-cloud-mirror} and references therein). 

Consider a particle moving along a given field line. A small perpendicular displacement $d_0$ (induced e.g. by field gradients and scattering) brings the particle to a new close-by field line.  After traveling a distance 
\begin{equation}\label{eq:LRR}
   L_{RR} \approx L_k \ln (L_c /d_0) 
\end{equation}
 along this new field line, the particle would reach a  perpendicular distance $L_c$ from its initial field line. 
The \textit{Rechester-Rosenbluth} distance $L_{RR}$ \citep{Rechester-Rosenbluth-1978, Chandran-2000-cloud-mirror} represents the distance along the new field lines beyond which  the particle trajectory is no more correlated with its initial field line. This meas that beyond $L_{RR}$, even if the particle is scattered back, it will not  retrace its original path and it will drift across field lines moving on an new path, uncorrelated with the original one. Such loss of correlation allows particles to  escape the regime of compound diffusion.

The effective diffusion coefficient perpendicular to the mean field is given by
\begin{equation}
    D_{\rm eff, \perp} = \frac{1}{2}\frac{(\Delta R)^2}{\Delta t},
\end{equation}
where $(\Delta R)^2$ is the rms displacement perpendicular to the average field during each \textit{statistically independent} random step, corresponding to a parallel motion by $L_{RR}$.  If the scattering mean-free path, $\lambda_{\rm mfp}$ (see Eq.~\ref{eq:lmfp}), associated to $D_{\parallel}$ is $\lambda_{\rm mfp} \ll L_{RR}$, namely in the case of effective scattering, we get
\begin{equation*}
    \begin{cases}
    (\Delta R)^2 = 2D_m L_{RR}\\
    \Delta t = \frac{L_{RR}^2}{D_{\parallel}}
\end{cases}   
\end{equation*}
namely 
\begin{equation}
    D^{\rm scattering}_{\rm eff, \perp} =  D_{\parallel}\frac{D_m}{L_{RR}}.
\end{equation}

If the scattering mean-free path, $\lambda_{\rm mfp} \gtrsim L_{RR} $, namely in the case of inefficient  scattering, we get
\begin{equation*}
    \begin{cases}
    (\Delta R)^2 = 2D_m L_{RR}\\
   \Delta t = \frac{L_{RR}}{v}
\end{cases}   
\end{equation*}
namely 
\begin{equation}\label{eq:Deff-perp-collisionless}
    D^{\rm collisionless}_{\rm eff, \perp} =  D_m\, v \approx 3\times 10^{28}\left(\frac{D_m}{\rm pc}\right)\, \rm cm^2/s,
\end{equation}
which is the same value obtained in the assumption of a pure gyromotion (see Eq.~\ref{eq:D-FLRW-perp-gyro}).\\

The value of $L_{RR}$ depends on a variety of parameters, and in particular  on the type and level of turbulence, and is typically found to be of the order of a few times $L_c$ \citep{Casse-Lemoine-Pelletier-2001, Chandran-2000-cloud-mirror}. \\
 
When particles are released in a turbulent field, an effective isotropic diffusion process, with $D = D_{\parallel}/3$ \citep{Casse-Lemoine-Pelletier-2001, Subedi-Blasi-2017} may be reached due to a combination of parallel diffusion and perpendicular motion, on scales much larger than the field coherence length(s)  and on sufficiently long timescales. 
Such effective isotropic diffusion is likely achieved by CRs in the GH, a region whose size  is typically much larger than $L_c$ and where particles can be effectively scattered with 
little damping of plasma waves (due e.g to a fully ionized medium). This is also well in agreement with the standard picture of Galactic CR propagation described in the introduction. \\

As for the GD, the situation is much less clear. Indeed, on the one end,  effective damping processes \citep{Recchia-2022-grammage} may make scattering in the GD ineffective, either preventing the cascade to resonant scales and/or  hampering  turbulence that is produced already at such scales. This was recognized in pioneering works  on CR propagation \citep{Skilling-1971}  where the near-disk region was assumed to contain little resonant waves. Particles were assumed to free-stream in such region, while the ionized GH was considered as the actual scattering region. On the other hand, SNe explosion, massive star winds and other astrophysical phenomena may be able to keep some level of resonant waves, 
or even  fill the GD with highly turbulent regions where the diffusion coefficient can be well smaller than in the GH \citep{Mertsch-2023-Be10-low-diff}.
In general, the 
actual particle transport in the GD  is ignored in the Galactic CR propagation, and is  either assumed to proceed  with the  diffusion coefficient of the GH, either the disk is treated as a passive target where particles undergo free-streaming.\\

Here we propose a novel picture for the CR transport in the GD 
based on the following assumptions: 
\begin{enumerate}
    \item particle scattering is ineffective in the GD due to strong wave damping (and/or to a small power of compressive modes, \citealt{Fornieri-2021-MHD, Lazarian-Xu-2021-mirror-diffusion}), 
    so that the scattering mean free path  along the GP is $\lambda{\rm mfp} \gtrsim L_{RR}$ ;
    \item particle transport perpendicular to the GD mostly takes place perpendicular to the average Galactic magnetic field and is dominated by turbulence on scales much larger than $r_L$.
\end{enumerate}
The latter hypothesis is motivated by the observation that large scale Galactic magnetic field in the disk is mainly parallel to the GP  while in the halo is preferentially perpendicular to the GP \citep{Cerri-2017-gradient-problem-anisotropic}. With a mean field preferentially parallel to the  GP plane, CRs in the GD have to move perpendicular to field lines in order to enter the GH. 

As shown in the discussion above, in such scenario the global CR propagation in the GD perpendicular to the GP can be treated as an effective \textit{rigidity-independent} diffusion, regulated by the field line diffusion coefficient $D_m$, as illustrated in Eq.~(\ref{eq:Deff-perp-collisionless}).
As reported in Table~\ref{tab:param}, in Eq.~(\ref{eq:Dm-estimate}) we adopt $L_c = 10\,\rm pc$ and $b^2 = 0.4$, which corresponds to $D_m = 1\,\rm pc$, namely 
$D_{\rm eff, \perp} \sim 3 \times 10^{28}\, \rm cm^2/s$. Considering that $L_{RR}$ is few times $L_c$, we assume as referene values $L_{RR}\sim 50-100\, \rm pc$.

Interestingly, such effective diffusion would correspond to a residence time in the GD of the order
\begin{equation}\label{eq:Tmin}
    \tau_{d}^{\rm min} \sim \frac{h^2}{D_{\rm eff, \perp}} \approx 2\times 10^5 \left(\frac{h}{150\, {\rm pc}}\right)^2 \left(\frac{\rm pc}{D_m}\right)\, \rm yr,
\end{equation}
a value close to that found in Eq.~(\ref{eq:restime-TeV-BC}) from the B/C at $\approx 1\,\rm TeV/n$. 

The superscript $min$ has been added in the definition of residence time provided in Eq.~(\ref{eq:Tmin}) to indicate that this is indeed a {\it minimum} value for such physical quantity.
It corresponds to the residence time of particles that cross diffusively the disk once, and after entering the halo they diffusively escape from the Galaxy without ever returning to the disk.
The corresponding minimum (and rigidity-independent)  grammage is then:
\begin{equation}\label{eq:Xmin}
    X_{\rm min} \approx 0.4\, n_d \left(\frac{h}{150\, {\rm pc}}\right)^2 \left(\frac{\rm pc}{D_m}\right)\, \rm g/cm^2.
\end{equation}
This is the grammage experienced by particles of large enough rigidity, which diffuse fast enough in the halo
(the diffusion coefficient in the GH increases with $R$) to never come back to the disk.\\

At the same time, CRs experience a much larger diffusion coefficient along the GP, with $\lambda_{\rm mfp} \gtrsim L_{RR}$. Taking  
$L_{RR} \sim 50-100\, \rm pc$, $\lambda_{\rm mfp} \gtrsim 50-100\, \rm pc$ implies $D_{\parallel} \gtrsim 1.5-3 \times 10^{30}\, \rm cm^2/s$. 
Such values are compatible  with scattering on compressive modes characterized by an injection scale $L\sim 10\, \rm pc$,  $\delta B/B_0 \lesssim 0.1$ and a plasma $\beta \sim 0.1$ (see \citealt{Fornieri-2021-MHD} and references therein).
Thus, during the time $\tau_d^{\rm min}$, CRs would cover a distance $d_{\parallel}$ along the plane 
\begin{equation}\label{eq:par-distance-GD}
    d_{\parallel} = \sqrt{ \frac{D_{\parallel}}{D_{\rm eff, \perp}}} \, h \gtrsim 1\, \sqrt{\frac{\lambda_{\rm mfp}}{50\,\rm pc}} \sqrt{\frac{1\,\rm pc}{D_m}} \frac{h}{150\, \rm pc}\, \rm kpc.
\end{equation}

\section{Propagation setup}
\label{sec:prop-setup}

In this Section we present steady-state solutions of the transport equation for CR nuclei in the GD and GH.
We consider a 1D problem along the coordinate $z$ directed orthogonally to the GD and centered on it.
In the GH such description is justified by 
our assumption that field lines are preferentially perpendicular to the GP. Moreover, a weak dependence on the radial coordinate is also confirmed by the small CR anisotropy and by the small CR  gradient along the GP, as inferred from $\gamma-$ray observation (see e.g. \citealt{Evoli-2019-ams02}).
In the GD, a 1D description is justified by our assumption that the particle motion perpendicular to the GP is much slower than  along the plane, in line with the small radial CR gradient.
%%%%%
\begin{figure}
\includegraphics[width=\columnwidth]{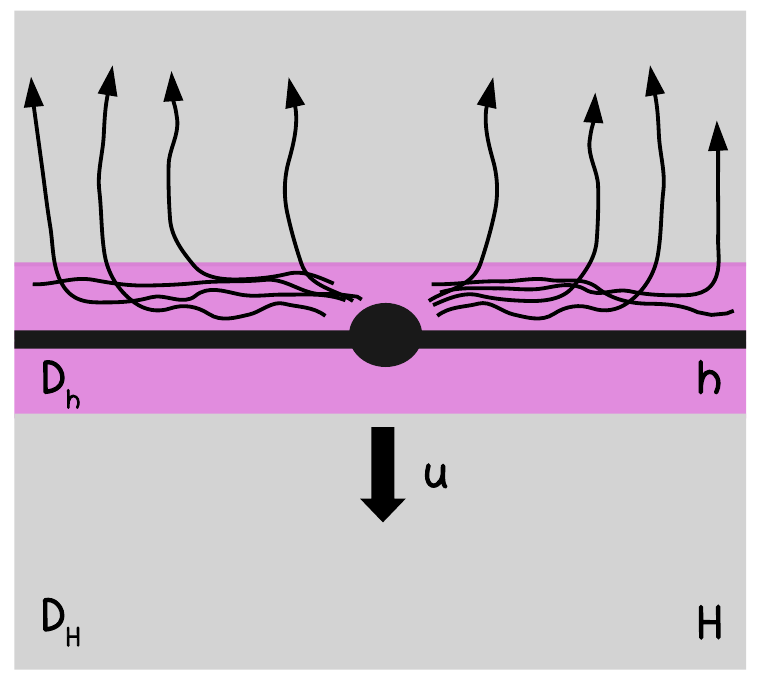}
\caption{Schematic representation of the propagation setup. The size of the \textbf{GD} (violet region) is  $h\sim 150\, \rm pc$ and magnetic field lines are preferentially parallel to the GP. The CR transport perpendicular to the plane is dominated by the large scale turbulence and encompassed in the rigidity-independent diffusion coefficient $D_h$; the size of the \textbf{GH} (grey region) is  $H\sim 4\, \rm kpc$ and magnetic field lines are preferentially perpendicular to the GP. The CR transport perpendicular to the plane is dominated by resonant scattering on plasma waves, encompassed in the rigidity-dependent diffusion coefficient $D_H$, plus the contribution of advection in a wind of velocity $u$.}
\label{fig:double-halo}
\end{figure}
%%%%%%%%%%%%%%%%%%
We provide in the following a detailed description of the propagation setup, which is also illustrated in Fig.~\ref{fig:double-halo}.

\subsection*{Transport equation}
Consider the phase-space
density of CRs $f_{\alpha}(z, p)$, as a function of the particle momentum $p$ and position $z$,  for a nucleus of species $\alpha$ of mass number $A_{\alpha}$ and atomic number $Z_{\alpha}$. Since  the kinetic energy per nucleon, $E_k$, is roughly conserved in the spallation processes, we express the transport equation  in terms of the particle flux  $I_{\alpha}(z, E_k) = A_{\alpha}\, c\,p^2 f_{\alpha}(z, p)$,
where $p = A_{\alpha} \sqrt{E_k^2 +2 m_p E_k}$ \citep{Bresci-2019-src-grammage}: 
\begin{align}
\label{eq:transp}
&    -\frac{\partial}{\partial z} \left[D_{\alpha} 
 \frac{\partial I_{\alpha}}{\partial z} - u I_{\alpha}\right] - \frac{d\,u}{d\,z} \frac{I_{\alpha}}{3} \frac{\partial\ln (I_{\alpha}\,p)}{\partial \ln p} = \\ \nonumber
 & -\, \frac{I_{\alpha}}{\tau_{r}} \,-\, 2\,h_d\,n_d v(E_k)\sigma_{\alpha}(E_k)\delta(z) I_{\alpha} \,+\,
c A p^2 q_{0 \alpha}\delta(z) \\ \nonumber
 & +\,  \sum_{\beta > \alpha} 2\,h_d\,n_d v(E_k) \sigma_{\beta\alpha}(E_k) \delta(z)I_{\beta}(E_k, z).
\end{align}
The first term in the LHS describes diffusive-advective transport while the second represents the adiabatic losses. 
The first term in the RHS describes the radioactive decay (relevant  for \isotope{10}{Be}), the second term the spallation of nucleus $\alpha$ to lighter nuclei, the third the injection from SNRs and other astrophysical sources, the fourth the 
contribution to nucleus $\alpha$ from the spallation of heavier nuclei $\beta$. 
In order to simplify the solution of the transport equation, we assume that the particle injection from sources and the spallation/production of nuclei take place in a very thin gaseous disk (hydrogen density $n_d = 1\, \rm cm^{-3}$, half-height $h_d = 150\,\rm pc$, radius $R_d = 15\, \rm kpc$), so that the corresponding terms in the transport equation can be approximated with a delta function, $\propto \delta(z)$.
In Eq.~\ref{eq:transp}, we neglect ionization losses since they are relevant only at $R\lesssim$ a few GV (\citealt{Evoli-2019-ams02} and references therein), where the observed spectrum is largely affected by the solar modulation (see Sec.~\ref{sec:results}). 

\subsection*{Injection spectrum}
The injection spectrum $q^{i}_{0\alpha}$ of a nucleus $\alpha$ from a given class $i$ of astrophysical sources is modeled as a power-law (in rigidity-momentum) plus cut-off at given rigidity $R_{{\rm max}, i}$
\begin{equation}\label{eq:inj-spec-general}
q^{i}_{0\alpha} = \frac{\epsilon_{\alpha} L_i}{\pi R_d^2\, \mathcal{G}(\gamma) c(m_p c)^4} \left(\frac{p}{m_pc}\right)^{-\gamma} e^{-R/R_{{\rm max}, i}}, 
\end{equation}
where $\epsilon_{\alpha} L_i$  the fraction of average source luminosity channeled  to CR nuclei of type $\alpha$, and $\mathcal{G}(\gamma) = 4\pi\int_0^{\infty} dx x^{2-\gamma}\exp^{-x/x_{max}}\left[ \sqrt{x^2+1} -1 \right]$ is a normalization factor.
The quantity $\epsilon_{\alpha}$ is determined by fitting the AMS02 flux for the given nucleus  at a reference rigidity, e.g. $R = 50 \, \rm GV$.\\

Following the discussion of Sec.~\ref{sec:intro}, we assume that the bulk 
of CRs are accelerated in \textit{typical} SNRs,  up to a typical cut-off rigidity $R^{\rm bulk}_{\rm max} = \, 50\,\rm TV$,  chosen in order to reproduce the spectral steepening found in the 
DAMPE data \citep{DAMPE-2019-H, DAMPE-2021-He} in the multi-TV range.
The injection spectrum for such population is given by
\begin{equation}\label{eq:inj-spec-bulk}
q^{\rm bulk}_{0 \alpha} = \frac{\epsilon_{\alpha} E_{SN} \mathcal{R}_{SN}}{\pi R_d^2\, \mathcal{G}(\gamma) c(m_p c)^4} \left(\frac{p}{m_pc}\right)^{-\gamma} e^{-R/R^{\rm bulk}_{\rm max}}, 
\end{equation}
where $E_{SN} = 10^{51}\, \rm erg$ and $\mathcal{R}_{SN} = 1/30 \, \rm yr^{-1}$ are the total explosion energy  and  rate of SNe, respectively.

We then assume that only a subclass of sources can accelerate particles up to PeV energies. 
We model the spectrum of CRs injected by such population as 
\begin{equation}\label{eq:inj-spec-PeV}
q^{\rm PeV}_{0\alpha} = \epsilon_{\rm bulk}^{\rm PeV}\left[ \frac{\epsilon_{\alpha} E_{SN} \mathcal{R}_{SN}}{\pi R_d^2\, \mathcal{G}(\gamma) c(m_p c)^4} \left(\frac{p}{m_pc}\right)^{-\gamma} \right]e^{-R/R^{\rm PeV}_{\rm max}}, 
\end{equation}
namely, we assume the  slope to be the same for the two populations but we here set $R^{\rm PeV}_{\rm max} = \rm 5\,PV $ and we add a normalisation factor equal to $\epsilon_{\rm bulk}^{\rm PeV}  \approx 0.15$   of the overall energy input of typical SNRs.

The total injection spectrum is then given by 
\begin{equation}\label{eq:inj-spec-tot}
q_{0\alpha} = q^{\rm bulk}_{0\alpha} + q^{\rm PeV}_{0\alpha}.
\end{equation}

\subsection*{Spallation and production cross-section}
We assume that CR nuclei interact with the ISM medium,  made mostly of hydrogen (H) and a fraction $f_{\rm He}\approx 10\%$ of helium (He). 

For the total spallation cross-section, $\sigma_{\alpha}(E_k) = \sigma^{\rm H}_{\alpha} + f_{\rm He}\sigma^{\rm He}_{\alpha}$, of CR nuclei heavier than He we adopt the parametrization 
by \citet{Tripathi-1999}, described in detail by \citet{Evoli-2018-DRAGON-II} and also implemented 
in DRAGON-II\footnote{\url{https://github.com/cosmicrays/DRAGON2-Beta_version}} and USINE\footnote{\url{https://dmaurin.gitlab.io/USINE/}} \citep{USINE-2020}.
For He nuclei we adopt the cross-sections reported by \citet{Coste-2012-light-nucl}.

As for the production cross-section,  $\sigma_{\beta\alpha}(E_k) = \sigma^{\rm H}_{\beta \alpha} + f_{\rm He}\sigma^{\rm He}_{\beta \alpha}$, in general we adopt the OPT22 parametrization available for the GALPROP code\footnote{\url{http://galprop.stanford.edu}} \citep{GALPROP-Porter-2022}, making use of the tables provide in the USINE code.   
For the channel \isotope{4}{He} $\rightarrow$ \isotope{3}{He} we use the cross-section reported by \citet{Coste-2012-light-nucl}. Limited to the production of Li, Be and B by CNO nuclei, we adopt the results by \citet{Evoli-2019-ams02}, using the tables  reported in the DRAGON-II code.

In is important to remark that the energy-dependent production of a given nucleus by the spallation of heavier nuclei is affected by rather large uncertainties, typically of the order of $\sim 20\%$, but often even larger. Such uncertainties are related to the quality and amount of data (if available) for a given production cross-section, and to the difficulty in  assessing the contribution  of all possible production channels, including that of \textit{ghost} nuclei (short lived isotopes).
A detailed discussion of such issue is beyond the scope of the present paper and we refer the reader to excellent works by \citet{ Genolini-2018-xsec-prod, Evoli-2019-ams02, Genolini-2023-xsec-prod}. 

\subsection*{Antiprotons} 
The production of secondary $\Bar{p}$ in the interaction of CR protons and He with the ISM (H and He target) is treated following  \citet{Korsmeier-2018-pbar-production}.
In this case, the last term of Eq.~(\ref{eq:transp}) can rewritten as 
\begin{equation}
    2h_d n_d\,v(E_k)\,\delta(z)\, Q_{\Bar{p}}(E_k).
\end{equation}
Here $Q_{\Bar{p}}(E_k)$ is the $\Bar{p}$ source term
\begin{equation}\label{eq:pbar-src-term}
    Q_{\Bar{p}}(E_k) = \sum_{i = {\rm p, He} \atop j= {\rm H, He}}\int_{T_{\rm th}}^{\infty} d\,E_{ki} \;  f_j\; I_i(E_{ki}) \frac{d\sigma_{ij}}{d\,E_k}(E_{ki}, E_k),
\end{equation}
where $E_k$ and $E_{ki}$ are the kinetic energy per nucleon of the $\Bar{p}$ and of the incoming CR, respectively,  $I_i(E_{ki})$ the incoming CR flux, $f_j = 1(0.1)$ for H(He) target, and $\frac{d\sigma_{ij}}{d\,E_k}$ is the differential production cross section, provided in tabular form in the supplemental material of \citet{Korsmeier-2018-pbar-production}.

As for the quantity $\sigma_{\alpha}(E_k)$ of Eq.~(\ref{eq:transp}), we use the total inelastic cross section for the $\Bar{p}-p$ interaction, that can be computed as the difference between the total and elastic cross sections provided by \citet{rpp-Workman-2022}.  
 
\subsection*{Diffusion and advection} 
We define the diffusion coefficient and advection velocity in the GD (subscript $h$) and GH (subscript $H$) as:
\begin{equation}\label{eq:D-def}
        D = \begin{cases}
      D_h & \text{if $0 < z \leq h $ }\\
      D_H & \text{if $ h < z \leq h+H $ }
    \end{cases}   
    \end{equation}
    \begin{equation}\label{eq:u-def}
        u = \begin{cases}
      0 & \text{if $0 < z \leq h $ }\\
      u & \text{if $ h < z \leq h+H $ }.
    \end{cases} 
    \end{equation}
    
The GD is assumed to have half-height $h=h_d = 150\, \rm pc$ and the transport of particles there is rigidity-independent, with diffusion coefficient $D_h = D_{\rm eff, \perp}^{\rm collisionless}$ defined in Eq.~(\ref{eq:Deff-perp-collisionless}). Note that the size of the GD and of the gaseous disk do not need necessarily to coincide, but we make this assumption here for simplicity. 

The GH is assumed to have half-height $H= 4\, \rm kpc$, the Galactic wind speed there is $u= 40\, \rm km/s$, and the particle (rigidity-dependent) diffusion coefficient is set equal to
\begin{equation}
    D_H(R) = D_0 \left( \frac{R}{\rm GV}\right)^{\delta}, 
\end{equation}
with $D_0 = 10^{28} \rm cm^2/s$ and $\delta = 0.7$. \\
Beyond a distance $z=H+h \equiv H^{*}$ from the GP, CRs are assumed to free escape, namely 
\begin{equation}\label{eq:free-esc}
    I_{\alpha}(H^{*}, E_k) = 0.
\end{equation}

\subsection{Analytic
solution for stable nuclei}
\label{sec:solution-stable}
In Appendix~\ref{sec:appendix-stable-nuclei} and~\ref{sec:appendix-unstable-nuclei} 
we derive the analytic solution of Eq.~(\ref{eq:transp}) in the disk,  $I_{\alpha 0}(E_k)$,
in the case of stable and unstable nuclei, with the conditions specified in 
Eqs.~(\ref{eq:D-def}, \ref{eq:u-def}, \ref{eq:free-esc}), and neglecting adiabatic losses (see Appendix~\ref{sec:appendix-adiabatic} for a discussion).

Here we report the solution for stable nuclei and discuss its main characteristics:
\begin{equation}\label{eq:solution-stable-I0}
   I_{\alpha 0}(E_k)  = \frac{\tau_{\alpha}^{hH}}{1 + n_d\,\frac{h_d}{h}\, v(E_k)\sigma_{\alpha}\tau_{\alpha}^{hH}} \times
    \left[\frac{1}{2h}\mathcal{Q}_{\alpha, \rm src} + n_d \frac{h_d}{h}\mathcal{Q}_{\alpha, \rm spall} \right],
\end{equation}
where 
\begin{equation}
    \begin{cases}
    \mathcal{Q}_{\alpha, \rm src} \equiv c A p^2 q_{0\alpha}\\
    \mathcal{Q}_{\alpha, \rm spall} \equiv \sum_{\beta > \alpha} \, v(E_k) \sigma_{\beta\alpha} (E_k)I_{\beta}(E_k)
\end{cases}  
\end{equation}
are the injection from sources and the injection from spallation of heavier nuclei (per ISM atom),  respectively.

The term $\tau_{\alpha}^{hH}$ can be interpreted as the residence time in the GD, as due to diffusion in the GD itself and  to the repeated crossings of the region induced by scattering in the GH, and reads
\begin{equation}\label{eq:tau-hH-def}
    \tau_{\alpha}^{hH} \equiv \frac{h^2}{D_h} + \frac{h\,H}{D_H} \frac{1-\exp^{-\frac{u\,H}{D_H}}}{\frac{u\,H}{D_H}}. 
\end{equation}
We also define the grammage 
\begin{equation}\label{eq:gramm-def}
    X_{\alpha}(E_k) = \left(n_d \frac{h_d}{h} \right)\, \mu v(E_k)\tau_{\alpha}^{hH},
\end{equation}
and  the critical grammage $X_{{\rm cr}\,\alpha} \equiv \mu/\sigma_{\alpha}$, where $\mu = 1.4 m_p$ is the mean mass of the ISM gas. The term $n_d\,h_d/h$ represents the ISM density averaged over the GD.\\
In terms of the grammage, Eq.~(\ref{eq:solution-stable-I0}) can be rewritten as
\begin{equation}\label{eq:solution-stable-I0-gramm}
   I_{\alpha 0}(E_k)  = \frac{1}{1 + \frac{X_{\alpha}(E_k)}{X_{{\rm cr} \alpha}}} \times
   \left[\frac{\tau_{\alpha}^{hH}}{2\,h}\mathcal{Q}_{\alpha, \rm src} + \frac{X_{\alpha}(E_k)}{\mu v(E_k)}\mathcal{Q}_{\alpha, \rm spall} \right]
\end{equation}

%%%%%%%%%%%%%%%%%%%%%%%%%%%%%%%%%%%
\begin{figure}
\includegraphics[width=\linewidth]{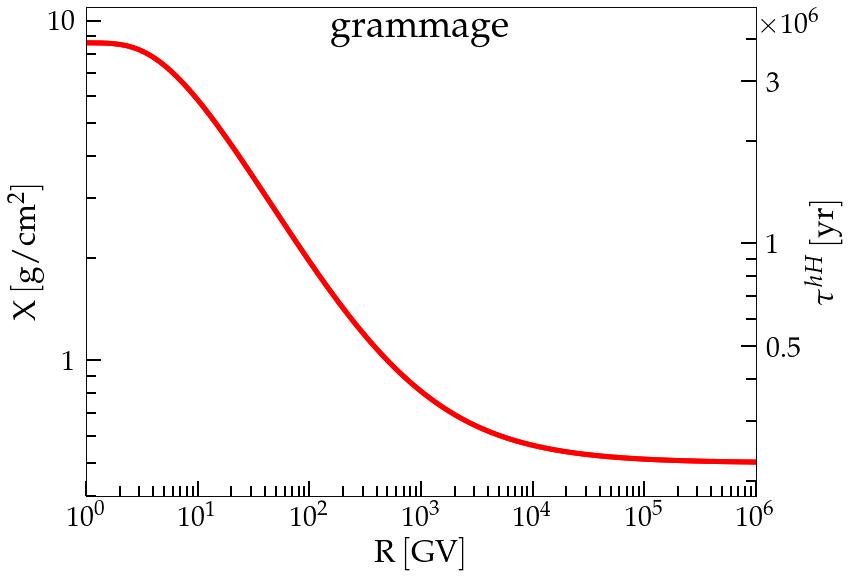}
\caption{Cosmic ray grammage as defined by \protect{Eq.~(\ref{eq:gramm-def})}.}
\label{fig:grammage}
\end{figure}

%%%%%%%%%%%%%%%%%%%%%%%%%%%%%%%%%%%%

Such solution can be understood as follows. Let us consider the case of CR protons, which are purely primary and for which $\sigma_{\alpha} = 0$.
The flux of Eq.~(\ref{eq:solution-stable-I0}) becomes
\begin{equation}
    I_{\alpha 0}(E_k) = \frac{\mathcal{Q}_{\alpha, \rm src}}{2} \left[\frac{h}{D_h} + \frac{H}{D_H} \frac{1-\exp^{-\frac{u\,H}{D_H}}}{\frac{u\,H}{D_H}}\right],
\end{equation}
namely the same solution found in Eq.~(\ref{eq:appendix-f0-approx}) when adiabatic losses are neglected.
Considering that $D_H \propto R^{\delta}$ increases with rigidity while $D_h$ is assumed to be constant, the residence time in the GD of Eq.~(\ref{eq:tau-hH-def}) is expected to be dominated by diffusion in the GD at high enough rigidity, such that $h/D_h > H/D_H$, namely above the transition rigidity
\begin{equation}\label{eq:R-trans}
    R^{*} = \left[80 \frac{H}{4\, \rm kpc} \frac{h}{150\, \rm pc} \frac{D_m}{\rm pc} \frac{10^{28}\rm cm^2/s}{D_0} \right]^{1/\delta}\, \rm GV.
\end{equation}
The value of $R^*$ depends on the relative size of the GD and GH, on  $D_m$ and on the normalization and slope of $D_H$. Taking the reference values reported in Eq.~(\ref{eq:R-trans}), the transition is expected at 
\begin{description}
    \item{\makebox[4.5cm][l]{ $\approx$ 6 TV \hfill} ($\delta = 0.5$)}
    \item{\makebox[4.5cm][l]{ $\approx$ 2 TV \hfill} ($\delta = 0.6$)}
    \item{\makebox[4.5cm][l]{ $\approx$ 500 GV \hfill} ($\delta = 0.7$)}
    \item{\makebox[4.5cm][l]{ $\approx$ 200 GV \hfill} ($\delta = 0.8$)}
\end{description}
Thus, if the slope of the diffusion coefficient in the GH is  $\delta \gtrsim 0.6$ \citep{Evoli-2019-ams02}, as typically found  in models of self-generated diffusion \citep{Aloisio-Balsi-2013-self}, at $R \gtrsim \rm 100s\,GV-TV$ the effect of propagation on the CR distribution in the disk  smoothly becomes  independent on rigidity.

Correspondingly, the grammage would tend to the constant minimum  value reported in Eq.~(\ref{eq:Xmin}) (in the assumption that $h=h_d$)
\begin{equation}
    X_{\alpha}(R \gg R^{*}) \quad\longrightarrow \quad n_d\, \mu \, c \frac{h^2}{D_h}\; = \;X_{\rm min}.
\end{equation}

Below $R\sim R^{*}$, the residence time in the GD
is dominated by  the repeated crossings due to diffusion in the GH, with a contribution of advection, which typically dominates at $R\lesssim 10 \,\rm GV $. The advective ($uH \gg D_H$) and diffusive ($uH \ll D_H$) limiting cases read
\begin{equation}
I_{\alpha 0} \longrightarrow  \begin{cases}
\frac{\mathcal{Q}_{\alpha, \rm src}}{2\,u} & \qquad\text{advection limit}\\
\frac{\mathcal{Q}_{\alpha, \rm src} H}{2\,D_H} & \qquad \text{diffusion limit,}
\end{cases}   
\end{equation}
similar to what found in Sect.~\ref{sec:appendix-adiabatic}.\\

As a final remark, it is important to point out that the diffusion equation in the GH is valid only up to a rigidity,  $R_H^{\rm mfp}$, such that the diffusive mean free path is smaller than the size of the  diffusive halo, namely  $\lambda_{\rm mfp} \lesssim H$, where
\begin{equation}\label{eq:lmfp}
    \lambda_{\rm mfp} \equiv \frac{3\,D_H(R)}{c} \approx 0.3\, \left(\frac{D_0}{\rm 10^{28}\, cm^2/s}\right) \left(\frac{R}{\rm GV}\right)^{\delta} \,\rm pc,
\end{equation}
which gives  
\begin{equation}\label{eq:RH-mfp}
    R_H^{\rm mfp} = \left[12\times 10^3 \frac{H}{4\, \rm kpc} \frac{10^{28}\rm cm^2/s}{D_0} \right]^{1/\delta}\, \rm GV.
\end{equation}
For $\delta = 0.5-0.8$ the value of $R_{{\rm mfp}, H}$ reads
\begin{description}
    \item{\makebox[4.5cm][l]{ $\approx$ 150 PV \hfill} ($\delta = 0.5$)}
    \item{\makebox[4.5cm][l]{ $\approx$ 6 PV \hfill} ($\delta = 0.6$)}
    \item{\makebox[4.5cm][l]{ $\approx$ 700 TV \hfill} ($\delta = 0.7$)}
    \item{\makebox[4.5cm][l]{ $\approx$ 100 TV \hfill} ($\delta = 0.8$)}.
\end{description}
Thus,  at high enough rigidity, particles in the GH do not really diffuse, but rather free-escape. We treat this case by performing the limit $H \rightarrow 0$ in Eq.~(\ref{eq:solution-stable-I0}), which corresponds to moving the free-escape boundary $ H^* \rightarrow h$. On the other hand, for the cases treated in this work $R^* \ll R_H^{\rm mfp}$, so that solution at $R \gtrsim R_H^{\rm mfp} $ would be largely dominated by the diffusion in the GD even without performing the limit $H \rightarrow 0$.

\section{Results}
\label{sec:results}
In this section, we describe the main results of our calculations. The CR fluxes are evaluated using Eq.~(\ref{eq:solution-stable-I0}), with the setup illustrated in Sect.~\ref{sec:prop-setup}. 

For the injection spectrum of Eq.~(\ref{eq:inj-spec-tot})
we assume the slope $\gamma_p = 4.35$ for protons, $\gamma_{\rm He} = 4.3$ for He and $\gamma_{n} = 4.33$ for all the other nuclei.
Notice that different slopes for the injection spectra of $p$, He and heavier nuclei, with $\gamma_p > \gamma_n > \gamma_{\rm He}$ were also used in other works aimed at fitting CR spectra (see e.g. \citealt{Aloisio-Balsi-2013-self, Bresci-2019-src-grammage, Evoli-2019-ams02}). While 
possible explanations based on  subtleties of the DSA process have been put forward (see e.g. \citealt{Malkov-2012-p-He-inj-spectra}), such difference is currently an open issue and discussing that goes beyond the  scope of the present work. 

The effect of the solar modulation is accounted for in the force-field approximation \citep{Gleeson-Axford-1968-solar-mod}, with a modulation potential $\Phi_{\rm mod} = 450\, \rm MV$.
The relevant physical parameters adopted in the calculations are summarized in Table~\ref{tab:param}.
%%%%
\begin{table}
\caption{Benchmark parameters. Solar modulation potential $\Phi_{\rm mod} = 450\, \rm MV$.}
\label{tab:param}
\begin{tabular}{ccccc}
\hline
\hline
\multicolumn{5}{c}{\textbf{gaseous disk}}\\
\hline
$n_d$   & $f_{\rm He}$ & $h_d$ & $R_d$  &  \\
$\rm 1\, cm^{-3}$   & $0.1$ & $\rm 150\,pc$ & $\rm 15\,kpc$  & \\
\hline
\hline
\multicolumn{5}{c}{\textbf{Galactic disk (GD)}}\\
\hline
$h$   & $L_c$ & $b^2$ & $D_m$ & $L_{RR}$   \\
$\rm 150\,pc$   & $\rm 10\,pc$ & $0.4$ & $\rm 1\,pc$ & $\rm 50-100\,pc$ \\
\hline
\hline
\multicolumn{5}{c}{\textbf{Galactic halo (GH)}}\\
\hline
\multicolumn{1}{c}{$H$} &
\multicolumn{1}{c}{$D_0$}&  \multicolumn{1}{c}{$\delta$} & 
\multicolumn{1}{c}{$u$}\\
\multicolumn{1}{c}{$\rm 4\,kpc$} &\multicolumn{1}{c}{$\rm 10^{28}\,cm^2/s$}&  \multicolumn{1}{c}{$0.7$} &
\multicolumn{1}{c}{$\rm 40\, km/s$} \\
\hline
\hline
\multicolumn{5}{c}{\textbf{bulk of SNRs}}\\
\hline
\multicolumn{1}{c}{$\gamma_p$} &
\multicolumn{1}{c}{$\gamma_{\rm He}$}&  \multicolumn{1}{c}{$\gamma_n$} & 
\multicolumn{1}{c}{$R_{\rm max }^{\rm bulk}$} 
\\
\multicolumn{1}{c}{$4.35$} &\multicolumn{1}{c}{$4.30$}&  \multicolumn{1}{c}{$4.33$} &
\multicolumn{1}{c}{$\rm 50\, TV$} \\
\hline
\hline  
\multicolumn{5}{c}{\textbf{PeV sources}}\\
\hline
\multicolumn{1}{c}{$\gamma_p$} &
\multicolumn{1}{c}{$\gamma_{\rm He}$}&  \multicolumn{1}{c}{$\gamma_n$} & 
\multicolumn{1}{c}{$R_{\rm max }^{\rm PeV}$} &
\multicolumn{1}{c}{$\epsilon_{\rm bulk}^{\rm PeV}$}
\\
\multicolumn{1}{c}{$4.35$} &\multicolumn{1}{c}{$4.30$}&  \multicolumn{1}{c}{$4.33$} &
\multicolumn{1}{c}{$\rm 5\, PV$}
& \multicolumn{1}{c}{$ 0.15$}
\\
\hline
\hline  
\end{tabular}
\end{table}
%%%%%

Before proceeding to a detailed description of our results, a remark is in order. While, in general, we obtain  a very good agreement with data at rigidity $R\gtrsim 10\, \rm GV$, in the low rigidity domain the agreement tends to be poorer, although quite decent in most cases. A better fit to the low rigidity data may be obtained with the addition of ionization losses, and with  a more refine modeling of advection, with the inclusion adiabatic losses.
Moreover, in order to keep the number of free parameters limited,  we calculated the solar-modulated spectra  with the same $\Phi_{\rm mod}$ for all nuclei. However different data sets were
collected at different times, so that  using a slightly different value of the  modulation potential for different data sets may further  improve the agreement at low rigidity. 
On the other hand, the scenario proposed in this work substantially differ from the standard CR propagation parading beyond $R \gtrsim 100\, \rm GV$, thus seeking a very good agreement with low rigidity CR data goes beyond  the scope of the present paper.
 
\subsection*{Proton, helium and antiprotons}
\begin{figure*}
\begin{multicols}{2}
\includegraphics[width=\linewidth]{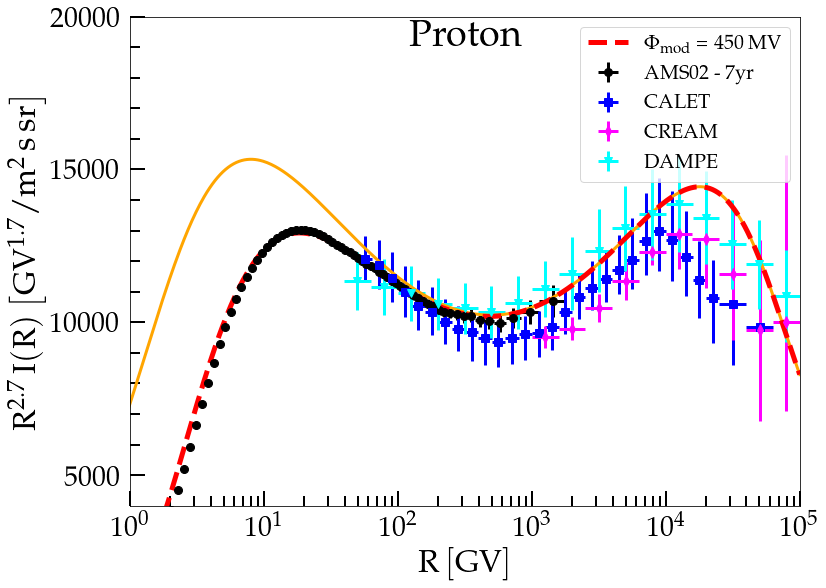}\par 
\includegraphics[width=\linewidth]{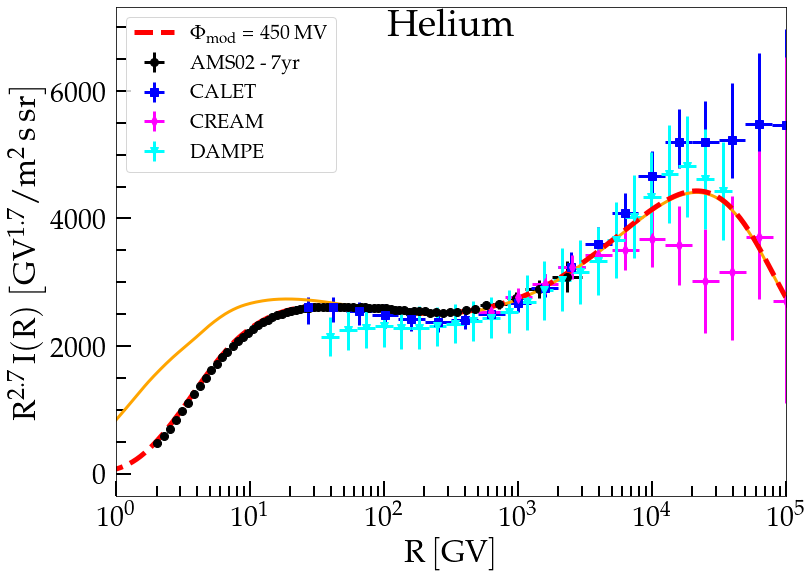}\par 
\end{multicols}
\centering
\includegraphics[width=0.5\linewidth]{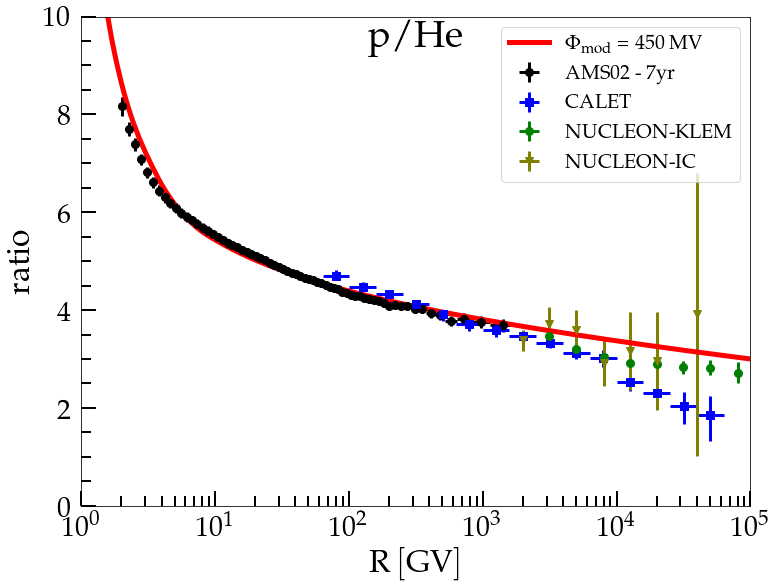}
\caption{\textit{Top left panel}: unmodulated (solid orange) and modulated (dashed red) proton flux compared to the data of AMS02 (\protect{\citealt{AMS02-7yr}}, black dots), CALET (\protect{\citealt{CALET-2022-H}}, blue squares), CREAM (\protect{\citealt{CREAM-2017-H-He}}, magenta diamonds) and DAMPE (\protect{\citealt{DAMPE-2019-H}}, cyan triangles). 
\textit{Top right panel}: unmodulated (solid orange) and modulated (dashed red) He flux compared to the data of AMS02 (\protect{\citealt{AMS02-7yr}}, black dots), CALET (\protect{\citealt{CALET-2023-He}}, blue squares), CREAM (\protect{\citealt{CREAM-2017-H-He}}, magenta diamonds) and DAMPE (\protect{\citealt{DAMPE-2021-He}}, cyan triangles).
\textit{Bottom panel}: p/He flux compared to the data of AMS02 (\protect{\citealt{AMS02-7yr}}, black dots), CALET (\protect{\citealt{CALET-2023-He}}, blue squares), NUCLEON (\protect{\citealt{NUCLEON-2020-HHe}}, green points).
}
\label{fig:p-He}
\end{figure*}

%%%%%%%%%%%%%%%%%%
In Fig.~\ref{fig:p-He} we show the proton and He (made mostly of \isotope{4}{He} with a fraction of secondary \isotope{3}{He} produced by the spallation of \isotope{4}{He}, see e.g \citealt{Coste-2012-light-nucl}) fluxes  in the range $1-10^5\, \rm GV$ and the corresponding $p/{\rm He}$ ratio.  
Such spectra can be understood with the help of Fig.~\ref{fig:grammage}.
At $R\lesssim 10\, \rm GV$ the (unmodulated) solution is dominated by advection, as indicated by the flattening in the grammage (see Fig.~\ref{fig:grammage}). As illustratd in Appendix~\ref{sec:appendix-adiabatic}, this results in a hardening of the CR flux at low $R$.

In the range $R\approx 10-100s$ GV the CR flux is mostly affected by the steep rigidity-dependent diffusion in the GH, $D_H \propto R^{0.7}$, which is reflected in the power-law drop of the grammage. On the other hand, above a few hundreds GV, the grammage smoothly begins to flatten, as described in Sect.~\ref{sec:solution-stable},
leading to a gentle, progressive hardening of the CR spectrum, in good agreement with data. 
Such smooth flattening of the spectrum is the result of the $h/D_h$ term  in Eq.~(\ref{eq:tau-hH-def})  becoming  comparable and then larger than $H/D_H$ with the increase of $R$, a transition that takes place in $\approx 1-2$ decades in $R$ around $\approx 1\, \rm TV$.  

The spectral steepening at $\approx 15\, \rm TV$ reported by DAMPE is the result of the majority of SNRs accelerating particle with a typical cut-off rigidity $R^{\rm bulk}_{\rm max} \approx 50\, \rm TV$. Such steepening can be also identified in the CREAM and CALET data for the $p$ spectrum,  while in the case of He the situation is less clear, especially due to the quite large discrepancy between the spectra reported by the three experiments. 

In general, we obtain a very good  agreement with both AMS02  and DAMPE data for $p$ and He in the whole $1-10^5\, \rm GV$ range, and in particular at $R\gtrsim 10\, \rm GV$, where solar modulation is unimportant. 
Our prediction for the $p/{\rm He}$ ratio matches well the AMS02 data at $R\lesssim 1\,\rm TV$ and the NUCLEON data at higher $R$.  Also in this case the different data sets show relevant discrepancies in the multi-TV range. \\
 
\begin{figure*}
\begin{multicols}{2}
    \includegraphics[width=\linewidth]{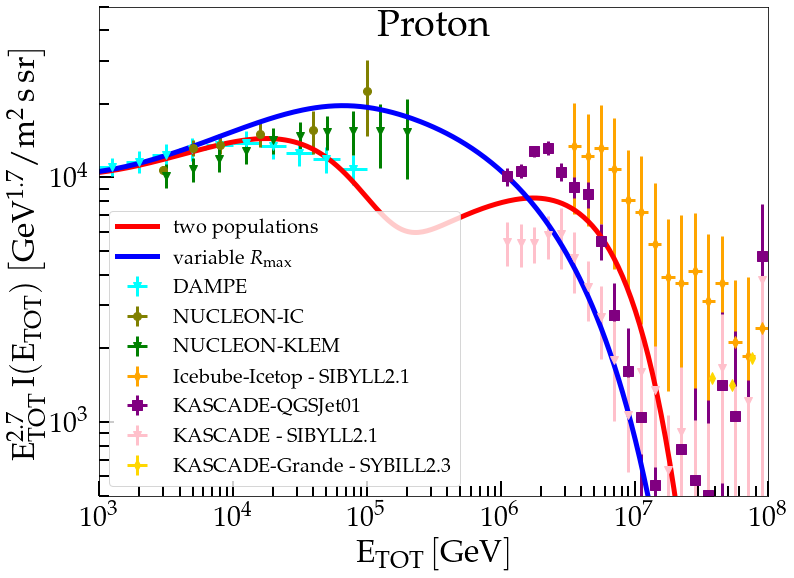}\par 
    \includegraphics[width=\linewidth]{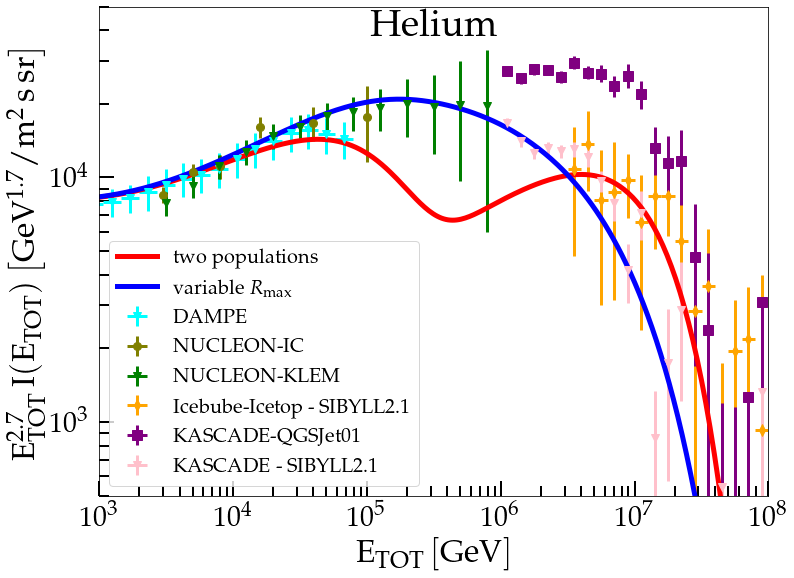}\par 
\end{multicols}
\caption{Proton (left panel) and He (right panel) fluxes compared to the data of DAMPE (\protect{\citealt{DAMPE-2019-H, DAMPE-2021-He}},  cyan triangles),  NUCLEON (\protect{\citealt{NUCLEON-2019}}, green points), Icecube-Icetop (\protect{\citealt{Icecube-Icetop-2019}}, orange diamonds), KASCADE  (\protect{\citealt{KASCADE-2005-H-He---}}, pink and violet points), KASCADE-Grande  (\protect{\citealt{KASCADE-Grande-2017}}, yellow points). 
The red curves correspond to the case with two source populations. The blue curves correspond to  the case of a source CR luminosity  decreasing with the increase of maximum rigidity. 
}
\label{fig:p-He-knee}
\end{figure*}

In Fig.~\ref{fig:p-He-knee} we show the  proton and He  fluxes of Fig.~\ref{fig:p-He}, as a function of the total energy $E_{\rm TOT}$, in the range $10^3-10^8\, \rm GeV$. 
The red  curve corresponds to the expected contribution from the two classes of sources described in Sec.~\ref{sec:prop-setup}, i.e., typical SNRs with a cut-off rigidity at $R^{\rm bulk}_{\max} = \rm 50\, TV$ (Eq.~(\ref{eq:inj-spec-bulk})) and PeV sources characterised by $R^{\rm PeV}_{\max} = \rm 5\, PV$ (Eq.~(\ref{eq:inj-spec-PeV})), and represents the  very high energy part of the fluxes shown in Fig.~\ref{fig:p-He}.
The CR luminosity of the PeV population  is  $\epsilon_{\rm bulk}^{\rm PeV} = 0.15$  times that of the bulk population. 

As a comparison, we also show the case (blue curve) of a uniform distribution of  $\ln(R_{\rm max})$, in the range $\ln(R^{\rm bulk}_{\max})- \ln(R^{\rm PeV}_{\max})$, with an overall CR luminosity linearly decreasing with $\ln(R_{\rm max})$ compared to the bulk of sources, in the range $1-\epsilon_{\rm bulk}^{\rm PeV}$. 
The injection spectrum in this case can be written as
\begin{equation}
    q_{0, \alpha} = \int_{l^{\rm bulk}_{\rm max}}^{l^{\rm PeV}_{\rm max}} q_{0, \alpha}^i(L_i = \epsilon_i(l_{\rm max})\, L^{\rm bulk}, e^{l_{\rm max}}) \frac{d\,l_{\rm max}}{l^{\rm PeV}_{\rm max} - l^{\rm bulk}_{\rm max}},
\end{equation}
where $q_{0,\alpha }^i$ is defined in Eq.~(\ref{eq:inj-spec-general}), $l^{(\rm bulk, PeV)}_{\rm max} \equiv \ln(R^{(\rm bulk, PeV)}_{\rm max})$, $\epsilon_i(l_{\rm max})\, L_{\rm bulk}$ is the luminosity for a given $l_{\rm max}$, with
\begin{equation}
    \epsilon_i(l_{\rm max}) = \epsilon_{\rm bulk}^{\rm PeV} - \frac{\epsilon_{\rm bulk}^{\rm PeV} -1}{l^{\rm bulk}_{\rm max} - l^{\rm PeV}_{\rm max}}\left(l_{\rm max} - l^{\rm PeV}_{\rm max}\right).
\end{equation}

Keeping in mind the quite large systematics between the different data-sets, especially in the multi-PeV domain, one can see that our model is able to  reproduce  the data quite well, from $R\gtrsim 10\, \rm GV$ up to the multi-PV domain. Such result is the combination of the rigidity-independent propagation experienced by CRs beyond $\approx \rm 10\, \, TV$ (see Fig.~\ref{fig:grammage}) and of the smaller CR luminosity of sources able to reach very high energies compared to that of the majority of sources, although  current data do not allow for a clear discrimination between a two-population scenario 
and a distribution of $R_{\rm max}$ with a progressive decrease of the CR luminosity. Notice that, in the presented scenario, it is not necessary to assume a different slope for the two source populations (or a dependence of the slope on $R_{\rm max}$, and indeed we obtained the reported fluxes with a fixed slope for a given species ($p$ or He, see Table~\ref{tab:param}).  

If we assumed that the bulk of SNRs accelerate particles up the \textit{knee}, with a rigidity-independent propagation, at such high energies we would get a very hard CR spectrum that would largely overshoot the data. 
In order to explain the steepening in the DAMPE data without affecting the CR flux at lower energies,   we would be forced to introduce a change in the propagation regime at $R\approx \rm 20 \, TV$ in addition to that required to explain the hardening at $R\approx \rm 300\, GV$. 

Instead, if we assume a propagation model such that the rigidity dependence of the grammage is the same as in Fig.~\ref{fig:grammage} up to $R\approx 1\, \rm TV$ and steeper at higher $R$, e.g the typical $\propto R^{0.3}$ of a Kolmogorov turbulence invoked at high energies by previous works (see e.g \citealt{Aloisio-Balsi-2013-self, Evoli-2018-halo-formation}), the predicted  spectra  would lie below the DAMPE and NUCLEON data. 

\begin{figure*}
\begin{multicols}{2} \includegraphics[width=\linewidth]{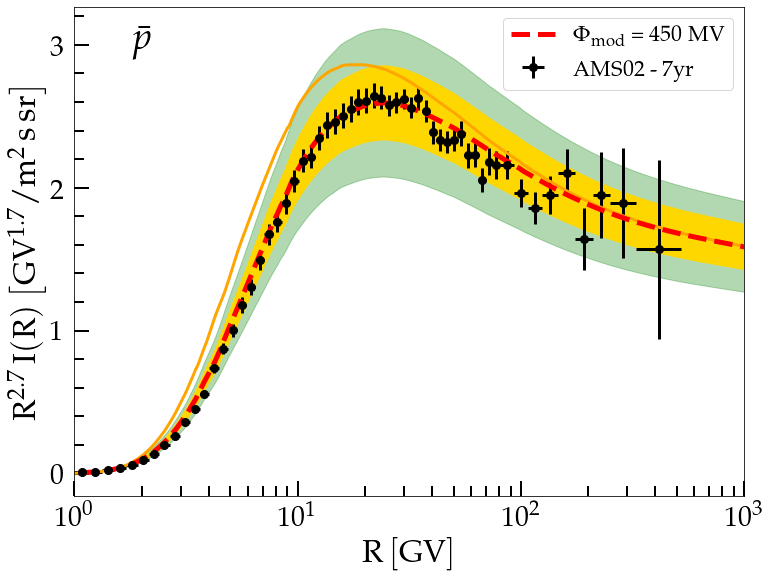}\par 
\includegraphics[width=\linewidth]{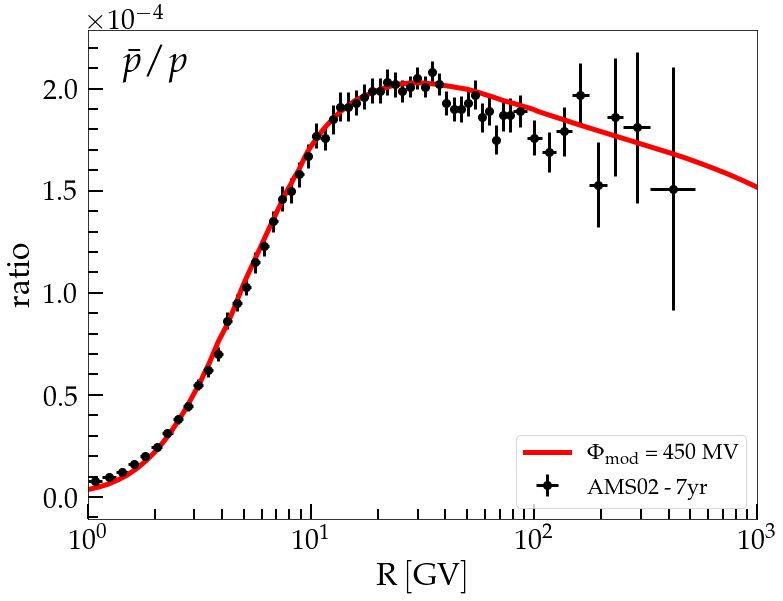}\par 
\end{multicols}
\begin{multicols}{2} \includegraphics[width=\linewidth]{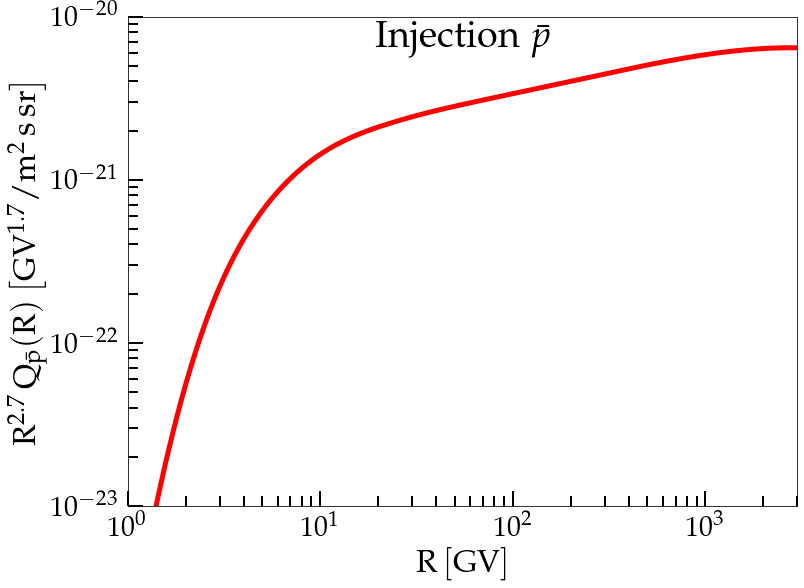}\par 
\includegraphics[width=\linewidth]{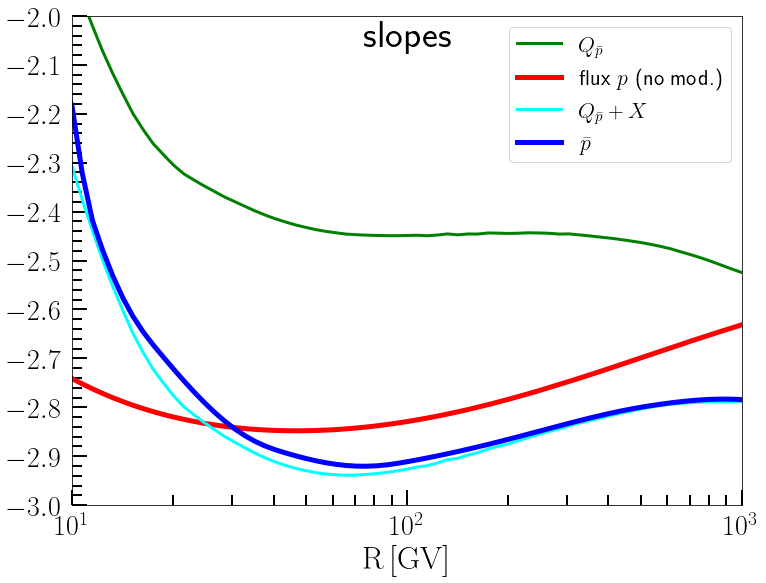}\par 
\end{multicols}
\caption{\textit{Top left panel}: unmodulated (solid orange) and modulated (dashed red) $\bar{p}$ flux compared to the data of AMS02 (\protect{\citealt{AMS02-7yr}}, black dots).  The yellow (green) shaded region corresponds a  $10\%$ ($20\%$) uncertainty in the  production cross-section.
\textit{Top right panel}: $\bar{p}/p$ ratio compared to the data of AMS02 (\protect{\citealt{AMS02-7yr}}, black dots).
\textit{Bottom left panel}: $\bar{p}$ source term, computed as in \protect{Eq.~(\ref{eq:pbar-src-term}}) and corresponding to the flux reported in the top left panel of the present figure. 
\textit{Bottom right panel}: slopes of the $\bar{p}$ source term (solid green line), of $\bar{p}$ flux (blue solid line) and of the proton flux (red solid line) reported in \protect{Fig.~\ref{fig:p-He}}. The solid cyan curve corresponds to the sum of the slope of $Q_{\bar{p}}$ and of the grammage reported in \protect{Fig.~\ref{fig:grammage}}.
}
\label{fig:pbar}
\end{figure*}

In Fig.~\ref{fig:pbar} we show our prediction for the $\Bar{p}$ flux and for the $\Bar{p}/p$ ratio, as compared to the AMS02 data. In order to highlight the typical uncertainty in the production cross-section of $\Bar{p}$ \citep{Korsmeier-2018-pbar-production}, we placed a $10\%$ (yellow) and $20\%$ (green) band around our predicted flux.
For both quantities  we obtain an excellent agreement with data. In particular, we reproduce well the quite flat $\bar{p}/p$ ratio that typically fails to be explained in standard propagation scenarios. 

Such result can be explained as follows: \textit{i)} the antiprotons produced in p-p collisions (with a non-negligible contribution from p-He, He-p  He-He channels, see \citealt{Korsmeier-2018-pbar-production}) have broad energy distributions and the kind of  kinematics involved in the process 
significantly reshapes the $\bar{p}$ production spectrum $Q_{\bar{p}}$, making it appreciably harder than the parent $p$ (and He) flux.
This can be seen in Fig.~\ref{fig:pbar}, where, in the bottom left panel, we show the $Q_{\bar{p}}$ corresponding to the $\bar{p}$ flux shown in the upper left panel of the same figure, and in the bottom right panel we report their slopes; \textit{ii)} antiprotons propagate in the ISM as protons (apart for the role played by the spallation  of $\bar{p}$) so that the  slope of the equilibrium spectrum of antiprotons we expect $\gamma_{\bar{p}} \approx \gamma_{Q_{\bar{p}}} + \gamma_X$, as it is indeed observed (see bottom right panel of Fig.~\ref{fig:pbar}). 
The combination of a hard  $Q_{\bar{p}}$ and of the progressive flattening of the grammage $X$ keeps 
$\vert \gamma_{\bar{p}} - \gamma_p \vert \lesssim 0.15$ in the range $R= 30-1000\, \rm GV$, which translates in a very mild decrease of the $\bar{p}/p$ ratio in such rigidity range.  
 
\subsection*{LiBeB and CNO nuclei}
\begin{figure*}
\begin{multicols}{2} \includegraphics[width=0.95\linewidth]{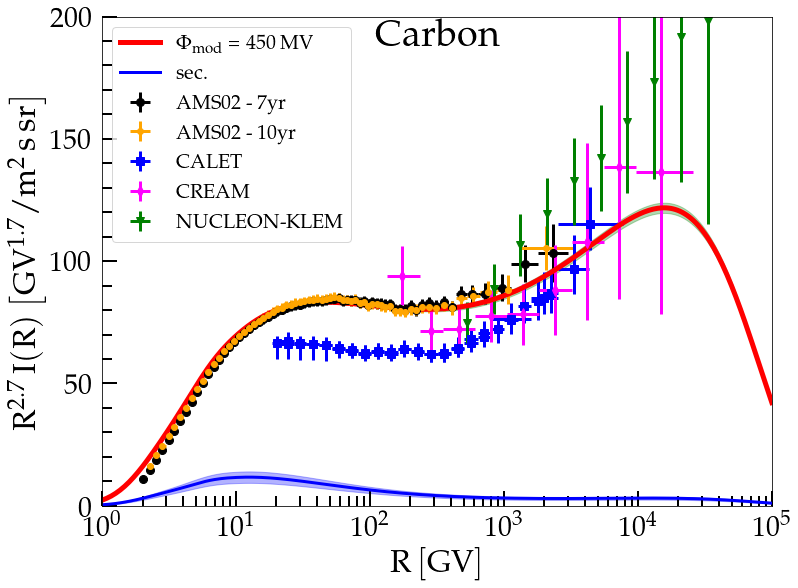}\par 
\includegraphics[width=0.95\linewidth]{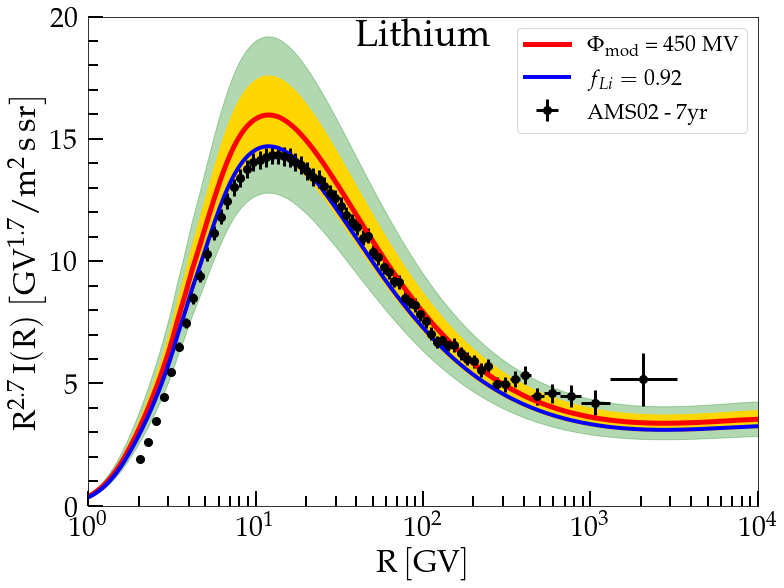}\par 
\end{multicols}
\begin{multicols}{2} \includegraphics[width=0.95\linewidth]{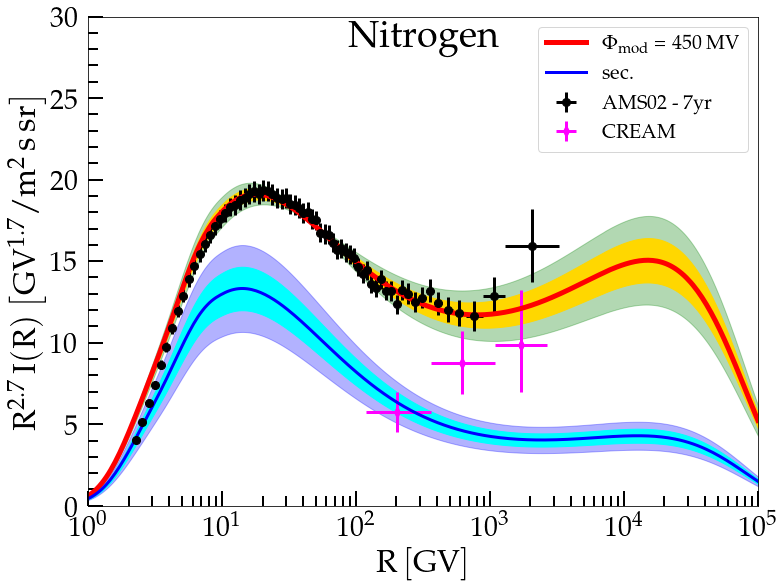}\par 
\includegraphics[width=0.95\linewidth]{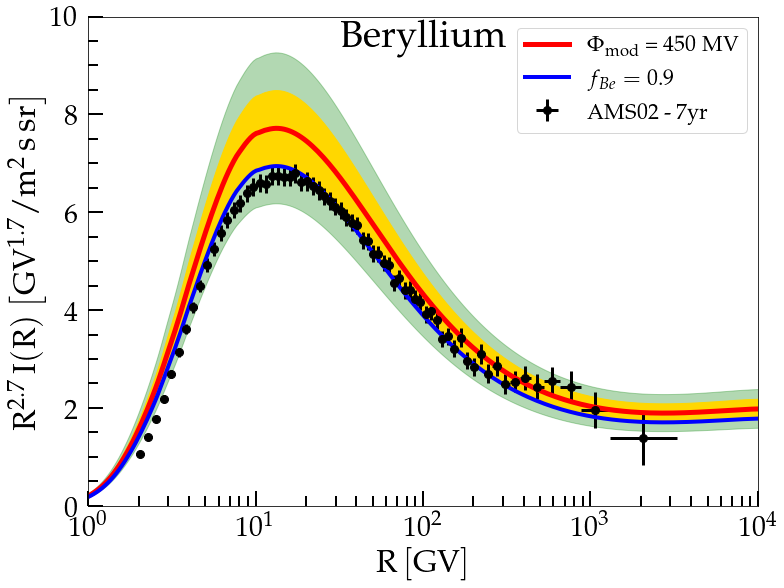}\par 
\end{multicols}
\begin{multicols}{2} \includegraphics[width=0.95\linewidth]{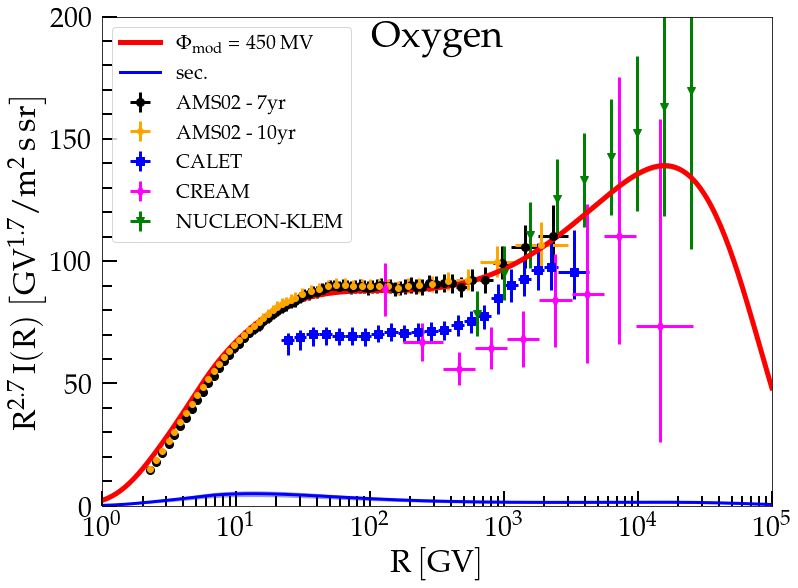}\par 
\includegraphics[width=0.95\linewidth]{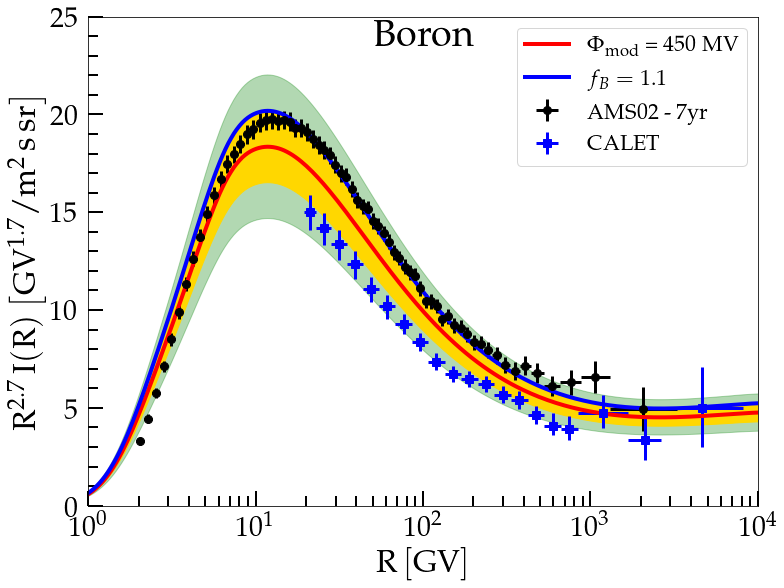}\par 
\end{multicols}
\caption{\textit{Top left panel}: C flux compared to the data of AMS02 (\protect{\citealt{AMS02-7yr}}, black dots), CALET (\protect{\citealt{CALET-2020-C-O}}, blue squares), CREAM (\protect{\citealt{CREAM-2009-CNO-heavy}}, magenta diamonds) and NUCLEON (\protect{\citealt{NUCLEON-2019}}, green triangles). 
\textit{Center left panel}: N flux compared to the data of AMS02 (\protect{\citealt{AMS02-7yr}}, black dots), CREAM (\protect{\citealt{CREAM-2009-CNO-heavy}}, magenta diamonds).
\textit{Bottom left panel}: O flux compared to the data of AMS02 (\protect{\citealt{AMS02-7yr, AMS02-10yr-heavy}}, black dots and yellow diamonds), CALET (\protect{\citealt{CALET-2020-C-O}}, blue squares), CREAM (\protect{\citealt{CREAM-2009-CNO-heavy}}, magenta diamonds) and NUCLEON (\protect{\citealt{NUCLEON-2019}}, green triangles).
The solid red line corresponds to the total modulated flux, while the solid blue line to the (modulated) secondary contribution to the given nucleus. The cyan (blue) shaded region corresponds a  $10\%$ ($20\%$) uncertainty in the secondary production cross-section, which translates in the yellow (green) shaded region in the total flux.
\textit{Right panels}: Li, Be and B fluxes compared to the data of AMS02 (\protect{\citealt{AMS02-7yr}}, black dots) and  CALET (\protect{\citealt{CALET-2020-C-O}}, blue squares).
The solid red line corresponds to the total modulated flux. The yellow (green) shaded region corresponds a  $10\%$ ($20\%$) uncertainty in the secondary production cross-section. The solid blue line corresponds to the total flux multiplied by $f_{Li} = 0.92$ (Li), $f_{Be} = 0.9$ (Be) and $f_{B} = 1.1$ (B).
}
\label{fig:CNO-LiBeB}
\end{figure*}

%%%
In the left panels of Fig.~\ref{fig:CNO-LiBeB} we show the C, N and O fluxes.

\begin{figure*}
\begin{multicols}{2}
\includegraphics[width=\linewidth]{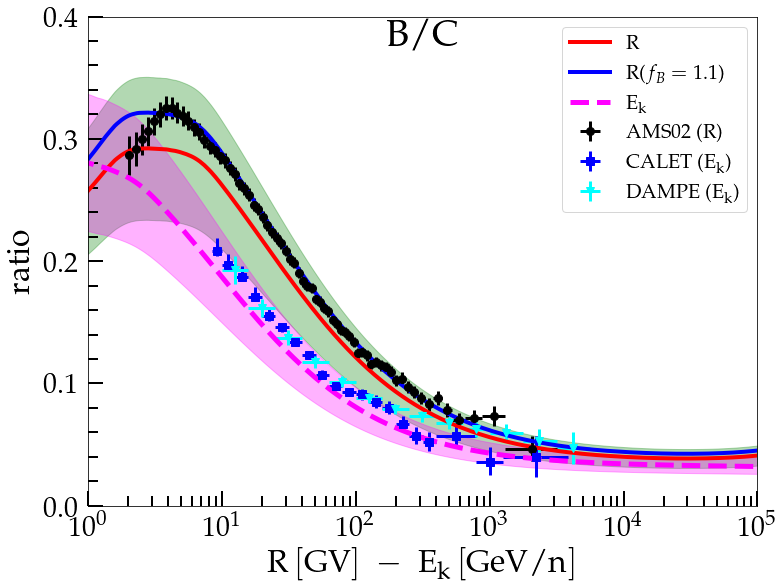}\par 
\includegraphics[width=\linewidth]{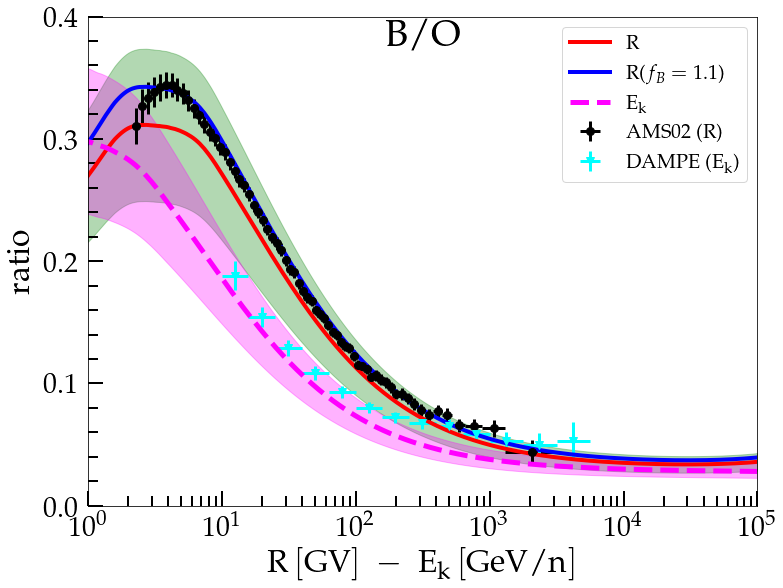}\par 
\end{multicols}
\centering
\includegraphics[width=0.5\linewidth]{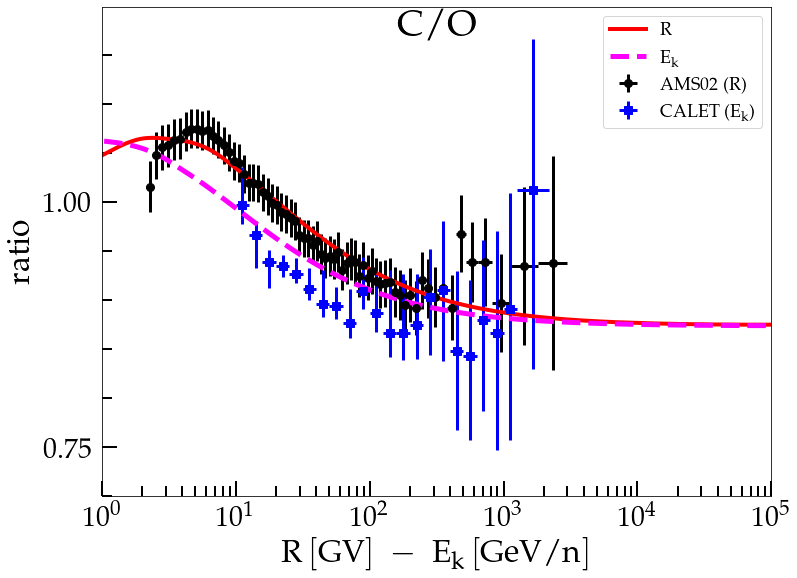}
\caption{B/C (top left panel), B/O (top right panel) and C/O (bottom  panel) compared to the data  of AMS02  (\protect{\citealt{AMS02-7yr}}, black dots), CALET (\protect{\citealt{CALET-2020-C-O}}, blue squares) and DAMPE  (\protect{\citealt{DAMPE-2022-BC-BO}}, cyan triangles). 
The solid red (dashed magenta) line corresponds to the ratio as a function of rigidity (kinetic energy per nucleon) and is compared to the AMS02 (CALET and DAMPE) data. The green (pink) shaded region corresponds a  $20\%$ uncertainty in the secondary production cross-section. The solid blue line corresponds to the B/C or B/O ratio when the B flux is multiplied by $f_{B} =  1.1$ as in \protect{Fig.~\ref{fig:CNO-LiBeB}}.
}
\label{fig:B-C-O-ratio}
\end{figure*}

O (mostly \isotope{16}{O}, \isotope{17}{O},\isotope{18}{O}) and C (mostly \isotope{12}{C}, \isotope{13}{C}) are mainly primary, with a fraction of a few $\%$ and $\lesssim 20\%$, respectively, of secondary contribution from heavier nuclei at low rigidity. Most of the secondary C is due to the spallation of O.
For both nuclei we reproduce  well the AMS02 data. In addition, the progressive transition to a flat grammage, described above, produces a rather hard spectrum  beyond $\sim \rm TV$ rigidity, in agreement with the CREAM and NUCLEON experiments. Note however, the sizeable discrepancies between data-sets at high-$R$. The $\rm C/O$ ratio is also in good agreement with the AMS02 data, as shown in Fig.~\ref{fig:B-C-O-ratio}.

N (\isotope{14}{N}, \isotope{15}{N}) is partially primary and partially secondary (with a major contribution from the spallation of O). Also in this case we get a good agreement with the AMS02 data, and in particular we reproduce the rather sharp high-$R$ hardening of the spectrum. Notice also that the multi-TV flux is affected by the (typically $\approx 20\%$) uncertainty in the secondary production cross-section. Indeed we normalize the injection spectrum of N from sources in order to reproduce the observed AMS02 flux at $50\, \rm TV$. A $10\%$ (cyan) and  $20\%$ (violet) uncertainty in the secondary production translates in a re-normalization of the primary contribution in order to get the same flux at  $50\, \rm TV$, which clearly appears in the yellow and green bands at multi-TV (where the primary part largely dominates). \\

In the right panels of Fig.~\ref{fig:CNO-LiBeB} we show the Li (\isotope{6}{Li},  \isotope{7}{Li}), Be (\isotope{7}{Be}, \isotope{9}{Be}, \isotope{10}{Be}) and B (\isotope{10}{B},  \isotope{11}{B}) fluxes. The contribution \isotope{11}{B} from the shot-lived isotope \isotope{11}{C} is also included (see \citealt{Evoli-2019-ams02}).  

The red curves are the fluxes obtained with the production cross-sections specified in Sect.~\ref{sec:prop-setup}, with the yellow and green band representing a $10\%$ and $20\%$ uncertainty in the production cross sections, respectively. 
The main production channels ($\gtrsim 80\%$ at $E_k \gtrsim 10 \, \rm GeV/n$) come from the spallation of CNO nuclei. As discussed in detail by \citet{Evoli-2019-ams02}, such cross sections are quite uncertain, especially at $E_k \gtrsim 10\, \rm GeV/n$, due to the fact that most of experimental data  are usually at hundreds of MeV/n, and just some channels are probed at few GeV/n’s. Looking at the channels CNO $\rightarrow$ LiBeB, reported in the appendix of \citet{Evoli-2019-ams02}, typical uncertainty are of $20-30\%$, and in some channels can exceed $\sim 50\%$, such as \isotope{16}{O} $\rightarrow$ \isotope{7}{Li}, \isotope{9}{Be}.  Moreover the channels \isotope{14}{N} $\rightarrow$ \isotope{6}{Li}, \isotope{7}{Li}, \isotope{9}{Be}, \isotope{10}{B}, \isotope{11}{B}, have only 1-2 of data points  below $\sim 1\, \rm GeV/n$,  thus making   the high-energy cross sections even more difficult to  determine. Additional uncertainties come from the LiBeB production from heavier nuclei.  Finally, the systematic uncertainties related to the available  measurements and the contribution of \textit{ghost} nuclei are difficult to asses. 

With the adopted cross sections we get Li, Be and B fluxes whose shape is good in agreement with AMS02 data, in particular we reproduce quite well the progressive  flattening around $\approx \rm TV$.  However, we get Li and Be fluxes systematically higher than the data and a B flux systematically lower.   
Keeping in mind the large uncertainties in the production cross sections, we also show with blue lines the Li flux multiplied by $f_{Li} = 0.92$, the Be flux multiplied by $f_{Be} = 0.9$ and the B flux multiplied by $f_{B} = 1.1$, which matches the AMS02 data well.  

\begin{figure*}
\begin{multicols}{2}
\includegraphics[width=\linewidth]{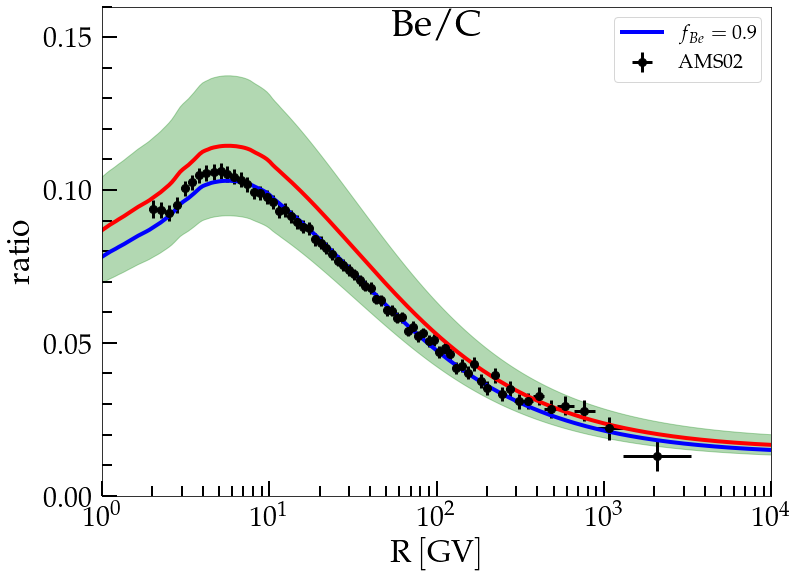}\par 
\includegraphics[width=\linewidth]{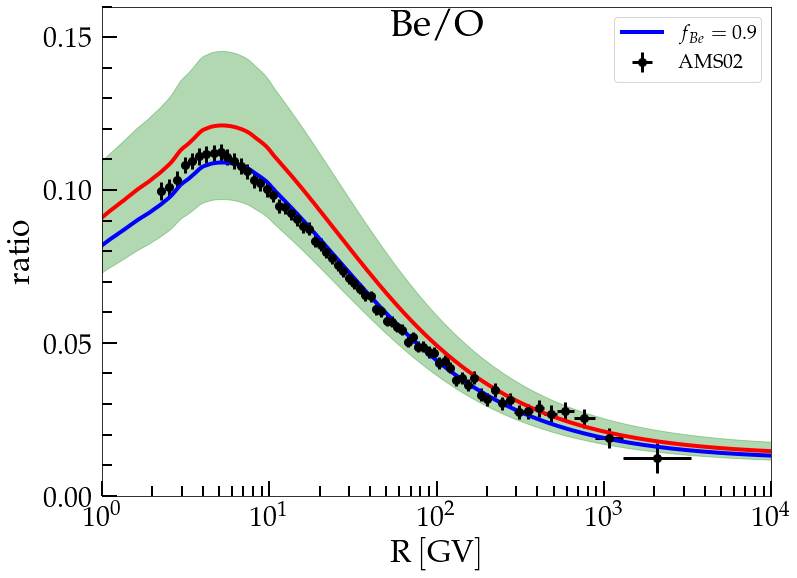}\par 
\end{multicols}
\begin{multicols}{2}
\includegraphics[width=\linewidth]{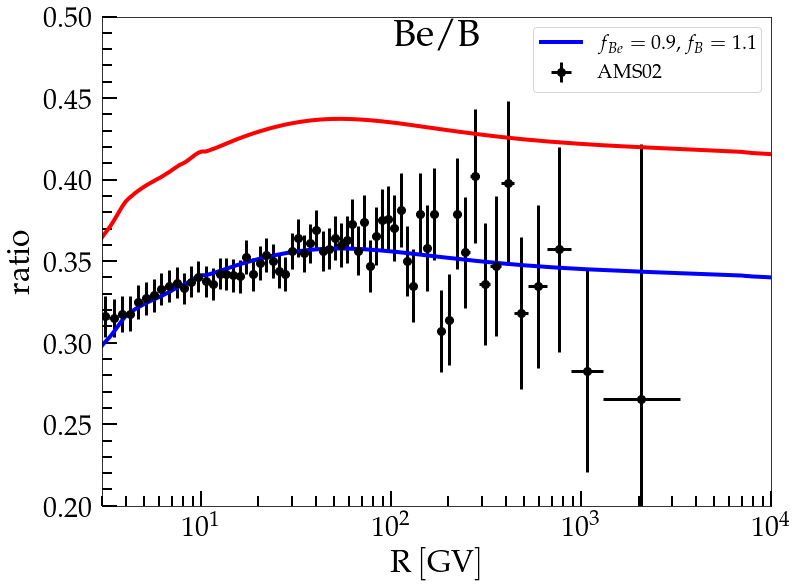}\par 
\includegraphics[width=\linewidth]{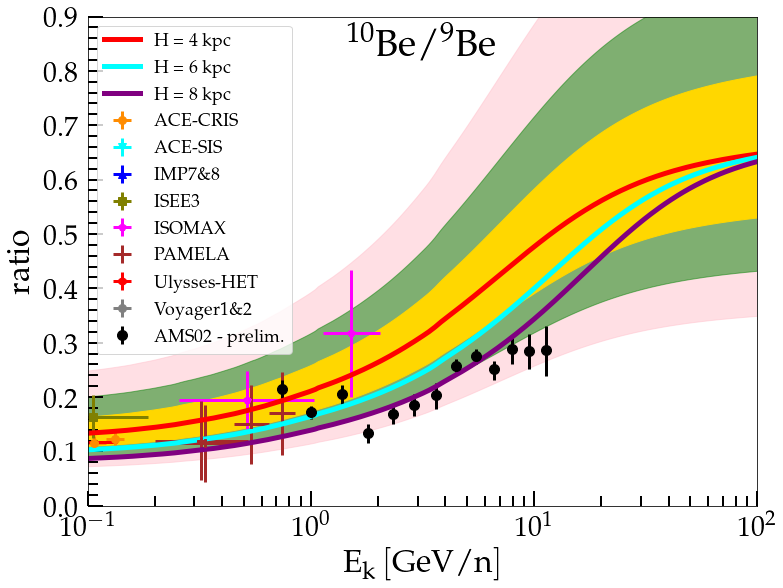}\par 
\end{multicols}
\caption{
Be/C (top left panel), Be/O (top right panel) and Be/B (bottom left panel) ratio (red solid line) compared to the data  of AMS02  (\protect{\citealt{AMS02-5yr-LiBeB, AMS02-7yr}}, black dots).
The green shaded region corresponds a  $20\%$ uncertainty in the secondary production cross-section. The solid blue line corresponds to the Be/C, Be/O or Be/B ratio when the Be flux is multiplied by $f_{Be} =  0.9$  and the B flux by $f_{B} =  1.9$ as in \protect{Fig.~\ref{fig:CNO-LiBeB}}. The \isotope{10}{Be}/\isotope{9}{Be} ratio (
bottom right panel) is compared to preliminary AMS02 data (black dots) and to the data of a variety of other experiments, collected from the Cosmic-Ray Database \citep{Maurin-2023-CRDB}. The solid red line corresponds to a GH size of $H= 4\, \rm kpc$. The yellow, green and pink shaded regions corresponds to a $10\%$, $20\%$ and $30\%$ uncertainty in the production cross-section, respectively. The solid cyan and purple lines corresponds to $H= 6\, \rm kpc$  and $H= 8\, \rm kpc$, respectively.
}
\label{fig:Be}
\end{figure*}

With such re-normalization we also find a very good agreement with AMS02 data for the $\rm B/C$ and $\rm B/O$ ratio (Fig.~\ref{fig:B-C-O-ratio}) and with AMS02 data for the $\rm Be/C$, $\rm Be/O$ and $\rm Be/B$  ratio (Fig.~\ref{fig:Be}). Notice in particular the high-$R$ flattening of the $\rm B/C$ related to the flattening of the grammage. \\

\isotope{10}{Be} is a long-lived unstable isotope (half-life of 1.39 Myr) which is of great importance in determining the CR residence time  in the Galaxy. 
Indeed, as shown in Appendix~\ref{sec:appendix-unstable-nuclei}, where we derive the solution of the transport equation in the case of unstable isotopes, limiting for simplicity the discussion to a diffusive propagation in a single halo,
the flux of unstable  isotopes scales as: 
\begin{equation}
    I_{\alpha 0} \propto 
    \begin{cases}
         \frac{H}{D} & \qquad\text{for $\tau_{\rm diff} \ll \tau_{r} $ }\\
    \frac{\tau_r}{\sqrt{D\, \tau_r}} & \qquad \text{for $\tau_{\rm diff} \gg \tau_{r} $ },
    \end{cases}
\end{equation}
where $\tau_{\rm diff} = H^2/D$ is the diffusion timescale and $\tau_r$ is the decay timescale. Note that the former limit corresponds to the case of stable isotopes, and is reached at high enough energy, since the decay time increases with the total energy, $\tau_r \propto E$. Thus, the flux of stable isotopes is sensitive to the ratio $H/D$ and not to the two quantities separately. 
Instead, measuring unstable isotopes, or the ratio between stable and unstable isotopes of the same nucleus provides an additional information that allows to determine the value of $H$.  The ratio between \isotope{10}{Be} and the stable isotope \isotope{9}{Be} tends to
\begin{equation}
    \frac{\rm ^{10}Be}{\rm ^{9}Be} \propto 
    \begin{cases}
         \text{const} & \qquad\text{for $\tau_{\rm diff} \ll \tau_{r} $ }\\
         \\
    \frac{\sqrt{D\, \tau_r}}{H} & \qquad \text{for $\tau_{\rm diff} \gg \tau_{r} $ }.
    \end{cases}
\end{equation}
In Fig.~\ref{fig:Be} we show our prediction for the \isotope{10}{Be}/\isotope{9}{Be}, with the benchmark parameters reported in Table~\ref{tab:param}, and in particular for $H=4\,\rm kpc$ (red curve), as compared to preliminary recent AMS02 data and to a variety of other data \citep{Maurin-2023-CRDB}. We also consider the large uncertainties in the production cross sections discussed above, and show the change of the ratio when the cross section for each isotope separately are increase or decreased by  $10\%$(yellow), $20\%$(green) and $30\%$(pink) band.  Given the large spread of data at low $E_k$, our benchmark result is compatible with such data, but is substantially larger than the AMS02 data. While the uncertainty in the cross section may ease the mismatch, a better agreement with the AMS02 data can be achieved by increasing the halo size to $H = 6-8$ kpc, as shown by the cyan and purple curves (see also the discussion by \citealt{Mertsch-2023-Be10-low-diff}). 
All other fluxes will not be affected by a rescaling of $H \rightarrow x\,H$ provided that (keeping $h$ and $D_H$ fixed)
\begin{description}
    \item $n_d \rightarrow n_d/x$
    \item $D_h \rightarrow D_h/x$
    \item $u \rightarrow u/x$.
\end{description}

%%%%%%%%%%%%%%%%%%%%%%%%%

\subsection*{Heavy nuclei}
\begin{figure*}
\begin{multicols}{2}
\includegraphics[width=\linewidth]{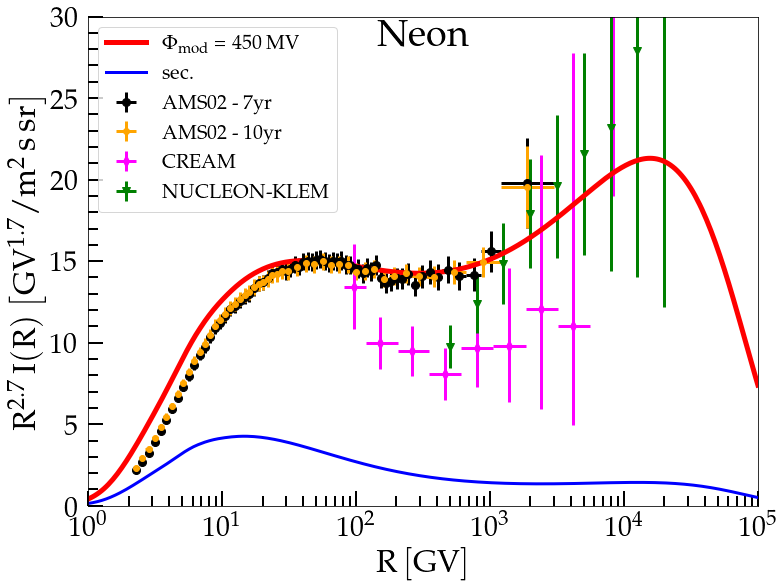}\par     \includegraphics[width=\linewidth]{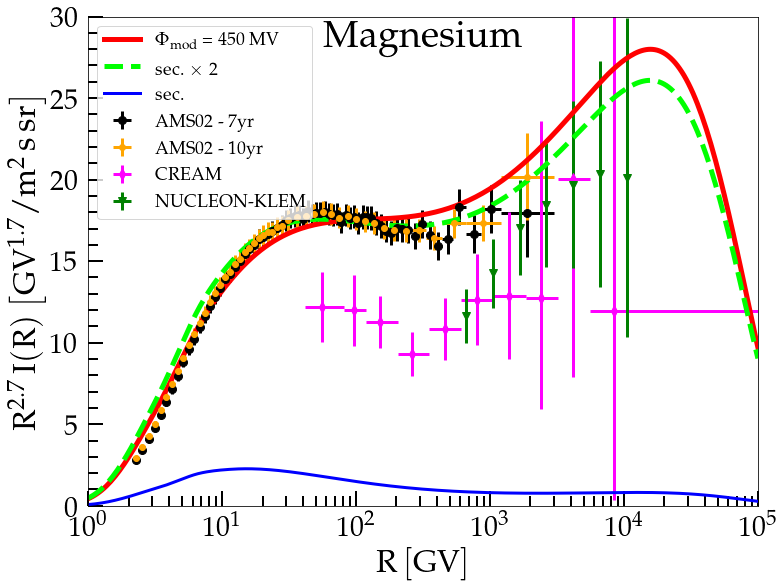}\par 
    \end{multicols}
\begin{multicols}{2}
\includegraphics[width=\linewidth]{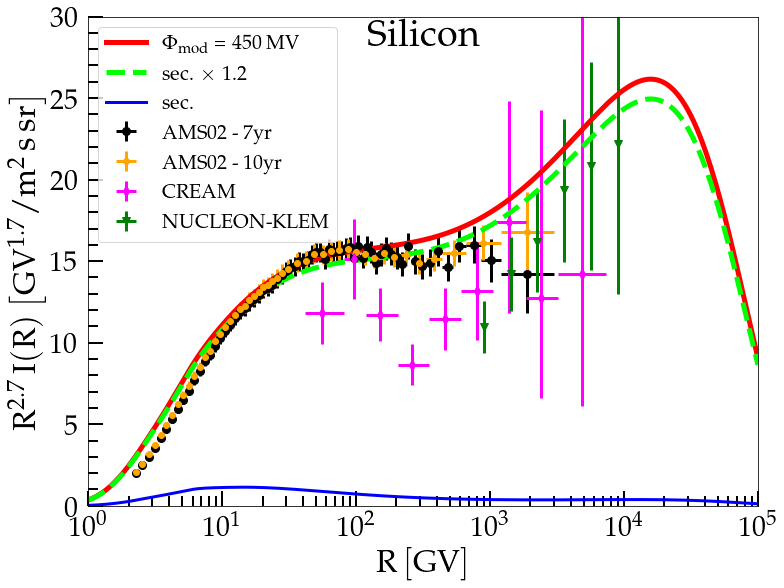}\par
\includegraphics[width=\linewidth]{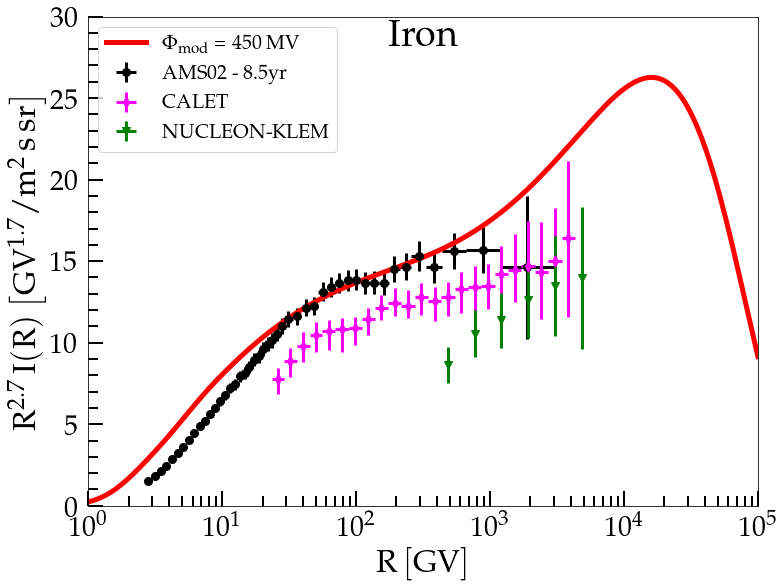}\par
\end{multicols}
\caption{Ne, Mg, Si and Fe modulated fluxes (red solid line) and their secondary component (blue solid line). 
%%%
\textit{Top left panel}: Ne  flux compared to the data of AMS02 (\protect{\citealt{AMS02-7yr, AMS02-10yr-heavy}}, black  dots, yellow diamonds), CREAM (\protect{\citealt{CREAM-2009-CNO-heavy}}, magenta diamonds) and NUCLEON (\protect{\citealt{NUCLEON-2019}}, green triangles). 
%%%%
\textit{Top right panel}: Mg modulated flux compared to the data of AMS02 (\protect{\citealt{AMS02-7yr, AMS02-10yr-heavy}}, black  dots, yellow diamonds), CREAM (\protect{\citealt{CREAM-2009-CNO-heavy}}, magenta diamonds) and NUCLEON (\protect{\citealt{NUCLEON-2019}}, green triangles). 
The dashed green line is obtained by multiplying the secondary component by $2$.
%%%%
\textit{Bottom left panel}: Si  flux compared to the data of AMS02 (\protect{\citealt{AMS02-7yr, AMS02-10yr-heavy}}, black  dots, yellow diamonds), CREAM (\protect{\citealt{CREAM-2009-CNO-heavy}}, magenta diamonds) and NUCLEON (\protect{\citealt{NUCLEON-2019}}, green triangles). 
The dashed green line is obtained by multiplying the secondary component by $1.2$.
%%%%
\textit{Bottom right panel}: Fe  flux compared to the data of AMS02 (\protect{\citealt{AMS02-8.5yr-Fe}}, black  dots), CALET (\protect{\citealt{CALET-2021-Fe}}, magenta diamonds) and NUCLEON (\protect{\citealt{NUCLEON-2019}}, green triangles). 
}
\label{fig:Ne-Mg-Si-Fe}
\end{figure*}

In Fig.~\ref{fig:Ne-Mg-Si-Fe} we show our results for the Ne, Mg, Si and Fe  fluxes.  For Fe we consider only the isotope \isotope{56}{Fe} and we treat it as purely primary.  
In general, always keeping in mind the  discrepancies between data sets,  we obtain a quite good agreement with both the AMS02 and NUCLEON data, except for the  case of Fe, where the NUCLEON data are systematically well below the AMS02 data. 
A $\approx 20\%$ uncertainty on the production cross sections should also be accounted for \citep{Genolini-2023-xsec-prod}. Moreover, the secondary contribution to these nuclei (blue curve in the plots) may be somewhat underestimated, considering that many fragmentation channels are involved starting from \isotope{56}{Fe} and the possible (unknown) systematics related to each channel. For instance, a better agreement with data can be obtained in the case of Si and Mg by multiplying the secondary flux by 1.2 and 2, respectively, and changing the primary flux accordingly (green dashed line). 

\begin{figure*}
\begin{multicols}{2}
    \includegraphics[width=\linewidth]{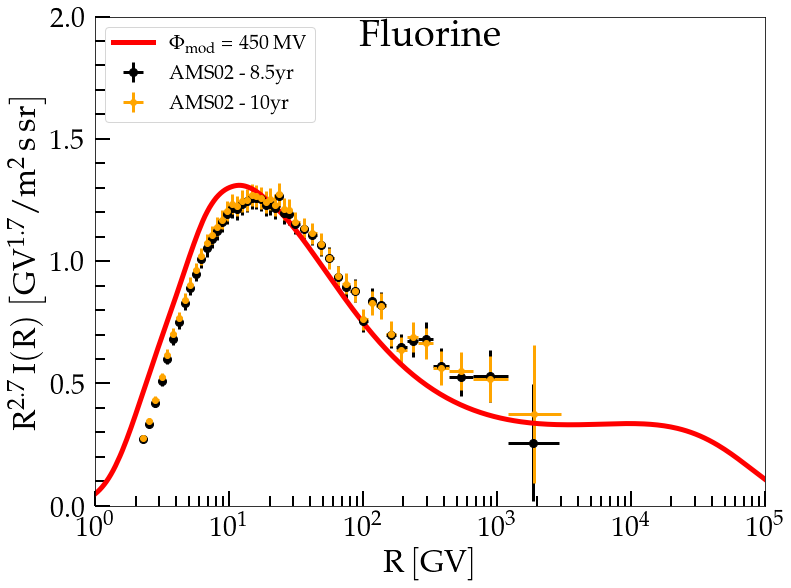}\par 
    \includegraphics[width=\linewidth]{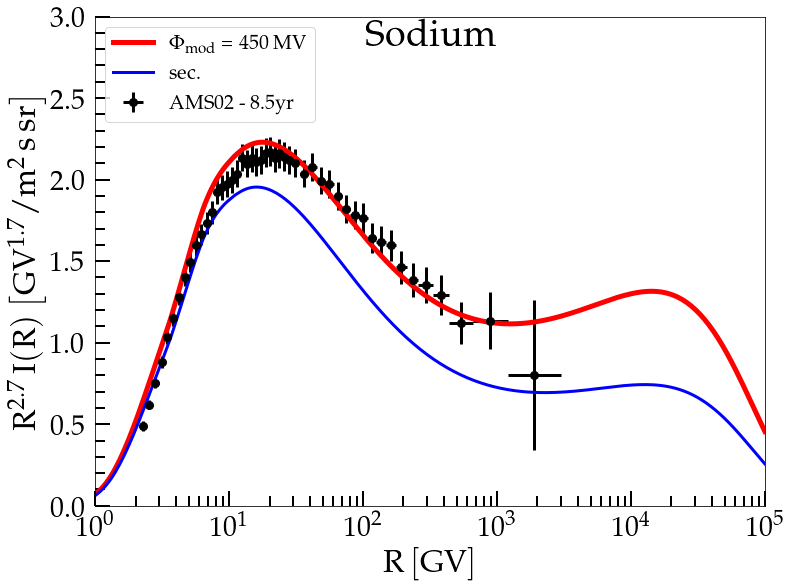}\par 
    \end{multicols}
\begin{multicols}{2}
    \includegraphics[width=\linewidth]{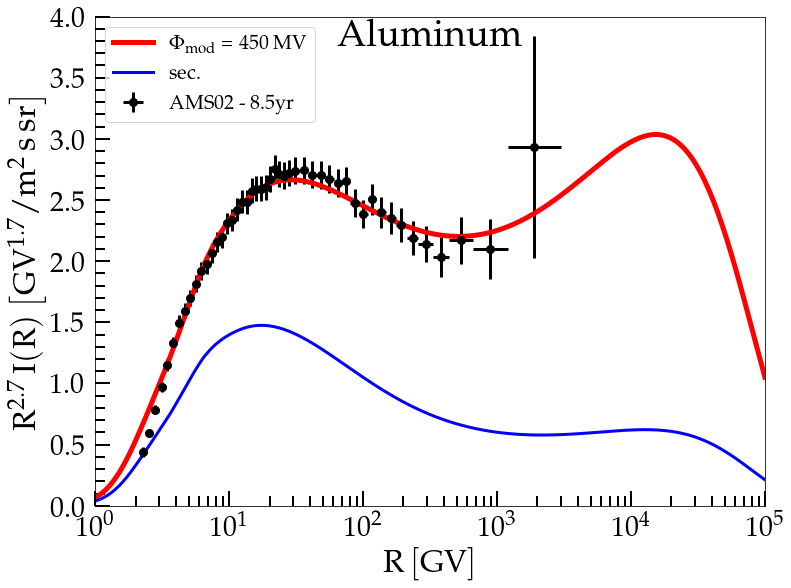}\par
    \includegraphics[width=\linewidth]{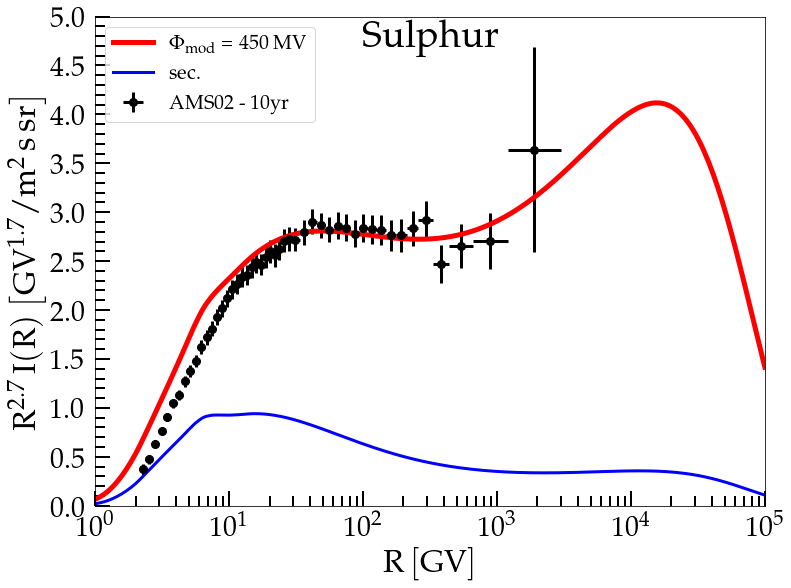}\par
\end{multicols}
\caption{
\textit{Top left panel}: F modulated flux (red solid line) compared to the data of AMS02 (\protect{\citealt{AMS02-8.5yr-F, AMS02-10yr-heavy}}, black  dots, yellow diamonds).
%%%%
\textit{Top right panel}: Na modulated flux (red solid line) and its secondary component (blue solid line) compared to the data of AMS02 (\protect{\citealt{AMS02-8.5yr-NaAlN}})
%%%%
\textit{Bottom left panel}: Al modulated flux (red solid line) and its secondary component (blue solid line) compared to the data of AMS02 (\protect{\citealt{AMS02-8.5yr-NaAlN}})
%%%%
\textit{Bottom right panel}: S modulated flux (red solid line) and its secondary component (blue solid line) compared to the data of AMS02 (\protect{\citealt{AMS02-10yr-heavy}}).
}
\label{fig:F-Na-Al-S}
\end{figure*}

In Fig.~\ref{fig:F-Na-Al-S}
we show the F , Na, Al and S fluxes, compared to AMS02 data. Considering the uncertainties in the production cross sections, particularly relevant for the predominantly secondary F, we reproduce the data quite well.\\ 

As a consistency check of our approach and of the considered production channels,   in Fig.~\ref{fig:Ar-K} and Fig.~\ref{fig:Mn-V} we show our predictions for the Ar, Ca, Cl, Cr, K, Mn, P, Sc, Ti, V fluxes. For such nuclei data are mostly  available at $E_k \lesssim  50 \, \rm GeV/n$, while higher energy data, if present, show large uncertainties \citep{Maurin-2023-CRDB}. 

\section{Conclusions}
\label{sec:conclusion}

Motivated by the recent report by DAMPE  of a  spectral steepening in the proton and He spectra at  $\rm \approx 15~TV$, and by the flattening in the  B/C ratio at $\approx$ 1 TeV/n reported by AMS02, CALET and DAMPE,  we proposed a novel global framework for the interpretation of current CR data, from $\sim$ GV up to multi-PV rigidity. \\

The first ingredient of our model, is the idea that the DAMPE break may reflect the 
maximum CR rigidity reached in the bulk of SNRs, while 
only a fraction of CR sources (e.g. peculiar SNRs, star clusters, ...) are assumed to accelerate particles up multi-PV, in agreement with state-of-the-art  models of CR acceleration and $\gamma-$ ray observations.
This is rather in disagreement with the standard paradigm of the origin of CRs, where SNRs as a population are thought to be able to accelerate particle up to the \textit{knee} and the DAMPE break is thought to reflect either the peculiarity of some local source either a change in the propagation regime.\\

The second pillar of our 
scenario is a  new perspective on the role played by the GD in the propagation of CRs,  in particular in the TV-PV range. Indeed, in the standard scenario of Galactic CR transport, the GH is considered as the diffusion region, and the time spent by CRs in the GD is supposed to be due to the repeated crossings induced by scattering in the GH. In such view, the GD is mostly considered as   thin target for spallation reactions, but it is a passive entity for what concerns CR transport.
Instead,  motivated by the possibility, previously discussed in the literature, that severe damping processes may significantly suppress  turbulence at the resonant scale in the  GD, we propose a model where particles in such region  experience a very weak scattering along magnetic field lines, which mostly lie parallel to the GP. At the same time, large scale turbulence are expected to induce a diffusive motion perpendicular to the average Galactic field (and to the GP) which is energy-independent (or depends very weakly on the particle energy). 

The intriguing result of such setup is that beyond $\approx $ TV the CR spectrum is shaped by the energy-independent propagation in the GD rather than by the repeated crossings of the disk induced by the CR diffusion in the GH. This naturally results in a smooth progressive hardening of the spectra of (mostly) primary species (p, He, C, O, Ne, Mg, Si, Fe) above $\approx$ hundreds GV, and in a flattening of the spectra of (mostly) secondary species (Li, Be, B, F) and of quantities such as the  B/C ratio  above $\approx 1$  TV, in agreement with data. Moreover, such diffusive motion in the GD perpendicular to the GP, provides a minimum, energy-independent residence time (which translates in a minimum grammage), which corresponds to the residence time experienced by  very high energy CRs, for which scattering in the GH becomes  ineffective (since the diffusive mean free path becomes larger than the halo size).

Interestingly, in order to fit CR data, we found an effective  diffusion coefficient in the GD (perpendicular to the GP) of the order of $\approx 10^{28}\, \rm cm^2/s$, which is  close to the diffusion coefficient of $\approx 10-100$ TeV leptons inferred from pulsar TeV halos, and is $\sim$ 100 times smaller that the diffusion coefficient in the GH at those energies. This, on the one hand, provides some observational support to our model, and on the other hand, clearly shows that the investigation of pulsar TeV halos provides a unique tool to study the  propagation of multi-TV CRs, for which information from quantities such as the B/C ratio is still lacking.\\

The third central ingredient of our approach is the fact that, while CRs in the GD move  diffusively perpendicular to the GP, they are assumed to move rather quickly along the GP. This means that, even high energy particles, that are little scattered in the GH, may come from a region around us $\gg h$. 
This may have a substantial impact in keeping the CR anisotropy small, since  the number of contributing sources may remain relatively large even at PV rigidity. 

Indeed, let us assume  a typical rate of SNe explosion in the GD (radius $10\, \rm kpc$) of $\nu_{\rm SNe} = 1/30\, \rm yr$, that $\xi_{\rm PV} \sim 15\%$ of them accelerate to PV and that the residence time in the GH is  as in Eq.~(\ref{eq:Tmin}). If in such time  particles travel along the GP a distance $d \approx 1\, \rm kpc$ (see Eq.~\ref{eq:par-distance-GD} in Sec.~\ref{sec:prop-disk}), the average number of contributing sources is
\begin{equation}
    \langle N_{\rm PV} \rangle \approx 10 \left(\frac{\xi_{\rm PV}}{0.15}\right) \left(\frac{\nu_{\rm SNe}}{1/30\, \rm yr}\right) \left(\frac{h}{150\, \rm pc}\right)^2 \left(\frac{\rm pc}{D_m}\right) \left(\frac{d}{\rm kpc}\right)^2,
\end{equation}
which  becomes $ \langle N_{\rm PV} \rangle \approx 40$ for $d= 2\, \rm kpc$.

Explaining the CR anisotropy is a crucial test for our approach, and we defer its investigation to a forthcoming paper.

\begin{acknowledgements} \end{acknowledgements}


\bibliographystyle{aa}
\bibliography{biblio} 

\begin{thebibliography}{89}
\expandafter\ifx\csname natexlab\endcsname\relax\def\natexlab#1{#1}\fi

\bibitem[{{Aartsen} {et~al.}(2019){Aartsen}, {Ackermann}, {Adams}, {Aguilar},
  {Ahlers}, {Ahrens}, {Alispach}, {Andeen}, {Anderson}, {Ansseau}, {Anton},
  {Arg{\"u}elles}, {Auffenberg}, {Axani}, {Backes}, {Bagherpour}, {Bai},
  {Barbano}, {Barwick}, {Baum}, {Baur}, {Bay}, {Beatty}, {Becker}, {Becker
  Tjus}, {BenZvi}, {Berley}, {Bernardini}, {Besson}, {Binder}, {Bindig},
  {Blaufuss}, {Blot}, {Bohm}, {B{\"o}rner}, {B{\"o}ser}, {Botner},
  {B{\"o}ttcher}, {Bourbeau}, {Bourbeau}, {Bradascio}, {Braun}, {Bretz},
  {Bron}, {Brostean-Kaiser}, {Burgman}, {Buscher}, {Busse}, {Carver}, {Chen},
  {Cheung}, {Chirkin}, {Clark}, {Classen}, {Collin}, {Conrad}, {Coppin},
  {Correa}, {Cowen}, {Cross}, {Dave}, {de Andr{\'e}}, {De Clercq}, {DeLaunay},
  {Dembinski}, {Deoskar}, {De Ridder}, {Desiati}, {de Vries}, {de Wasseige},
  {de With}, {DeYoung}, {Diaz}, {D{\'\i}az-V{\'e}lez}, {Dujmovic}, {Dunkman},
  {Dvorak}, {Eberhardt}, {Ehrhardt}, {Eller}, {Evenson}, {Fahey}, {Fazely},
  {Felde}, {Feusels}, {Filimonov}, {Finley}, {Franckowiak}, {Friedman},
  {Fritz}, {Gaisser}, {Gallagher}, {Ganster}, {Garrappa}, {Gerhardt},
  {Ghorbani}, {Glauch}, {Gl{\"u}senkamp}, {Goldschmidt}, {Gonzalez}, {Grant},
  {Griffith}, {G{\"u}nder}, {G{\"u}nd{\"u}z}, {Haack}, {Hallgren}, {Halve},
  {Halzen}, {Hanson}, {Hebecker}, {Heereman}, {Heix}, {Helbing}, {Hellauer},
  {Henningsen}, {Hickford}, {Hignight}, {Hill}, {Hoffman}, {Hoffmann},
  {Hoinka}, {Hokanson-Fasig}, {Hoshina}, {Huang}, {Huber}, {Hultqvist},
  {H{\"u}nnefeld}, {Hussain}, {In}, {Iovine}, {Ishihara}, {Jacobi},
  {Japaridze}, {Jeong}, {Jero}, {Jones}, {Jonske}, {Joppe}, {Kang}, {Kappes},
  {Kappesser}, {Karg}, {Karl}, {Karle}, {Katz}, {Kauer}, {Kelley},
  {Kheirandish}, {Kim}, {Kintscher}, {Kiryluk}, {Kittler}, {Klein}, {Koirala},
  {Kolanoski}, {K{\"o}pke}, {Kopper}, {Kopper}, {Koskinen}, {Kowalski},
  {Krings}, {Kr{\"u}ckl}, {Kulacz}, {Kunwar}, {Kurahashi}, {Kyriacou},
  {Labare}, {Lanfranchi}, {Larson}, {Lauber}, {Lazar}, {Leonard}, {Leuermann},
  {Liu}, {Lohfink}, {Lozano Mariscal}, {Lu}, {Lucarelli}, {L{\"u}nemann},
  {Luszczak}, {Madsen}, {Maggi}, {Mahn}, {Makino}, {Mallik}, {Mallot},
  {Mancina}, {Mari{\c{s}}}, {Maruyama}, {Mase}, {Maunu}, {Meagher}, {Medici},
  {Medina}, {Meier}, {Meighen-Berger}, {Menne}, {Merino}, {Meures}, {Miarecki},
  {Micallef}, {Moment{\'e}}, {Montaruli}, {Moore}, {Morse}, {Moulai}, {Muth},
  {Nagai}, {Nahnhauer}, {Nakarmi}, {Naumann}, {Neer}, {Niederhausen},
  {Nowicki}, {Nygren}, {Obertacke Pollmann}, {Olivas}, {O'Murchadha},
  {O'Sullivan}, {Palczewski}, {Pandya}, {Pankova}, {Park}, {Peiffer},
  {P{\'e}rez de los Heros}, {Philippen}, {Pieloth}, {Pinat}, {Pizzuto}, {Plum},
  {Porcelli}, {Price}, {Przybylski}, {Raab}, {Raissi}, {Rameez}, {Rauch},
  {Rawlins}, {Rea}, {Reimann}, {Relethford}, {Renzi}, {Resconi}, {Rhode},
  {Richman}, {Robertson}, {Rongen}, {Rott}, {Ruhe}, {Ryckbosch}, {Rysewyk},
  {Safa}, {Sanchez Herrera}, {Sandrock}, {Sandroos}, {Santander}, {Sarkar},
  {Sarkar}, {Satalecka}, {Schaufel}, {Schlunder}, {Schmidt}, {Schneider},
  {Schneider}, {Schumacher}, {Sclafani}, {Seckel}, {Seunarine}, {Shefali},
  {Silva}, {Snihur}, {Soedingrekso}, {Soldin}, {Song}, {Spiczak}, {Spiering},
  {Stachurska}, {Stamatikos}, {Stanev}, {Stasik}, {Stein}, {Stettner},
  {Steuer}, {Stezelberger}, {Stokstad}, {St{\"o}{\ss}l}, {Strotjohann},
  {St{\"u}rwald}, {Stuttard}, {Sullivan}, {Sutherland}, {Taboada}, {Tenholt},
  {Ter-Antonyan}, {Terliuk}, {Tilav}, {Tomankova}, {T{\"o}nnis}, {Toscano},
  {Tosi}, {Tselengidou}, {Tung}, {Turcati}, {Turcotte}, {Turley}, {Ty},
  {Unger}, {Unland Elorrieta}, {Usner}, {Vandenbroucke}, {Van Driessche}, {van
  Eijk}, {van Eijndhoven}, {Vanheule}, {van Santen}, {Vraeghe}, {Walck},
  {Wallace}, {Wallraff}, {Wandkowsky}, {Watson}, {Weaver}, {Weiss}, {Weldert},
  {Wendt}, {Werthebach}, {Westerhoff}, {Whelan}, {Whitehorn}, {Wiebe},
  {Wiebusch}, {Wille}, {Williams}, {Wills}, {Wolf}, {Wood}, {Wood},
  {Woschnagg}, {Wrede}, {Xu}, {Xu}, {Xu}, {Yanez}, {Yodh}, {Yoshida}, {Yuan},
  {Z{\"o}cklein}, \& {IceCube Collaboration}}]{Icecube-Icetop-2019}
{Aartsen}, M.~G., {Ackermann}, M., {Adams}, J., {et~al.} 2019, \prd, 100,
  082002

\bibitem[{{Abeysekara} {et~al.}(2017){Abeysekara}, {Albert}, {Alfaro},
  {Alvarez}, {{\'A}lvarez}, {Arceo}, {Arteaga-Vel{\'a}zquez}, {Avila Rojas},
  {Ayala Solares}, {Barber}, {Bautista-Elivar}, {Becerril}, {Belmont-Moreno},
  {BenZvi}, {Berley}, {Bernal}, {Braun}, {Brisbois}, {Caballero-Mora},
  {Capistr{\'a}n}, {Carrami{\~n}ana}, {Casanova}, {Castillo}, {Cotti},
  {Cotzomi}, {Couti{\~n}o de Le{\'o}n}, {De Le{\'o}n}, {De la Fuente},
  {Dingus}, {DuVernois}, {D{\'\i}az-V{\'e}lez}, {Ellsworth}, {Engel},
  {Enr{\'\i}quez-Rivera}, {Fiorino}, {Fraija}, {Garc{\'\i}a-Gonz{\'a}lez},
  {Garfias}, {Gerhardt}, {Gonz{\'a}lez Mu{\~n}oz}, {Gonz{\'a}lez}, {Goodman},
  {Hampel-Arias}, {Harding}, {Hern{\'a}ndez}, {Hern{\'a}ndez-Almada}, {Hinton},
  {Hona}, {Hui}, {H{\"u}ntemeyer}, {Iriarte}, {Jardin-Blicq}, {Joshi},
  {Kaufmann}, {Kieda}, {Lara}, {Lauer}, {Lee}, {Lennarz}, {Vargas},
  {Linnemann}, {Longinotti}, {Luis Raya}, {Luna-Garc{\'\i}a}, {L{\'o}pez-Coto},
  {Malone}, {Marinelli}, {Martinez}, {Martinez-Castellanos},
  {Mart{\'\i}nez-Castro}, {Mart{\'\i}nez-Huerta}, {Matthews},
  {Miranda-Romagnoli}, {Moreno}, {Mostaf{\'a}}, {Nellen}, {Newbold}, {Nisa},
  {Noriega-Papaqui}, {Pelayo}, {Pretz}, {P{\'e}rez-P{\'e}rez}, {Ren}, {Rho},
  {Rivi{\`e}re}, {Rosa-Gonz{\'a}lez}, {Rosenberg}, {Ruiz-Velasco}, {Salazar},
  {Salesa Greus}, {Sandoval}, {Schneider}, {Schoorlemmer}, {Sinnis}, {Smith},
  {Springer}, {Surajbali}, {Taboada}, {Tibolla}, {Tollefson}, {Torres},
  {Ukwatta}, {Vianello}, {Weisgarber}, {Westerhoff}, {Wisher}, {Wood},
  {Yapici}, {Yodh}, {Younk}, {Zepeda}, {Zhou}, {Guo}, {Hahn}, {Li}, \&
  {Zhang}}]{HAWC-2017-Geminga}
{Abeysekara}, A.~U., {Albert}, A., {Alfaro}, R., {et~al.} 2017, Science, 358,
  911

\bibitem[{{Adriani} {et~al.}(2020){Adriani}, {Akaike}, {Asano}, {Asaoka},
  {Bagliesi}, {Berti}, {Bigongiari}, {Binns}, {Bongi}, {Brogi}, {Bruno},
  {Buckley}, {Cannady}, {Castellini}, {Checchia}, {Cherry}, {Collazuol},
  {Ebisawa}, {Fuke}, {Gonzi}, {Guzik}, {Hams}, {Hibino}, {Ichimura}, {Ioka},
  {Ishizaki}, {Israel}, {Kasahara}, {Kataoka}, {Kataoka}, {Katayose}, {Kato},
  {Kawanaka}, {Kawakubo}, {Kobayashi}, {Kohri}, {Krawczynski}, {Krizmanic},
  {Link}, {Maestro}, {Marrocchesi}, {Messineo}, {Mitchell}, {Miyake},
  {Moiseev}, {Mori}, {Mori}, {Motz}, {Munakata}, {Nakahira}, {Nishimura}, {de
  Nolfo}, {Okuno}, {Ormes}, {Ospina}, {Ozawa}, {Pacini}, {Palma}, {Papini},
  {Rauch}, {Ricciarini}, {Sakai}, {Sakamoto}, {Sasaki}, {Shimizu}, {Shiomi},
  {Sparvoli}, {Spillantini}, {Stolzi}, {Sugita}, {Suh}, {Sulaj}, {Takita},
  {Tamura}, {Terasawa}, {Torii}, {Tsunesada}, {Uchihori}, {Vannuccini},
  {Wefel}, {Yamaoka}, {Yanagita}, {Yoshida}, {Yoshida}, \& {Calet
  Collaboration}}]{CALET-2020-C-O}
{Adriani}, O., {Akaike}, Y., {Asano}, K., {et~al.} 2020, \prl, 125, 251102

\bibitem[{{Adriani} {et~al.}(2023){Adriani}, {Akaike}, {Asano}, {Asaoka},
  {Berti}, {Bigongiari}, {Binns}, {Bongi}, {Brogi}, {Bruno}, {Buckley},
  {Cannady}, {Castellini}, {Checchia}, {Cherry}, {Collazuol}, {de Nolfo},
  {Ebisawa}, {Ficklin}, {Fuke}, {Gonzi}, {Guzik}, {Hams}, {Hibino}, {Ichimura},
  {Ioka}, {Ishizaki}, {Israel}, {Kasahara}, {Kataoka}, {Kataoka}, {Katayose},
  {Kato}, {Kawanaka}, {Kawakubo}, {Kobayashi}, {Kohri}, {Krawczynski},
  {Krizmanic}, {Maestro}, {Marrocchesi}, {Messineo}, {Mitchell}, {Miyake},
  {Moiseev}, {Mori}, {Mori}, {Motz}, {Munakata}, {Nakahira}, {Nishimura},
  {Okuno}, {Ormes}, {Ozawa}, {Pacini}, {Papini}, {Rauch}, {Ricciarini},
  {Sakai}, {Sakamoto}, {Sasaki}, {Shimizu}, {Shiomi}, {Spillantini}, {Stolzi},
  {Sugita}, {Sulaj}, {Takita}, {Tamura}, {Terasawa}, {Torii}, {Tsunesada},
  {Uchihori}, {Vannuccini}, {Wefel}, {Yamaoka}, {Yanagita}, {Yoshida},
  {Yoshida}, {Zober}, \& {Calet Collaboration}}]{CALET-2023-He}
{Adriani}, O., {Akaike}, Y., {Asano}, K., {et~al.} 2023, \prl, 130, 171002

\bibitem[{{Adriani} {et~al.}(2022){Adriani}, {Akaike}, {Asano}, {Asaoka},
  {Berti}, {Bigongiari}, {Binns}, {Bongi}, {Brogi}, {Bruno}, {Buckley},
  {Cannady}, {Castellini}, {Checchia}, {Cherry}, {Collazuol}, {Ebisawa},
  {Ficklin}, {Fuke}, {Gonzi}, {Guzik}, {Hams}, {Hibino}, {Ichimura}, {Ioka},
  {Ishizaki}, {Israel}, {Kasahara}, {Kataoka}, {Kataoka}, {Katayose}, {Kato},
  {Kawanaka}, {Kawakubo}, {Kobayashi}, {Kohri}, {Krawczynski}, {Krizmanic},
  {Maestro}, {Marrocchesi}, {Messineo}, {Mitchell}, {Miyake}, {Moiseev},
  {Mori}, {Mori}, {Motz}, {Munakata}, {Nakahira}, {Nishimura}, {de Nolfo},
  {Okuno}, {Ormes}, {Ozawa}, {Pacini}, {Papini}, {Rauch}, {Ricciarini},
  {Sakai}, {Sakamoto}, {Sasaki}, {Shimizu}, {Shiomi}, {Spillantini}, {Stolzi},
  {Sugita}, {Sulaj}, {Takita}, {Tamura}, {Terasawa}, {Torii}, {Tsunesada},
  {Uchihori}, {Vannuccini}, {Wefel}, {Yamaoka}, {Yanagita}, {Yoshida},
  {Yoshida}, {Zober}, \& {Calet Collaboration}}]{CALET-2022-H}
{Adriani}, O., {Akaike}, Y., {Asano}, K., {et~al.} 2022, \prl, 129, 101102

\bibitem[{{Adriani} {et~al.}(2021){Adriani}, {Akaike}, {Asano}, {Asaoka},
  {Berti}, {Bigongiari}, {Binns}, {Bongi}, {Brogi}, {Bruno}, {Buckley},
  {Cannady}, {Castellini}, {Checchia}, {Cherry}, {Collazuol}, {Ebisawa},
  {Fuke}, {Gonzi}, {Guzik}, {Hams}, {Hibino}, {Ichimura}, {Ioka}, {Ishizaki},
  {Israel}, {Kasahara}, {Kataoka}, {Kataoka}, {Katayose}, {Kato}, {Kawanaka},
  {Kawakubo}, {Kobayashi}, {Kohri}, {Krawczynski}, {Krizmanic}, {Link},
  {Maestro}, {Marrocchesi}, {Messineo}, {Mitchell}, {Miyake}, {Moiseev},
  {Mori}, {Mori}, {Motz}, {Munakata}, {Nakahira}, {Nishimura}, {de Nolfo},
  {Okuno}, {Ormes}, {Ospina}, {Ozawa}, {Pacini}, {Papini}, {Rauch},
  {Ricciarini}, {Sakai}, {Sakamoto}, {Sasaki}, {Shimizu}, {Shiomi},
  {Spillantini}, {Stolzi}, {Sugita}, {Sulaj}, {Takita}, {Tamura}, {Terasawa},
  {Torii}, {Tsunesada}, {Uchihori}, {Vannuccini}, {Wefel}, {Yamaoka},
  {Yanagita}, {Yoshida}, {Yoshida}, \& {Calet Collaboration}}]{CALET-2021-Fe}
{Adriani}, O., {Akaike}, Y., {Asano}, K., {et~al.} 2021, \prl, 126, 241101

\bibitem[{{Adriani} {et~al.}(2011){Adriani}, {Barbarino}, {Bazilevskaya},
  {Bellotti}, {Boezio}, {Bogomolov}, {Bonechi}, {Bongi}, {Bonvicini},
  {Borisov}, {Bottai}, {Bruno}, {Cafagna}, {Campana}, {Carbone}, {Carlson},
  {Casolino}, {Castellini}, {Consiglio}, {De Pascale}, {De Santis}, {De
  Simone}, {Di Felice}, {Galper}, {Gillard}, {Grishantseva}, {Jerse},
  {Karelin}, {Koldashov}, {Krutkov}, {Kvashnin}, {Leonov}, {Malakhov},
  {Malvezzi}, {Marcelli}, {Mayorov}, {Menn}, {Mikhailov}, {Mocchiutti},
  {Monaco}, {Mori}, {Nikonov}, {Osteria}, {Palma}, {Papini}, {Pearce},
  {Picozza}, {Pizzolotto}, {Ricci}, {Ricciarini}, {Rossetto}, {Sarkar},
  {Simon}, {Sparvoli}, {Spillantini}, {Stozhkov}, {Vacchi}, {Vannuccini},
  {Vasilyev}, {Voronov}, {Yurkin}, {Wu}, {Zampa}, {Zampa}, \&
  {Zverev}}]{PAMELA-2011-Hard}
{Adriani}, O., {Barbarino}, G.~C., {Bazilevskaya}, G.~A., {et~al.} 2011,
  Science, 332, 69

\bibitem[{Aguilar {et~al.}(2015)Aguilar, Aisa, Alpat, Alvino, Ambrosi, Andeen,
  Arruda, Attig, Azzarello, Bachlechner, Barao, Barrau, Barrin, Bartoloni,
  Basara, Battarbee, Battiston, Bazo, Becker, Behlmann, Beischer, Berdugo,
  Bertucci, Bigongiari, Bindi, Bizzaglia, Bizzarri, Boella, de~Boer, Bollweg,
  Bonnivard, Borgia, Borsini, Boschini, Bourquin, Burger, Cadoux, Cai, Capell,
  Caroff, Casaus, Cascioli, Castellini, Cernuda, Cerreta, Cervelli, Chae,
  Chang, Chen, Chen, Cheng, Chen, Cheng, Chou, Choumilov, Choutko, Chung,
  Clark, Clavero, Coignet, Consolandi, Contin, Corti, Gil, Coste, Creus,
  Crispoltoni, Cui, Dai, Delgado, Della~Torre, Demirk\"oz, Derome, Di~Falco,
  Di~Masso, Dimiccoli, D\'{\i}az, von Doetinchem, Donnini, Du, Duranti, D'Urso,
  Eline, Eppling, Eronen, Fan, Farnesini, Feng, Fiandrini, Fiasson, Finch,
  Fisher, Galaktionov, Gallucci, Garc\'{\i}a, Garc\'{\i}a-L\'opez, Gargiulo,
  Gast, Gebauer, Gervasi, Ghelfi, Gillard, Giovacchini, Goglov, Gong, Goy,
  Grabski, Grandi, Graziani, Guandalini, Guerri, Guo, Haas, Habiby, Haino, Han,
  He, Heil, Hoffman, Hsieh, Huang, Huh, Incagli, Ionica, Jang, Jinchi,
  Kanishev, Kim, Kim, Kirn, Kossakowski, Kounina, Kounine, Koutsenko, Krafczyk,
  La~Vacca, Laudi, Laurenti, Lazzizzera, Lebedev, Lee, Lee, Leluc, Levi, Li,
  Li, Li, Li, Li, Li, Li, Li, Li, Lim, Lin, Lipari, Lippert, Liu, Liu, Lolli,
  Lomtadze, Lu, Lu, Lu, Luebelsmeyer, Luo, Lv, Majka, Ma\~n\'a, Mar\'{\i}n,
  Martin, Mart\'{\i}nez, Masi, Maurin, Menchaca-Rocha, Meng, Mo, Morescalchi,
  Mott, M\"uller, Ni, Nikonov, Nozzoli, Nunes, Obermeier, Oliva, Orcinha,
  Palmonari, Palomares, Paniccia, Papi, Pauluzzi, Pedreschi, Pensotti, Pereira,
  Picot-Clemente, Pilo, Piluso, Pizzolotto, Plyaskin, Pohl, Poireau, Postaci,
  Putze, Quadrani, Qi, Qin, Qu, R\"aih\"a, Rancoita, Rapin, Ricol,
  Rodr\'{\i}guez, Rosier-Lees, Rozhkov, Rozza, Sagdeev, Sandweiss, Saouter,
  Sbarra, Schael, Schmidt, von Dratzig, Schwering, Scolieri, Seo, Shan, Shan,
  Shi, Shi, Shi, Siedenburg, Son, Spada, Spinella, Sun, Sun, Tacconi, Tang,
  Tang, Tang, Tao, Tescaro, Ting, Ting, Tomassetti, Torsti,
  T\"urko\ifmmode~\breve{g}\else \u{g}\fi{}lu, Urban, Vagelli, Valente,
  Vannini, Valtonen, Vaurynovich, Vecchi, Velasco, Vialle, Vitale, Vitillo,
  Wang, Wang, Wang, Wang, Wang, Wang, Weng, Whitman, Wienkenh\"over, Wu, Wu,
  Xia, Xie, Xie, Xiong, Xin, Xu, Xu, Yan, Yang, Yang, Ye, Yi, Yu, Yu, Zeissler,
  Zhang, Zhang, Zhang, Zhang, Zheng, Zhuang, Zhukov, Zichichi, Zimmermann,
  Zuccon, \& Zurbach}]{AMS02-2015-hard}
Aguilar, M., Aisa, D., Alpat, B., {et~al.} 2015, Phys. Rev. Lett., 114, 171103

\bibitem[{Aguilar {et~al.}(2013)Aguilar, Alberti, Alpat, Alvino, Ambrosi,
  Andeen, Anderhub, Arruda, Azzarello, Bachlechner, Barao, Baret, Barrau,
  Barrin, Bartoloni, Basara, Basili, Batalha, Bates, Battiston, Bazo, Becker,
  Becker, Behlmann, Beischer, Berdugo, Berges, Bertucci, Bigongiari, Biland,
  Bindi, Bizzaglia, Boella, de~Boer, Bollweg, Bolmont, Borgia, Borsini,
  Boschini, Boudoul, Bourquin, Brun, Bu\'enerd, Burger, Burger, Cadoux, Cai,
  Capell, Casadei, Casaus, Cascioli, Castellini, Cernuda, Cervelli, Chae,
  Chang, Chen, Chen, Chen, Cheng, Chen, Cheng, Chernoplyiokov, Chikanian,
  Choumilov, Choutko, Chung, Clark, Clavero, Coignet, Commichau, Consolandi,
  Contin, Corti, Costado~Dios, Coste, Crespo, Cui, Dai, Delgado, Della~Torre,
  Demirkoz, Dennett, Derome, Di~Falco, Diao, Diago, Djambazov, D\'{\i}az, von
  Doetinchem, Du, Dubois, Duperay, Duranti, D'Urso, Egorov, Eline, Eppling,
  Eronen, van Es, Esser, Falvard, Fiandrini, Fiasson, Finch, Fisher, Flood,
  Foglio, Fohey, Fopp, Fouque, Galaktionov, Gallilee, Gallin-Martel, Gallucci,
  Garc\'{\i}a, Garc\'{\i}a, Garc\'{\i}a-L\'opez, Garc\'{\i}a-Tabares, Gargiulo,
  Gast, Gebauer, Gentile, Gervasi, Gillard, Giovacchini, Girard, Goglov, Gong,
  Goy-Henningsen, Grandi, Graziani, Grechko, Gross, Guerri, de~la Gu\'{\i}a,
  Guo, Habiby, Haino, Hauler, He, Heil, Heilig, Hermel, Hofer, Huang,
  Hungerford, Incagli, Ionica, Jacholkowska, Jang, Jinchi, Jongmanns, Journet,
  Jungermann, Karpinski, Kim, Kim, Kirn, Kossakowski, Koulemzine, Kounina,
  Kounine, Koutsenko, Krafczyk, Laudi, Laurenti, Lauritzen, Lebedev, Lee, Lee,
  Leluc, Le\'on~Vargas, Lepareur, Li, Li, Li, Li, Li, Lipari, Lin, Liu, Liu,
  Lomtadze, Lu, Lucidi, L\"ubelsmeyer, Luo, Lustermann, Lv, Madsen, Majka,
  Malinin, Ma\~n\'a, Mar\'{\i}n, Martin, Mart\'{\i}nez, Masciocchi, Masi,
  Maurin, McInturff, McIntyre, Menchaca-Rocha, Meng, Menichelli, Mereu,
  Millinger, Mo, Molina, Mott, Mujunen, Natale, Nemeth, Ni, Nikonov, Nozzoli,
  Nunes, Obermeier, Oh, Oliva, Palmonari, Palomares, Paniccia, Papi, Park,
  Pauluzzi, Pauss, Pauw, Pedreschi, Pensotti, Pereira, Perrin, Pessina,
  Pierschel, Pilo, Piluso, Pizzolotto, Plyaskin, Pochon, Pohl, Poireau, Porter,
  Pouxe, Putze, Quadrani, Qi, Rancoita, Rapin, Ren, Ricol, Riihonen,
  Rodr\'{\i}guez, Roeser, Rosier-Lees, Rossi, Rozhkov, Rozza, Sabellek,
  Sagdeev, Sandweiss, Santos, Saouter, Sarchioni, Schael, Schinzel, Schmanau,
  Schwering, Schulz~von Dratzig, Scolieri, Seo, Shan, Shi, Shi, Siedenburg,
  Siedling, Son, Spada, Spinella, Steuer, Stiff, Sun, Sun, Sun, Tacconi, Tang,
  Tang, Tang, Tao, Tassan-Viol, Ting, Ting, Titus, Tomassetti, Toral, Torsti,
  Tsai, Tutt, Ulbricht, Urban, Vagelli, Valente, Vannini, Valtonen,
  Vargas~Trevino, Vaurynovich, Vecchi, Vergain, Verlaat, Vescovi, Vialle,
  Viertel, Volpini, Wang, Wang, Wang, Wang, Wang, Wang, Wallraff, Weng,
  Willenbrock, Wlochal, Wu, Wu, Wu, Xiao, Xie, Xiong, Xin, Xu, Xu, Yan, Yang,
  Yang, Ye, Yi, Yu, Yu, Zeissler, Zhang, Zhang, Zhang, Zheng, Zhuang, Zhukov,
  Zichichi, Zuccon, \& Zurbach}]{AMS02-2013-pos-frac}
Aguilar, M., Alberti, G., Alpat, B., {et~al.} 2013, Phys. Rev. Lett., 110,
  141102

\bibitem[{Aguilar {et~al.}(2016)Aguilar, Ali~Cavasonza, Alpat, Ambrosi, Arruda,
  Attig, Aupetit, Azzarello, Bachlechner, Barao, Barrau, Barrin, Bartoloni,
  Basara, Ba\ifmmode \mbox{\c{s}}\else \c{s}\fi{}e\ifmmode \check{g}\else
  \v{g}\fi{}mez-du Pree, Battarbee, Battiston, Bazo, Becker, Behlmann,
  Beischer, Berdugo, Bertucci, Bindi, Boella, de~Boer, Bollweg, Bonnivard,
  Borgia, Boschini, Bourquin, Bueno, Burger, Cadoux, Cai, Capell, Caroff,
  Casaus, Castellini, Cernuda, Cervelli, Chae, Chang, Chen, Chen, Chen, Cheng,
  Chou, Choumilov, Choutko, Chung, Clark, Clavero, Coignet, Consolandi, Contin,
  Corti, Coste, Creus, Crispoltoni, Cui, Dai, Delgado, Della~Torre, Demirk\"oz,
  Derome, Di~Falco, Dimiccoli, D\'{\i}az, von Doetinchem, Dong, Donnini,
  Duranti, D'Urso, Egorov, Eline, Eronen, Feng, Fiandrini, Finch, Fisher,
  Formato, Galaktionov, Gallucci, Garc\'{\i}a, Garc\'{\i}a-L\'opez, Gargiulo,
  Gast, Gebauer, Gervasi, Ghelfi, Giovacchini, Goglov, G\'omez-Coral, Gong,
  Goy, Grabski, Grandi, Graziani, Guerri, Guo, Habiby, Haino, Han, He, Heil,
  Hoffman, Hsieh, Huang, Huang, Huh, Incagli, Ionica, Jang, Jinchi, Kang,
  Kanishev, Kim, Kim, Kirn, Konak, Kounina, Kounine, Koutsenko, Krafczyk,
  La~Vacca, Laudi, Laurenti, Lazzizzera, Lebedev, Lee, Lee, Leluc, Li, Li, Li,
  Li, Li, Li, Li, Li, Lim, Lin, Lipari, Lippert, Liu, Liu, Lu, Lu,
  Luebelsmeyer, Luo, Luo, Lv, Majka, Ma\~n\'a, Mar\'{\i}n, Martin,
  Mart\'{\i}nez, Masi, Maurin, Menchaca-Rocha, Meng, Mo, Morescalchi, Mott,
  Nelson, Ni, Nikonov, Nozzoli, Nunes, Oliva, Orcinha, Palmonari, Palomares,
  Paniccia, Pauluzzi, Pensotti, Pereira, Picot-Clemente, Pilo, Pizzolotto,
  Plyaskin, Pohl, Poireau, Putze, Quadrani, Qi, Qin, Qu, R\"aih\"a, Rancoita,
  Rapin, Ricol, Rodr\'{\i}guez, Rosier-Lees, Rozhkov, Rozza, Sagdeev,
  Sandweiss, Saouter, Schael, Schmidt, Schulz~von Dratzig, Schwering, Seo,
  Shan, Shi, Siedenburg, Son, Song, Sun, Tacconi, Tang, Tang, Tao, Tescaro,
  Ting, Ting, Tomassetti, Torsti, T\"urko\ifmmode~\breve{g}\else \u{g}\fi{}lu,
  Urban, Vagelli, Valente, Vannini, Valtonen, V\'azquez~Acosta, Vecchi,
  Velasco, Vialle, Vitale, Vitillo, Wang, Wang, Wang, Wang, Wang, Wang, Wei,
  Weng, Whitman, Wienkenh\"over, Willenbrock, Wu, Wu, Xia, Xiong, Xu, Yan,
  Yang, Yang, Yang, Yi, Yu, Yu, Zeissler, Zhang, Zhang, Zhang, Zhang, Zhang,
  Zhang, Zheng, Zhu, Zhuang, Zhukov, Zichichi, Zimmermann, \&
  Zuccon}]{AMS02-2016-pbar}
Aguilar, M., Ali~Cavasonza, L., Alpat, B., {et~al.} 2016, Phys. Rev. Lett.,
  117, 091103

\bibitem[{{Aguilar} {et~al.}(2023){Aguilar}, {Ali Cavasonza}, {Alpat},
  {Ambrosi}, {Arruda}, {Attig}, {Bagwell}, {Barao}, {Barrin}, {Bartoloni},
  {Ba{\c{s}}e{\v{g}}mez-du Pree}, {Battiston}, {Belyaev}, {Berdugo},
  {Bertucci}, {Bindi}, {Bollweg}, {Bolster}, {Borchiellini}, {Borgia},
  {Boschini}, {Bourquin}, {Bueno}, {Burger}, {Burger}, {Cai}, {Capell},
  {Casaus}, {Castellini}, {Cervelli}, {Chang}, {Chen}, {Chen}, {Chen}, {Chen},
  {Chen}, {Cheng}, {Chou}, {Chouridou}, {Choutko}, {Chung}, {Clark}, {Coignet},
  {Consolandi}, {Contin}, {Corti}, {Cui}, {Dadzie}, {Dass}, {Delgado}, {Della
  Torre}, {Demirk{\"o}z}, {Derome}, {Di Falco}, {Di Felice}, {D{\'\i}az},
  {Dimiccoli}, {von Doetinchem}, {Dong}, {Donnini}, {Duranti}, {Egorov},
  {Eline}, {Faldi}, {Feng}, {Fiandrini}, {Fisher}, {Formato}, {G{\'a}mez},
  {Garc{\'\i}a-L{\'o}pez}, {Gargiulo}, {Gast}, {Gervasi}, {Giovacchini},
  {G{\'o}mez-Coral}, {Gong}, {Goy}, {Grabski}, {Grandi}, {Graziani}, {Guracho},
  {Haino}, {Han}, {Hashmani}, {He}, {Heber}, {Hsieh}, {Hu}, {Huang}, {Incagli},
  {Jang}, {Jia}, {Jinchi}, {Karag{\"o}z}, {Khiali}, {Kim}, {Kirn}, {Kounina},
  {Kounine}, {Koutsenko}, {Krasnopevtsev}, {Kuhlman}, {Kulemzin}, {La Vacca},
  {Laudi}, {Laurenti}, {LaVecchia}, {Lazzizzera}, {Lee}, {Lee}, {Li}, {Li},
  {Li}, {Li}, {Li}, {Li}, {Li}, {Li}, {Li}, {Li}, {Li}, {Liang}, {Liang},
  {Lin}, {Lippert}, {Liu}, {Lu}, {Lu}, {Luebelsmeyer}, {Luo}, {Luo}, {Luo},
  {Machate}, {Ma{\~n}{\'a}}, {Mar{\'\i}n}, {Marquardt}, {Martin},
  {Mart{\'\i}nez}, {Masi}, {Maurin}, {Medvedeva}, {Menchaca-Rocha}, {Meng},
  {Mikhailov}, {Molero}, {Mott}, {Mussolin}, {Negrete}, {Nikonov}, {Nozzoli},
  {Ocampo-Peleteiro}, {Oliva}, {Orcinha}, {Ottupara}, {Palermo}, {Palmonari},
  {Paniccia}, {Pashnin}, {Pauluzzi}, {Pensotti}, {Plyaskin}, {Poluianov},
  {Qin}, {Qu}, {Quadrani}, {Rancoita}, {Rapin}, {Reina Conde}, {Robyn},
  {Romaneehsen}, {Rozhkov}, {Rozza}, {Sagdeev}, {Schael}, {Schultz von
  Dratzig}, {Schwering}, {Seo}, {Shan}, {Siedenburg}, {Song}, {Song},
  {Sonnabend}, {Strigari}, {Su}, {Sun}, {Sun}, {Tacconi}, {Tang}, {Tang},
  {Tian}, {Tian}, {Ting}, {Ting}, {Tomassetti}, {Torsti}, {Urban}, {Usoskin},
  {Vagelli}, {Vainio}, {Valencia-Otero}, {Valente}, {Valtonen}, {V{\'a}zquez
  Acosta}, {Vecchi}, {Velasco}, {Vialle}, {Wang}, {Wang}, {Wang}, {Wang},
  {Wang}, {Wang}, {Wang}, {Wang}, {Wang}, {Wei}, {Weng}, {Wu}, {Wu}, {Xiao},
  {Xiong}, {Xiong}, {Xu}, {Yan}, {Yang}, {Yang}, {Yashin}, {Yelland}, {Yi},
  {You}, {Yu}, {Yu}, {Zannoni}, {Zhang}, {Zhang}, {Zhang}, {Zhang}, {Zhang},
  {Zhang}, {Zhao}, {Zheng}, {Zheng}, {Zhuang}, {Zhukov}, {Zichichi}, {Zuccon},
  \& {AMS Collaboration}}]{AMS02-10yr-heavy}
{Aguilar}, M., {Ali Cavasonza}, L., {Alpat}, B., {et~al.} 2023, \prl, 130,
  211002

\bibitem[{{Aguilar} {et~al.}(2018){Aguilar}, {Ali Cavasonza}, {Ambrosi},
  {Arruda}, {Attig}, {Aupetit}, {Azzarello}, {Bachlechner}, {Barao}, {Barrau},
  {Barrin}, {Bartoloni}, {Basara}, {Ba{\c{s}}e{\v{g}}mez-du Pree}, {Battarbee},
  {Battiston}, {Becker}, {Behlmann}, {Beischer}, {Berdugo}, {Bertucci},
  {Bindel}, {Bindi}, {de Boer}, {Bollweg}, {Bonnivard}, {Borgia}, {Boschini},
  {Bourquin}, {Bueno}, {Burger}, {Burger}, {Cadoux}, {Cai}, {Capell}, {Caroff},
  {Casaus}, {Castellini}, {Cervelli}, {Chae}, {Chang}, {Chen}, {Chen}, {Chen},
  {Cheng}, {Chou}, {Choumilov}, {Choutko}, {Chung}, {Clark}, {Clavero},
  {Coignet}, {Consolandi}, {Contin}, {Corti}, {Creus}, {Crispoltoni}, {Cui},
  {Dadzie}, {Dai}, {Datta}, {Delgado}, {Della Torre}, {Demirk{\"o}z}, {Derome},
  {Di Falco}, {Dimiccoli}, {D{\'\i}az}, {von Doetinchem}, {Dong}, {Donnini},
  {Duranti}, {D'Urso}, {Egorov}, {Eline}, {Eronen}, {Feng}, {Fiandrini},
  {Fisher}, {Formato}, {Galaktionov}, {Gallucci}, {Garc{\'\i}a-L{\'o}pez},
  {Gargiulo}, {Gast}, {Gebauer}, {Gervasi}, {Ghelfi}, {Giovacchini},
  {G{\'o}mez-Coral}, {Gong}, {Goy}, {Grabski}, {Grandi}, {Graziani}, {Guo},
  {Haino}, {Han}, {He}, {Heil}, {Hsieh}, {Huang}, {Huang}, {Huh}, {Incagli},
  {Ionica}, {Jang}, {Jia}, {Jinchi}, {Kang}, {Kanishev}, {Khiali}, {Kim},
  {Kim}, {Kirn}, {Konak}, {Kounina}, {Kounine}, {Koutsenko}, {Kulemzin}, {La
  Vacca}, {Laudi}, {Laurenti}, {Lazzizzera}, {Lebedev}, {Lee}, {Lee}, {Leluc},
  {Li}, {Li}, {Li}, {Li}, {Li}, {Li}, {Li}, {Lim}, {Lin}, {Lipari}, {Lippert},
  {Liu}, {Liu}, {Lordello}, {Lu}, {Lu}, {Luebelsmeyer}, {Luo}, {Luo}, {Lyu},
  {Machate}, {Ma{\~n}{\'a}}, {Mar{\'\i}n}, {Martin}, {Mart{\'\i}nez}, {Masi},
  {Maurin}, {Menchaca-Rocha}, {Meng}, {Mikuni}, {Mo}, {Mott}, {Nelson}, {Ni},
  {Nikonov}, {Nozzoli}, {Oliva}, {Orcinha}, {Palermo}, {Palmonari},
  {Palomares}, {Paniccia}, {Pauluzzi}, {Pensotti}, {Perrina}, {Phan},
  {Picot-Clemente}, {Pilo}, {Pizzolotto}, {Plyaskin}, {Pohl}, {Poireau},
  {Quadrani}, {Qi}, {Qin}, {Qu}, {R{\"a}ih{\"a}}, {Rancoita}, {Rapin}, {Ricol},
  {Rosier-Lees}, {Rozhkov}, {Rozza}, {Sagdeev}, {Schael}, {Schmidt}, {Schulz
  von Dratzig}, {Schwering}, {Seo}, {Shan}, {Shi}, {Siedenburg}, {Son}, {Song},
  {Tacconi}, {Tang}, {Tang}, {Tescaro}, {Ting}, {Ting}, {Tomassetti}, {Torsti},
  {T{\"u}rko{\v{g}}lu}, {Urban}, {Vagelli}, {Valente}, {Valtonen}, {V{\'a}zquez
  Acosta}, {Vecchi}, {Velasco}, {Vialle}, {Vitale}, {Wang}, {Wang}, {Wang},
  {Wang}, {Wang}, {Wang}, {Wei}, {Weng}, {Whitman}, {Wu}, {Wu}, {Xiong}, {Xu},
  {Yan}, {Yang}, {Yang}, {Yang}, {Yi}, {Yu}, {Yu}, {Zannoni}, {Zeissler},
  {Zhang}, {Zhang}, {Zhang}, {Zhang}, {Zhang}, {Zhang}, {Zheng}, {Zhuang},
  {Zhukov}, {Zichichi}, {Zimmermann}, {Zuccon}, \& {AMS
  Collaboration}}]{AMS02-5yr-LiBeB}
{Aguilar}, M., {Ali Cavasonza}, L., {Ambrosi}, G., {et~al.} 2018, \prl, 120,
  021101

\bibitem[{Aguilar {et~al.}(2021)Aguilar, {Ali Cavasonza}, Ambrosi, Arruda,
  Attig, Barao, Barrin, Bartoloni, {Başeğmez-du Pree}, Bates, Battiston,
  Behlmann, Beischer, Berdugo, Bertucci, Bindi, {de Boer}, Bollweg, Borgia,
  Boschini, Bourquin, Bueno, Burger, Burger, Burmeister, Cai, Capell, Casaus,
  Castellini, Cervelli, Chang, Chen, Chen, Chen, Cheng, Chou, Chouridou,
  Choutko, Chung, Clark, Coignet, Consolandi, Contin, Corti, Cui, Dadzie, Dai,
  Delgado, {Della Torre}, Demirköz, Derome, {Di Falco}, {Di Felice}, Díaz,
  Dimiccoli, {von Doetinchem}, Dong, Donnini, Duranti, Egorov, Eline, Feng,
  Fiandrini, Fisher, Formato, Freeman, Galaktionov, Gámez, García-López,
  Gargiulo, Gast, Gebauer, Gervasi, Giovacchini, Gómez-Coral, Gong, Goy,
  Grabski, Grandi, Graziani, Guo, Haino, Han, Hashmani, He, Heber, Hsieh, Hu,
  Huang, Hungerford, Incagli, Jang, Jia, Jinchi, Kanishev, Khiali, Kim, Kirn,
  Konyushikhin, Kounina, Kounine, Koutsenko, Kuhlman, Kulemzin, {La Vacca},
  Laudi, Laurenti, Lazzizzera, Lebedev, Lee, Lee, Leluc, Li, Li, Li, Li, Li,
  Li, Light, Lin, Lippert, Liu, Lu, Lu, Luebelsmeyer, Luo, Lyu, Machate,
  Mañá, Marín, Marquardt, Martin, Martínez, Masi, Maurin, Menchaca-Rocha,
  Meng, Mo, Molero, Mott, Mussolin, Ni, Nikonov, Nozzoli, Oliva, Orcinha,
  Palermo, Palmonari, Paniccia, Pashnin, Pauluzzi, Pensotti, Phan, Plyaskin,
  Pohl, Porter, Qi, Qin, Qu, Quadrani, Rancoita, Rapin, {Reina Conde},
  Rosier-Lees, Rozhkov, Rozza, Sagdeev, Schael, Schmidt, {Schulz von Dratzig},
  Schwering, Seo, Shan, Shi, Siedenburg, Solano, Song, Sonnabend, Sun, Sun,
  Tacconi, Tang, Tang, Tian, Ting, Ting, Tomassetti, Torsti, Tüysüz, Urban,
  Usoskin, Vagelli, Vainio, Valente, Valtonen, {Vázquez Acosta}, Vecchi,
  Velasco, Vialle, Wang, Wang, Wang, Wang, Wang, Wang, Wei, Weng, Wu, Xiong,
  Xu, Yan, Yang, Yi, Yu, Yu, Zannoni, Zhang, Zhang, Zhang, Zhang, Zhang, Zhao,
  Zheng, Zhuang, Zhukov, Zichichi, Zimmermann, \& Zuccon}]{AMS02-7yr}
Aguilar, M., {Ali Cavasonza}, L., Ambrosi, G., {et~al.} 2021, Physics Reports,
  894, 1, the Alpha Magnetic Spectrometer (AMS) on the International Space
  Station: Part II - Results from the First Seven Years

\bibitem[{{Aguilar} {et~al.}(2021{\natexlab{a}}){Aguilar}, {Cavasonza},
  {Allen}, {Alpat}, {Ambrosi}, {Arruda}, {Attig}, {Barao}, {Barrin},
  {Bartoloni}, {Ba{\c{s}}e{\v{g}}mez-du Pree}, {Battiston}, {Behlmann},
  {Beischer}, {Berdugo}, {Bertucci}, {Bindi}, {de Boer}, {Bollweg}, {Borgia},
  {Boschini}, {Bourquin}, {Bueno}, {Burger}, {Burger}, {Burmeister}, {Cai},
  {Capell}, {Casaus}, {Castellini}, {Cervelli}, {Chang}, {Chen}, {Chen},
  {Chen}, {Chen}, {Cheng}, {Chou}, {Chouridou}, {Choutko}, {Chung}, {Clark},
  {Coignet}, {Consolandi}, {Contin}, {Corti}, {Cui}, {Dadzie}, {Delgado},
  {Della Torre}, {Demirk{\"o}z}, {Derome}, {Di Falco}, {Di Felice},
  {D{\'\i}az}, {Dimiccoli}, {von Doetinchem}, {Dong}, {Donnini}, {Duranti},
  {Egorov}, {Eline}, {Feng}, {Fiandrini}, {Fisher}, {Formato}, {Freeman},
  {Galaktionov}, {G{\'a}mez}, {Garc{\'\i}a-L{\'o}pez}, {Gargiulo}, {Gast},
  {Gervasi}, {Giovacchini}, {G{\'o}mez-Coral}, {Gong}, {Goy}, {Grabski},
  {Grandi}, {Graziani}, {Haino}, {Han}, {Hashmani}, {He}, {Heber}, {Hsieh},
  {Hu}, {Incagli}, {Jang}, {Jia}, {Jinchi}, {Kanishev}, {Khiali}, {Kim},
  {Kirn}, {Konyushikhin}, {Kounina}, {Kounine}, {Koutsenko}, {Kuhlman},
  {Kulemzin}, {La Vacca}, {Laudi}, {Laurenti}, {Lazzizzera}, {Lebedev}, {Lee},
  {Lee}, {Li}, {Li}, {Li}, {Li}, {Li}, {Li}, {Liang}, {Light}, {Lin},
  {Lippert}, {Liu}, {Liu}, {Lu}, {Lu}, {Luebelsmeyer}, {Luo}, {Luo}, {Lyu},
  {Machate}, {Ma{\~n}{\'a}}, {Mar{\'\i}n}, {Marquardt}, {Martin},
  {Mart{\'\i}nez}, {Masi}, {Maurin}, {Menchaca-Rocha}, {Meng}, {Mikhailov},
  {Mo}, {Molero}, {Mott}, {Mussolin}, {Negrete}, {Nikonov}, {Nozzoli}, {Oliva},
  {Orcinha}, {Palermo}, {Palmonari}, {Paniccia}, {Pashnin}, {Pauluzzi},
  {Pensotti}, {Phan}, {Piandani}, {Plyaskin}, {Poluianov}, {Qin}, {Qu},
  {Quadrani}, {Rancoita}, {Rapin}, {Conde}, {Robyn}, {Rosier-Lees}, {Rozhkov},
  {Rozza}, {Sagdeev}, {Schael}, {von Dratzig}, {Schwering}, {Seo}, {Shakfa},
  {Shan}, {Siedenburg}, {Solano}, {Song}, {Song}, {Sonnabend}, {Strigari},
  {Su}, {Sun}, {Sun}, {Tacconi}, {Tang}, {Tang}, {Tian}, {Ting}, {Ting},
  {Tomassetti}, {Torsti}, {T{\"u}ys{\"u}z}, {Urban}, {Usoskin}, {Vagelli},
  {Vainio}, {Valencia-Otero}, {Valente}, {Valtonen}, {V{\'a}zquez Acosta},
  {Vecchi}, {Velasco}, {Vialle}, {Wang}, {Wang}, {Wang}, {Wang}, {Wang},
  {Wang}, {Wang}, {Wang}, {Wang}, {Wei}, {Weng}, {Wu}, {Xiong}, {Xu}, {Yan},
  {Yang}, {Yashin}, {Yi}, {Yu}, {Yu}, {Zannoni}, {Zhang}, {Zhang}, {Zhang},
  {Zhang}, {Zhang}, {Zhao}, {Zheng}, {Zheng}, {Zhuang}, {Zhukov}, {Zichichi},
  {Zimmermann}, {Zuccon}, \& {AMS Collaboration}}]{AMS02-8.5yr-Fe}
{Aguilar}, M., {Cavasonza}, L.~A., {Allen}, M.~S., {et~al.} 2021{\natexlab{a}},
  \prl, 126, 041104

\bibitem[{{Aguilar} {et~al.}(2021{\natexlab{b}}){Aguilar}, {Cavasonza},
  {Allen}, {Alpat}, {Ambrosi}, {Arruda}, {Attig}, {Barao}, {Barrin},
  {Bartoloni}, {Ba{\c{s}}e{\v{g}}mez-du Pree}, {Battiston}, {Behlmann},
  {Beranek}, {Berdugo}, {Bertucci}, {Bindi}, {Bollweg}, {Borgia}, {Boschini},
  {Bourquin}, {Bueno}, {Burger}, {Burger}, {Burmeister}, {Cai}, {Capell},
  {Casaus}, {Castellini}, {Cervelli}, {Chang}, {Chen}, {Chen}, {Chen}, {Chen},
  {Cheng}, {Chou}, {Chouridou}, {Choutko}, {Chung}, {Clark}, {Coignet},
  {Consolandi}, {Contin}, {Corti}, {Cui}, {Dadzie}, {Delgado}, {Della Torre},
  {Demirk{\"o}z}, {Derome}, {Di Falco}, {Di Felice}, {D{\'\i}az}, {Dimiccoli},
  {von Doetinchem}, {Dong}, {Donnini}, {Duranti}, {Egorov}, {Eline}, {Feng},
  {Fiandrini}, {Fisher}, {Formato}, {Freeman}, {Galaktionov}, {G{\'a}mez},
  {Garc{\'\i}a-L{\'o}pez}, {Gargiulo}, {Gast}, {Gervasi}, {Giovacchini},
  {G{\'o}mez-Coral}, {Gong}, {Goy}, {Grabski}, {Grandi}, {Graziani}, {Haino},
  {Han}, {Hashmani}, {He}, {Heber}, {Hsieh}, {Hu}, {Incagli}, {Jang}, {Jia},
  {Jinchi}, {Kanishev}, {Khiali}, {Kim}, {Kirn}, {Konyushikhin}, {Kounina},
  {Kounine}, {Koutsenko}, {Kuhlman}, {Kulemzin}, {La Vacca}, {Laudi},
  {Laurenti}, {Lazzizzera}, {Lebedev}, {Lee}, {Lee}, {Li}, {Li}, {Li}, {Li},
  {Li}, {Li}, {Liang}, {Light}, {Lin}, {Lippert}, {Liu}, {Liu}, {Lu}, {Lu},
  {Luebelsmeyer}, {Luo}, {Luo}, {Lyu}, {Machate}, {Ma{\~n}{\'a}}, {Mar{\'\i}n},
  {Marquardt}, {Martin}, {Mart{\'\i}nez}, {Masi}, {Maurin}, {Menchaca-Rocha},
  {Meng}, {Mikhailov}, {Mo}, {Molero}, {Mott}, {Mussolin}, {Negrete},
  {Nikonov}, {Nozzoli}, {Oliva}, {Orcinha}, {Palermo}, {Palmonari}, {Paniccia},
  {Pashnin}, {Pauluzzi}, {Pensotti}, {Phan}, {Piandani}, {Plyaskin},
  {Poluianov}, {Qin}, {Qu}, {Quadrani}, {Rancoita}, {Rapin}, {Conde}, {Robyn},
  {Rosier-Lees}, {Rozhkov}, {Rozza}, {Sagdeev}, {Schael}, {Schulz von Dratzig},
  {Schwering}, {Seo}, {Shakfa}, {Shan}, {Siedenburg}, {Solano}, {Song}, {Song},
  {Sonnabend}, {Strigari}, {Su}, {Sun}, {Sun}, {Tacconi}, {Tang}, {Tang},
  {Tian}, {Ting}, {Ting}, {Tomassetti}, {Torsti}, {T{\"u}ys{\"u}z}, {Urban},
  {Usoskin}, {Vagelli}, {Vainio}, {Valencia-Otero}, {Valente}, {Valtonen},
  {V{\'a}zquez Acosta}, {Vecchi}, {Velasco}, {Vialle}, {Wang}, {Wang}, {Wang},
  {Wang}, {Wang}, {Wang}, {Wang}, {Wang}, {Wang}, {Wei}, {Weng}, {Wu}, {Xiong},
  {Xu}, {Yan}, {Yang}, {Yashin}, {Yi}, {Yu}, {Yu}, {Zannoni}, {Zhang}, {Zhang},
  {Zhang}, {Zhang}, {Zhang}, {Zhao}, {Zheng}, {Zheng}, {Zhuang}, {Zhukov},
  {Zichichi}, {Zuccon}, \& {AMS Collaboration}}]{AMS02-8.5yr-F}
{Aguilar}, M., {Cavasonza}, L.~A., {Allen}, M.~S., {et~al.} 2021{\natexlab{b}},
  \prl, 126, 081102

\bibitem[{{Aguilar} {et~al.}(2021{\natexlab{c}}){Aguilar}, {Cavasonza},
  {Alpat}, {Ambrosi}, {Arruda}, {Attig}, {Barao}, {Barrin}, {Bartoloni},
  {Ba{\c{s}}e{\v{g}}mez-du Pree}, {Battiston}, {Behlmann}, {Beranek},
  {Berdugo}, {Bertucci}, {Bindi}, {Bollweg}, {Borgia}, {Boschini}, {Bourquin},
  {Bueno}, {Burger}, {Burger}, {Burmeister}, {Cai}, {Capell}, {Casaus},
  {Castellini}, {Cervelli}, {Chang}, {Chen}, {Chen}, {Chen}, {Chen}, {Cheng},
  {Chou}, {Chouridou}, {Choutko}, {Chung}, {Clark}, {Coignet}, {Consolandi},
  {Contin}, {Corti}, {Cui}, {Dadzie}, {Delgado}, {Della Torre}, {Demirk{\"o}z},
  {Derome}, {Di Falco}, {Di Felice}, {D{\'\i}az}, {Dimiccoli}, {von
  Doetinchem}, {Dong}, {Donnini}, {Duranti}, {Egorov}, {Eline}, {Feng},
  {Fiandrini}, {Fisher}, {Formato}, {Freeman}, {G{\'a}mez},
  {Garc{\'\i}a-L{\'o}pez}, {Gargiulo}, {Gast}, {Gervasi}, {Giovacchini},
  {G{\'o}mez-Coral}, {Gong}, {Goy}, {Grabski}, {Grandi}, {Graziani}, {Haino},
  {Han}, {Hashmani}, {He}, {Heber}, {Hsieh}, {Hu}, {Incagli}, {Jang}, {Jia},
  {Jinchi}, {Khiali}, {Kim}, {Kirn}, {Konyushikhin}, {Kounina}, {Kounine},
  {Koutsenko}, {Krasnopevtsev}, {Kuhlman}, {Kulemzin}, {La Vacca}, {Laudi},
  {Laurenti}, {Lazzizzera}, {Lebedev}, {Lee}, {Lee}, {Li}, {Li}, {Li}, {Li},
  {Li}, {Li}, {Liang}, {Light}, {Lin}, {Lippert}, {Liu}, {Liu}, {Lu}, {Lu},
  {Luebelsmeyer}, {Luo}, {Luo}, {Machate}, {Ma{\~n}{\'a}}, {Mar{\'\i}n},
  {Marquardt}, {Martin}, {Mart{\'\i}nez}, {Masi}, {Maurin}, {Medvedeva},
  {Menchaca-Rocha}, {Meng}, {Mikhailov}, {Molero}, {Mott}, {Mussolin},
  {Negrete}, {Nikonov}, {Nozzoli}, {Oliva}, {Orcinha}, {Palermo}, {Palmonari},
  {Paniccia}, {Pashnin}, {Pauluzzi}, {Pensotti}, {Phan}, {Plyaskin}, {Pohl},
  {Poluianov}, {Qin}, {Qu}, {Quadrani}, {Rancoita}, {Rapin}, {Conde}, {Robyn},
  {Rosier-Lees}, {Rozhkov}, {Rozza}, {Sagdeev}, {Schael}, {von Dratzig},
  {Schwering}, {Seo}, {Shakfa}, {Shan}, {Siedenburg}, {Solano}, {Song}, {Song},
  {Sonnabend}, {Strigari}, {Su}, {Sun}, {Sun}, {Tacconi}, {Tang}, {Tang},
  {Tian}, {Ting}, {Ting}, {Tomassetti}, {Torsti}, {T{\"u}ys{\"u}z}, {Urban},
  {Usoskin}, {Vagelli}, {Vainio}, {Valencia-Otero}, {Valente}, {Valtonen},
  {V{\'a}zquez Acosta}, {Vecchi}, {Velasco}, {Vialle}, {Wang}, {Wang}, {Wang},
  {Wang}, {Wang}, {Wang}, {Wang}, {Wang}, {Wang}, {Wei}, {Weng}, {Wu}, {Xiong},
  {Xu}, {Yan}, {Yang}, {Yashin}, {Yi}, {Yu}, {Yu}, {Zannoni}, {Zhang}, {Zhang},
  {Zhang}, {Zhang}, {Zhang}, {Zhao}, {Zheng}, {Zheng}, {Zhuang}, {Zhukov},
  {Zichichi}, {Zuccon}, \& {AMS Collaboration}}]{AMS02-8.5yr-NaAlN}
{Aguilar}, M., {Cavasonza}, L.~A., {Alpat}, B., {et~al.} 2021{\natexlab{c}},
  \prl, 127, 021101

\bibitem[{{Ahn} {et~al.}(2009){Ahn}, {Allison}, {Bagliesi}, {Barbier},
  {Beatty}, {Bigongiari}, {Brandt}, {Childers}, {Conklin}, {Coutu}, {Du
  Vernois}, {Ganel}, {Han}, {Jeon}, {Kim}, {Lee}, {Maestro}, {Malinine},
  {Marrocchesi}, {Minnick}, {Mognet}, {Nam}, {Nutter}, {Park}, {Park}, {Seo},
  {Sina}, {Walpole}, {Wu}, {Yang}, {Yoon}, {Zei}, \&
  {Zinn}}]{CREAM-2009-CNO-heavy}
{Ahn}, H.~S., {Allison}, P., {Bagliesi}, M.~G., {et~al.} 2009, \apj, 707, 593

\bibitem[{{Alemanno} {et~al.}(2021){Alemanno}, {An}, {Azzarello}, {Barbato},
  {Bernardini}, {Bi}, {Cai}, {Catanzani}, {Chang}, {Chen}, {Chen}, {Chen},
  {Cui}, {Cui}, {Cui}, {Dai}, {D'Amone}, {de Benedittis}, {de Mitri}, {de
  Palma}, {Deliyergiyev}, {di Santo}, {Dong}, {Dong}, {Donvito}, {Droz},
  {Duan}, {Duan}, {D'Urso}, {Fan}, {Fan}, {Fang}, {Fang}, {Feng}, {Feng},
  {Fusco}, {Gao}, {Gargano}, {Gong}, {Gong}, {Guo}, {Guo}, {Guo}, {Han}, {Hu},
  {Huang}, {Huang}, {Huang}, {Ionica}, {Jiang}, {Kong}, {Kotenko}, {Kyratzis},
  {Lei}, {Li}, {Li}, {Li}, {Li}, {Liang}, {Liu}, {Liu}, {Liu}, {Liu}, {Liu},
  {Liu}, {Loparco}, {Luo}, {Ma}, {Ma}, {Ma}, {Ma}, {Marsella}, {Mazziotta},
  {Mo}, {Niu}, {Pan}, {Parenti}, {Peng}, {Peng}, {Perrina}, {Qiao}, {Rao},
  {Ruina}, {Salinas}, {Shang}, {Shen}, {Shen}, {Shen}, {Silveri}, {Song},
  {Stolpovskiy}, {Su}, {Su}, {Sun}, {Surdo}, {Teng}, {Tykhonov}, {Wang},
  {Wang}, {Wang}, {Wang}, {Wang}, {Wang}, {Wang}, {Wang}, {Wang}, {Wei}, {Wei},
  {Wei}, {Wen}, {Wu}, {Wu}, {Wu}, {Wu}, {Wu}, {Xia}, {Xu}, {Xu}, {Xu}, {Xu},
  {Xue}, {Yang}, {Yang}, {Yang}, {Yao}, {Yu}, {Yuan}, {Yuan}, {Yue}, {Zang},
  {Zhang}, {Zhang}, {Zhang}, {Zhang}, {Zhang}, {Zhang}, {Zhang}, {Zhang},
  {Zhang}, {Zhang}, {Zhao}, {Zhao}, {Zhao}, {Zhou}, {Zhu}, \& {Dampe
  Collaboration}}]{DAMPE-2021-He}
{Alemanno}, F., {An}, Q., {Azzarello}, P., {et~al.} 2021, \prl, 126, 201102

\bibitem[{Aloisio \& Blasi(2013)}]{Aloisio-Balsi-2013-self}
Aloisio, R. \& Blasi, P. 2013, Journal of Cosmology and Astroparticle Physics,
  2013, 001

\bibitem[{{An} {et~al.}(2019){An}, {Asfandiyarov}, {Azzarello}, {Bernardini},
  {Bi}, {Cai}, {Chang}, {Chen}, {Chen}, {Chen}, {Chen}, {Cui}, {Cui}, {Dai},
  {D'Amone}, {De Benedittis}, {De Mitri}, {Di Santo}, {Ding}, {Dong}, {Dong},
  {Dong}, {Donvito}, {Droz}, {Duan}, {Duan}, {D'Urso}, {Fan}, {Fan}, {Fang},
  {Feng}, {Feng}, {Fusco}, {Gallo}, {Gan}, {Gao}, {Gargano}, {Gong}, {Gong},
  {Guo}, {Guo}, {Guo}, {Han}, {Hu}, {Huang}, {Huang}, {Huang}, {Ionica},
  {Jiang}, {Jin}, {Kong}, {Lei}, {Li}, {Li}, {Li}, {Li}, {Li}, {Liang},
  {Liang}, {Liao}, {Liu}, {Liu}, {Liu}, {Liu}, {Liu}, {Liu}, {Loparco}, {Luo},
  {Ma}, {Ma}, {Ma}, {Ma}, {Ma}, {Marsella}, {Mazziotta}, {Mo}, {Niu}, {Pan},
  {Peng}, {Peng}, {Qiao}, {Rao}, {Salinas}, {Shang}, {Shen}, {Shen}, {Shen},
  {Song}, {Su}, {Su}, {Sun}, {Surdo}, {Teng}, {Tykhonov}, {Vitillo}, {Wang},
  {Wang}, {Wang}, {Wang}, {Wang}, {Wang}, {Wang}, {Wang}, {Wang}, {Wang},
  {Wang}, {Wang}, {Wang}, {Wei}, {Wei}, {Wei}, {Wen}, {Wu}, {Wu}, {Wu}, {Wu},
  {Wu}, {Xi}, {Xia}, {Xu}, {Xu}, {Xu}, {Xu}, {Xue}, {Yang}, {Yang}, {Yang},
  {Yang}, {Yao}, {Yu}, {Yuan}, {Yue}, {Zang}, {Zhang}, {Zhang}, {Zhang},
  {Zhang}, {Zhang}, {Zhang}, {Zhang}, {Zhang}, {Zhang}, {Zhang}, {Zhang},
  {Zhang}, {Zhang}, {Zhao}, {Zhao}, {Zhao}, {Zhou}, {Zhou}, {Zhu}, {Zhu}, \&
  {Zimmer}}]{DAMPE-2019-H}
{An}, Q., {Asfandiyarov}, R., {Azzarello}, P., {et~al.} 2019, Science Advances,
  5, eaax3793

\bibitem[{{Antoni} {et~al.}(2005){Antoni}, {Apel}, {Badea}, {Bekk}, {Bercuci},
  {Bl{\"u}mer}, {Bozdog}, {Brancus}, {Chilingarian}, {Daumiller}, {Doll},
  {Engel}, {Engler}, {Fe{\ss}ler}, {Gils}, {Glasstetter}, {Haungs}, {Heck},
  {H{\"o}randel}, {Kampert}, {Klages}, {Maier}, {Mathes}, {Mayer}, {Milke},
  {M{\"u}ller}, {Obenland}, {Oehlschl{\"a}ger}, {Ostapchenko}, {Petcu},
  {Rebel}, {Risse}, {Risse}, {Roth}, {Schatz}, {Schieler}, {Scholz}, {Thouw},
  {Ulrich}, {van Buren}, {Vardanyan}, {Weindl}, {Wochele}, \&
  {Zabierowski}}]{KASCADE-2005-H-He---}
{Antoni}, T., {Apel}, W.~D., {Badea}, A.~F., {et~al.} 2005, Astroparticle
  Physics, 24, 1

\bibitem[{{Arteaga-Vel{\'a}zquez} {et~al.}(2017){Arteaga-Vel{\'a}zquez},
  {Rivera-Rangel}, {Apel}, {Bekk}, {Bertaina}, {Bl{\"u}mer}, {Bozdog},
  {Brancus}, {Cantoni}, {Chiavassa}, {Cossavella}, {Daumiller}, {de Souza}, {di
  Pierro}, {Doll}, {Engel}, {Fuhrmann}, {Gherghel-Lascu}, {Gils},
  {Glasstetter}, {Grupen}, {Haungs}, {Heck}, {H{\"o}randel}, {Huber}, {Huege},
  {Kampert}, {Kang}, {Klages}, {Link}, {{\L}uczak}, {Mathes}, {Mayer}, {Milke},
  {Mitrica}, {Morello}, {Oehlschl{\"a}ger}, {Ostapchenko}, {Palmieri},
  {Pierog}, {Rebel}, {Roth}, {Schieler}, {Schoo}, {Schr{\"o}der}, {Sima},
  {Toma}, {Trinchero}, {Ulrich}, {Weindl}, {Wochele}, {Zabierowski}, \&
  {KASCADE-Grande Collaboration}}]{KASCADE-Grande-2017}
{Arteaga-Vel{\'a}zquez}, C.~J., {Rivera-Rangel}, D., {Apel}, W.~D., {et~al.}
  2017, in International Cosmic Ray Conference, Vol. 301, 35th International
  Cosmic Ray Conference (ICRC2017), 316

\bibitem[{{Bell} {et~al.}(2013){Bell}, {Schure}, {Reville}, \&
  {Giacinti}}]{Bell-2013-acc-escape}
{Bell}, A.~R., {Schure}, K.~M., {Reville}, B., \& {Giacinti}, G. 2013, \mnras,
  431, 415

\bibitem[{{Berezinskii} {et~al.}(1990){Berezinskii}, {Bulanov}, {Dogiel}, \&
  {Ptuskin}}]{Berezinkii-1990-bible}
{Berezinskii}, V.~S., {Bulanov}, S.~V., {Dogiel}, V.~A., \& {Ptuskin}, V.~S.
  1990, {Astrophysics of cosmic rays} (Amsterdam: North-Holland)

\bibitem[{{Blasi}(2013)}]{Blasi-2013-review}
{Blasi}, P. 2013, \aapr, 21, 70

\bibitem[{{Bresci} {et~al.}(2019){Bresci}, {Amato}, {Blasi}, \&
  {Morlino}}]{Bresci-2019-src-grammage}
{Bresci}, V., {Amato}, E., {Blasi}, P., \& {Morlino}, G. 2019, \mnras, 488,
  2068

\bibitem[{{Bykov} {et~al.}(2020){Bykov}, {Marcowith}, {Amato}, {Kalyashova},
  {Kruijssen}, \& {Waxman}}]{Bykov-2020-review-acc-star-forming}
{Bykov}, A.~M., {Marcowith}, A., {Amato}, E., {et~al.} 2020, \ssr, 216, 42

\bibitem[{{Casse} {et~al.}(2001){Casse}, {Lemoine}, \&
  {Pelletier}}]{Casse-Lemoine-Pelletier-2001}
{Casse}, F., {Lemoine}, M., \& {Pelletier}, G. 2001, \prd, 65, 023002

\bibitem[{{Cerri} {et~al.}(2017){Cerri}, {Gaggero}, {Vittino}, {Evoli}, \&
  {Grasso}}]{Cerri-2017-gradient-problem-anisotropic}
{Cerri}, S.~S., {Gaggero}, D., {Vittino}, A., {Evoli}, C., \& {Grasso}, D.
  2017, \jcap, 2017, 019

\bibitem[{{Chandran}(2000)}]{Chandran-2000-cloud-mirror}
{Chandran}, B. D.~G. 2000, \apj, 529, 513

\bibitem[{{Chernyshov} {et~al.}(2023){Chernyshov}, {Ivlev}, \&
  {Dogiel}}]{Chernyshov-2023-NLLD}
{Chernyshov}, D.~O., {Ivlev}, A.~V., \& {Dogiel}, V.~A. 2023, arXiv e-prints,
  arXiv:2309.04772

\bibitem[{{Coste} {et~al.}(2012){Coste}, {Derome}, {Maurin}, \&
  {Putze}}]{Coste-2012-light-nucl}
{Coste}, B., {Derome}, L., {Maurin}, D., \& {Putze}, A. 2012, \aap, 539, A88

\bibitem[{Cowsik \& Burch(2010)}]{Cowsik-2010-cocoon}
Cowsik, R. \& Burch, B. 2010, Phys. Rev. D, 82, 023009

\bibitem[{Cowsik \& Madziwa-Nussinov(2016)}]{Cowsik-2016-cocoon}
Cowsik, R. \& Madziwa-Nussinov, T. 2016, The Astrophysical Journal, 827, 119

\bibitem[{{Cristofari}(2021)}]{Cristofari-2021-review-PeV}
{Cristofari}, P. 2021, Universe, 7, 324

\bibitem[{{Cristofari} {et~al.}(2020){Cristofari}, {Blasi}, \&
  {Amato}}]{Cristofari-2020-low-rate-PeV}
{Cristofari}, P., {Blasi}, P., \& {Amato}, E. 2020, Astroparticle Physics, 123,
  102492

\bibitem[{{Dampe Collaboration}(2022)}]{DAMPE-2022-BC-BO}
{Dampe Collaboration}. 2022, Science Bulletin, 67, 2162

\bibitem[{{D'Angelo} {et~al.}(2016){D'Angelo}, {Blasi}, \&
  {Amato}}]{Dangelo-2016-grammage}
{D'Angelo}, M., {Blasi}, P., \& {Amato}, E. 2016, \prd, 94, 083003

\bibitem[{{Evoli} {et~al.}(2019){Evoli}, {Aloisio}, \&
  {Blasi}}]{Evoli-2019-ams02}
{Evoli}, C., {Aloisio}, R., \& {Blasi}, P. 2019, \prd, 99, 103023

\bibitem[{{Evoli} {et~al.}(2018{\natexlab{a}}){Evoli}, {Blasi}, {Morlino}, \&
  {Aloisio}}]{Evoli-2018-halo-formation}
{Evoli}, C., {Blasi}, P., {Morlino}, G., \& {Aloisio}, R. 2018{\natexlab{a}},
  \prl, 121, 021102

\bibitem[{{Evoli} {et~al.}(2018{\natexlab{b}}){Evoli}, {Gaggero}, {Vittino},
  {Di Mauro}, {Grasso}, \& {Mazziotta}}]{Evoli-2018-DRAGON-II}
{Evoli}, C., {Gaggero}, D., {Vittino}, A., {et~al.} 2018{\natexlab{b}}, \jcap,
  2018, 006

\bibitem[{{Fornieri} {et~al.}(2021{\natexlab{a}}){Fornieri}, {Gaggero},
  {Cerri}, {De La Torre Luque}, \& {Gabici}}]{Fornieri-2021-MHD}
{Fornieri}, O., {Gaggero}, D., {Cerri}, S.~S., {De La Torre Luque}, P., \&
  {Gabici}, S. 2021{\natexlab{a}}, \mnras, 502, 5821

\bibitem[{{Fornieri} {et~al.}(2021{\natexlab{b}}){Fornieri}, {Gaggero},
  {Guberman}, {Brahimi}, {Luque}, \&
  {Marcowith}}]{Fornieri-2021PhRvD-DAMPE-10TeV-loca-src}
{Fornieri}, O., {Gaggero}, D., {Guberman}, D., {et~al.} 2021{\natexlab{b}},
  \prd, 104, 103013

\bibitem[{{Gabici} {et~al.}(2019){Gabici}, {Evoli}, {Gaggero}, {Lipari},
  {Mertsch}, {Orlando}, {Strong}, \& {Vittino}}]{Gabici-2019-review}
{Gabici}, S., {Evoli}, C., {Gaggero}, D., {et~al.} 2019, International Journal
  of Modern Physics D, 28, 1930022

\bibitem[{{Gabici} {et~al.}(2016){Gabici}, {Gaggero}, \&
  {Zandanel}}]{Gabici-2016-SNR-PeVatrons}
{Gabici}, S., {Gaggero}, D., \& {Zandanel}, F. 2016, arXiv e-prints,
  arXiv:1610.07638

\bibitem[{{G{\'e}nolini} {et~al.}(2018){G{\'e}nolini}, {Maurin}, {Moskalenko},
  \& {Unger}}]{Genolini-2018-xsec-prod}
{G{\'e}nolini}, Y., {Maurin}, D., {Moskalenko}, I.~V., \& {Unger}, M. 2018,
  \prc, 98, 034611

\bibitem[{{G{\'e}nolini} {et~al.}(2023){G{\'e}nolini}, {Maurin}, {Moskalenko},
  \& {Unger}}]{Genolini-2023-xsec-prod}
{G{\'e}nolini}, Y., {Maurin}, D., {Moskalenko}, I.~V., \& {Unger}, M. 2023,
  arXiv e-prints, arXiv:2307.06798

\bibitem[{G\'enolini {et~al.}(2017)G\'enolini, Serpico, Boudaud, Caroff,
  Poulin, Derome, Lavalle, Maurin, Poireau, Rosier, Salati, \&
  Vecchi}]{Genolini-2017-hard-secondary}
G\'enolini, Y., Serpico, P.~D., Boudaud, M., {et~al.} 2017, Phys. Rev. Lett.,
  119, 241101

\bibitem[{{Gleeson} \& {Axford}(1968)}]{Gleeson-Axford-1968-solar-mod}
{Gleeson}, L.~J. \& {Axford}, W.~I. 1968, \apj, 154, 1011

\bibitem[{{Grebenyuk} {et~al.}(2019){Grebenyuk}, {Karmanov}, {Kovalev},
  {Kudryashov}, {Kurganov}, {Panov}, {Podorozhny}, {Tkachenko}, {Tkachev},
  {Turundaevskiy}, {Vasiliev}, \& {Voronin}}]{NUCLEON-2019}
{Grebenyuk}, V., {Karmanov}, D., {Kovalev}, I., {et~al.} 2019, Advances in
  Space Research, 64, 2546

\bibitem[{{Hooper} {et~al.}(2009){Hooper}, {Blasi}, \&
  {Serpico}}]{Hooper-2009-pulsar}
{Hooper}, D., {Blasi}, P., \& {Serpico}, P.~D. 2009, \jcap, 2009, 025

\bibitem[{{Jacobs} {et~al.}(2023){Jacobs}, {Mertsch}, \&
  {Phan}}]{Mertsch-2023-Be10-low-diff}
{Jacobs}, H., {Mertsch}, P., \& {Phan}, V. H.~M. 2023, \mnras, 526, 160

\bibitem[{{Jokipii} \& {Parker}(1969)}]{Jokipii-Parker-1969}
{Jokipii}, J.~R. \& {Parker}, E.~N. 1969, \apj, 155, 777

\bibitem[{{Kadomtsev} \& {Pogutse}(1979)}]{Kadomtsev-Pogutse-1979}
{Kadomtsev}, B.~B. \& {Pogutse}, O.~P. 1979, in Plasma Physics and Controlled
  Nuclear Fusion Research 1978, Volume 1, Vol.~1, 649--662

\bibitem[{{Kamijima} \& {Ohira}(2022)}]{Ohira-2022PhRvD-perp-shocks-10TeV}
{Kamijima}, S.~F. \& {Ohira}, Y. 2022, \prd, 106, 123025

\bibitem[{{Karmanov} {et~al.}(2020){Karmanov}, {Kovalev}, {Kudryashov},
  {Kurganov}, {Panov}, {Podorozhny}, {Turundaevskiy}, \&
  {Vasiliev}}]{NUCLEON-2020-HHe}
{Karmanov}, D.~E., {Kovalev}, I.~M., {Kudryashov}, I.~A., {et~al.} 2020, Soviet
  Journal of Experimental and Theoretical Physics Letters, 111, 363

\bibitem[{{Kirk} {et~al.}(1996){Kirk}, {Duffy}, \&
  {Gallant}}]{Kirk-Duffy-Gallant-1996}
{Kirk}, J.~G., {Duffy}, P., \& {Gallant}, Y.~A. 1996, \aap, 314, 1010

\bibitem[{{Korsmeier} {et~al.}(2018){Korsmeier}, {Donato}, \& {Di
  Mauro}}]{Korsmeier-2018-pbar-production}
{Korsmeier}, M., {Donato}, F., \& {Di Mauro}, M. 2018, \prd, 97, 103019

\bibitem[{{Lagutin} \& {Volkov}(2023)}]{Lagutin-2023-DAMPE-break-pHe-local-src}
{Lagutin}, A.~A. \& {Volkov}, N.~V. 2023, arXiv e-prints, arXiv:2309.07420

\bibitem[{{Lazarian} \& {Xu}(2021)}]{Lazarian-Xu-2021-mirror-diffusion}
{Lazarian}, A. \& {Xu}, S. 2021, \apj, 923, 53

\bibitem[{Lazarian {et~al.}(2023)Lazarian, Xu, \&
  Hu}]{Lazarian-2023-review-MHD-transport}
Lazarian, A., Xu, S., \& Hu, Y. 2023, Frontiers in Astronomy and Space
  Sciences, 10

\bibitem[{Lipari(2017)}]{Lipari-2017-prop-new}
Lipari, P. 2017, Phys. Rev. D, 95, 063009

\bibitem[{{Luo} {et~al.}(2022){Luo}, {Qiao}, {Liu}, {Cui}, \&
  {Guo}}]{Liu-2022-DAMPE-10TeV-local-src}
{Luo}, Q., {Qiao}, B.-q., {Liu}, W., {Cui}, S.-w., \& {Guo}, Y.-q. 2022, \apj,
  930, 82

\bibitem[{Malkov {et~al.}(2012)Malkov, Diamond, \&
  Sagdeev}]{Malkov-2012-p-He-inj-spectra}
Malkov, M.~A., Diamond, P.~H., \& Sagdeev, R.~Z. 2012, Phys. Rev. Lett., 108,
  081104

\bibitem[{{Malkov} \& {Moskalenko}(2022)}]{Malkov-2022-DAMPE-bump-10TeV}
{Malkov}, M.~A. \& {Moskalenko}, I.~V. 2022, \apj, 933, 78

\bibitem[{{Maurin}(2020)}]{USINE-2020}
{Maurin}, D. 2020, Computer Physics Communications, 247, 106942

\bibitem[{{Maurin} {et~al.}(2023){Maurin}, {Ahlers}, {Dembinski}, {Haungs},
  {Mangeard}, {Melot}, {Mertsch}, {Wochele}, \& {Wochele}}]{Maurin-2023-CRDB}
{Maurin}, D., {Ahlers}, M., {Dembinski}, H., {et~al.} 2023, European Physical
  Journal C, 83, 971

\bibitem[{{Mertsch}(2020)}]{Mertsch-2020-test-particle-transp-review}
{Mertsch}, P. 2020, \apss, 365, 135

\bibitem[{{Mertsch} {et~al.}(2021){Mertsch}, {Vittino}, \&
  {Sarkar}}]{Mertsch-2021-sec-src-old-SNR}
{Mertsch}, P., {Vittino}, A., \& {Sarkar}, S. 2021, \prd, 104, 103029

\bibitem[{Porter {et~al.}(2022)Porter, Jóhannesson, \&
  Moskalenko}]{GALPROP-Porter-2022}
Porter, T.~A., Jóhannesson, G., \& Moskalenko, I.~V. 2022, The Astrophysical
  Journal Supplement Series, 262, 30

\bibitem[{Ptuskin {et~al.}(2013)Ptuskin, Zirakashvili, \&
  Seo}]{Ptuskin-2013-hard-nonlin-acc}
Ptuskin, V., Zirakashvili, V., \& Seo, E.-S. 2013, The Astrophysical Journal,
  763, 47

\bibitem[{{Ptuskin} \& {Soutoul}(1998)}]{Ptuskin-Soutoul1998-review-CR-clocks}
{Ptuskin}, V.~S. \& {Soutoul}, A. 1998, \ssr, 86, 225

\bibitem[{{Recchia} \& {Gabici}(2018)}]{Recchia-2018-pamela}
{Recchia}, S. \& {Gabici}, S. 2018, \mnras, 474, L42

\bibitem[{{Recchia} {et~al.}(2022){Recchia}, {Galli}, {Nava}, {Padovani},
  {Gabici}, {Marcowith}, {Ptuskin}, \& {Morlino}}]{Recchia-2022-grammage}
{Recchia}, S., {Galli}, D., {Nava}, L., {et~al.} 2022, \aap, 660, A57

\bibitem[{{Rechester} \& {Rosenbluth}(1978)}]{Rechester-Rosenbluth-1978}
{Rechester}, A.~B. \& {Rosenbluth}, M.~N. 1978, \prl, 40, 38

\bibitem[{{Shalchi}(2020)}]{Shalchi-2020-perp-transp-review}
{Shalchi}, A. 2020, \ssr, 216, 23

\bibitem[{Shalchi(2021)}]{Shalchi-2021-FLRW-MHD}
Shalchi, A. 2021, Physics of Plasmas, 28, 120501

\bibitem[{{Skilling}(1971)}]{Skilling-1971}
{Skilling}, J. 1971, \apj, 170, 265

\bibitem[{{Skilling}(1975)}]{Skilling-1975-effect-waves}
{Skilling}, J. 1975, \mnras, 172, 557

\bibitem[{{Subedi} {et~al.}(2017){Subedi}, {Sonsrettee}, {Blasi}, {Ruffolo},
  {Matthaeus}, {Montgomery}, {Chuychai}, {Dmitruk}, {Wan}, {Parashar}, \&
  {Chhiber}}]{Subedi-Blasi-2017}
{Subedi}, P., {Sonsrettee}, W., {Blasi}, P., {et~al.} 2017, \apj, 837, 140

\bibitem[{{Tatischeff} {et~al.}(2021){Tatischeff}, {Raymond}, {Duprat},
  {Gabici}, \& {Recchia}}]{Tatischeff-2021-composition}
{Tatischeff}, V., {Raymond}, J.~C., {Duprat}, J., {Gabici}, S., \& {Recchia},
  S. 2021, \mnras, 508, 1321

\bibitem[{Thoudam \& Hörandel(2012)}]{Thoudam-2012}
Thoudam, S. \& Hörandel, J.~R. 2012, Monthly Notices of the Royal Astronomical
  Society, 421, 1209

\bibitem[{Tomassetti(2012)}]{Tomassetti-2012}
Tomassetti, N. 2012, The Astrophysical Journal Letters, 752, L13

\bibitem[{{Tomassetti} \& {Donato}(2012)}]{Tomassetti-Donato-2012-sec-src}
{Tomassetti}, N. \& {Donato}, F. 2012, \aap, 544, A16

\bibitem[{{Tripathi} {et~al.}(1999){Tripathi}, {Cucinotta}, \&
  {Wilson}}]{Tripathi-1999}
{Tripathi}, R.~K., {Cucinotta}, F.~A., \& {Wilson}, J.~W. 1999, {Universal
  Parameterization of Absorption Cross Sections}, Technical Report,
  NASA/TP-1999-209726; NAS 1.60:209726; L-17832

\bibitem[{{Vieu} \& {Reville}(2023)}]{Vieu-2023-PeV-massive-star-cluster}
{Vieu}, T. \& {Reville}, B. 2023, \mnras, 519, 136

\bibitem[{{Vink}(2012)}]{Vink-2012-SNR-X-review}
{Vink}, J. 2012, \aapr, 20, 49

\bibitem[{Workman {et~al.}(2022)}]{rpp-Workman-2022}
Workman, R.~L. {et~al.} 2022, PTEP, 2022, 083C01

\bibitem[{{Yoon} {et~al.}(2017){Yoon}, {Anderson}, {Barrau}, {Conklin},
  {Coutu}, {Derome}, {Han}, {Jeon}, {Kim}, {Kim}, {Lee}, {Lee}, {Lee}, {Lee},
  {Link}, {Menchaca-Rocha}, {Mitchell}, {Mognet}, {Nutter}, {Park},
  {Picot-Clemente}, {Putze}, {Seo}, {Smith}, \& {Wu}}]{CREAM-2017-H-He}
{Yoon}, Y.~S., {Anderson}, T., {Barrau}, A., {et~al.} 2017, \apj, 839, 5

\end{thebibliography}

%%%%%%%%%%%%%%%%%%%%%%%%%%%%%%%%%%%%%%%%%%%%%%%%%%

%%%%%%%%%%%%%%%%% APPENDICES %%%%%%%%%%%%%%%%%%%%%

\begin{appendix}

\section{Advection and adiabatic losses}
\label{sec:appendix-adiabatic}
In this section we motivate the choice to neglect adiabatic losses in the transport equation, showing that for typical CR spectra, such term can be absorbed in the value of the advection velocity $u$ required to fit CR data. 

To make the result more apparent  we limit to the propagation of protons (so that we can neglect spallation) in the GH, with a rigidity dependent diffusion coefficient and a constant advection speed pointing away from the GP, namely $du/dz = 2u\delta(z)$.
We solve the transport equation for the distribution function $f(p, z)$ (see e.g \citealt{Evoli-2019-ams02})
\begin{equation}\label{eq:transp-f-proton}
 - \frac{\partial}{\partial z} \left[D 
 \frac{\partial f}{\partial z} - u f \right] -2\,u\,\delta(z) \frac{1}{3p^2} \frac{\partial (f\,p^3)}{\partial p} =  q_0\,\delta(z), 
\end{equation}
with the  free-escape condition $f(p, H) = 0$
%%%%%%
\subsection*{Solution for $\mathbf{z \neq 0}$} 
Eq.~(\ref{eq:transp-f-proton}) becomes
\begin{equation}
D\frac{\partial f}{\partial z} - u f= \mathcal{K} = \, \rm const,
\end{equation}
which, using the condition $f(p, H) = 0$ and $f(p, 0) \equiv f_0$, gives
\begin{equation}\label{eq:appendix-fz}
f(p, z) = f_0 \frac{1-e^{-\frac{u}{D}(H-z)}}{1-e^{-\frac{uH}{D}}}.
\end{equation}
%%%%
\subsection*{Integration around $\mathbf{z=0}$}
Using the continuity of $f_0$ across $z=0$ and $\left.\frac{\partial f}{\partial z}\right\vert_{0^+} = - \left.\frac{\partial f}{\partial z}\right\vert_{0^-}$ we get
\begin{equation}\label{eq:appendix-diff-eq-f0}
    -2D \left. \frac{\partial f}{\partial z}\right\vert_{0^+} + 2\, u\, f_0  - \frac{2\,u}{3 p^2} \frac{d\, (f_0\,p^3)}{d\, p} = q_0.
\end{equation}
The term
\begin{equation}
    D\left.\frac{\partial f}{\partial z}\right\vert_{0^+} = -u\, f_0 \frac{e^{-\frac{uH}{D}}}{1-e^{-\frac{uH}{D}}}
\end{equation}
can be computed using Eq.~(\ref{eq:appendix-fz}) and plugged into Eq.~(\ref{eq:appendix-diff-eq-f0}) to get a differntial equation for $f_0$,
\begin{equation}
  2\,u\, f_0 \frac{e^{-\frac{uH}{D}}}{1-e^{-\frac{uH}{D}}} - \frac{2}{3}u\,p\, \frac{d\, f_0}{d\, p} = q_0,
\end{equation}
whose solution reads
\begin{equation}
    f_0(p) = 
    \frac{3}{2\,u}
    \int_p^{\infty} 
    \frac{d\,p'}{p'}
    q_0(p')\exp\left[-3\int_p^{p'}
    \frac{dp''}{p''} \frac{e^{-\frac{uH}{D(p'')}}}{1-e^{-\frac{uH}{D(p'')}}}
    \right].
\end{equation}
Two important limits can be obtained in the advection-dominated regime, relevant at low momenta, and diffusion-dominated regime, relevant at high momenta:
\begin{equation}
    f_0 = \begin{cases}
    f_0^{\rm adv} = \frac{3\,q_0}{2\,u\,\gamma} \propto p^{-\gamma} & \qquad\qquad \text{$(uH/D \gg 1)$ }\\
    f_0^{\rm diff} = \frac{q_0 \, H}{2\,D} \propto p^{-\gamma -\delta} & \qquad\qquad \text{$ (uH/D \ll 1) $ }
\end{cases}   
\end{equation}
where we assumed $q_0 \propto p^{-\gamma}$ and $D\propto p^{-\delta}$.
The quantity $uH/D$ controls the transition between the two regimes, where at low $p$ the steady-state spectrum resembles the injection spectrum and at high $p$ it is steeper due to the momentum-dependent diffusion.
\subsection*{Approximate solution}
In order to simplify the solution for $f_0$, let us rewrite Eq.~(\ref{eq:appendix-diff-eq-f0}) as
\begin{equation}\label{eq:appendix-f0-approx-eq}
   2\,u\,f_0\frac{e^{-\frac{uH}{D}}}{1-e^{-\frac{uH}{D}}} + 2\,u\,f_0 + 2\,u\,f_0\, s_0   = q_0, 
\end{equation}
where $s_0 \equiv - \frac{1}{3}\frac{d\ln (f_0 \,p^3)}{d\ln p}$ is minus  the slope of $f_0\,p^3$ divided by 3. Since generally CR spectra are close to perfect power-laws $\propto p^{-4-\alpha}$, with $\alpha \approx 0-1$, $s_0$ is expected to vary quite slowly with $p$.\\
Assuming $s_0$ to be constant we can solve Eq.~(\ref{eq:appendix-f0-approx-eq}) as
\begin{equation}\label{eq:appendix-f0-approx}
    f_0(p) = \frac{q_0\, H}{2\, D} \,\frac{1-e^{-\frac{uH}{D}}}{\frac{u\,H}{D} + s_0 \frac{u\,H}{D}\left[1-e^{-\frac{uH}{D}} \right]}.
\end{equation}
The advection-dominate and diffusion-dominated regimes read
\begin{equation}\label{eq:appendix-f0-limits-approx}
    f_0 = \begin{cases}
    f_{0, \rm approx}^{\rm adv} = \frac{q_0}{2\,u\,(1+s_0)}\\
    f_{0, \rm approx}^{\rm diff} = \frac{q_0 \, H}{2\,D(p)}
\end{cases}   
\end{equation}
which are the same of the general solution, if we consider that in the advection-dominated regime $s_0 \approx \gamma/3 -1$. \\ 

Notice that, posing $s_0 = 0$ we get 
\begin{equation}\label{eq:appendix-f0-tau-HH}
   f_0(p) \rightarrow \frac{Q_0}{2} \left[\frac{H}{D} \frac{1-e^{-\frac{uH}{D}}}{\frac{u\,H}{D}}\right].
\end{equation}   

This discussion shows that ignoring the adiabatic losses, or choosing a different spatial  profile for $u$ instead of $\frac{du}{dz} = 2\,u\,\delta(z)$ (for instance a smooth increase of the advection velocity from $u_{\rm adv}(z=0) = 0$ to the value $u_{\rm adv}(z>h) = u$  beyond the disk of size $h$), would  slightly change, in the propagated spectrum, the momentum-dependent transition between the advective and diffusive regimes.
To zero order, in fitting CR data,  such effect can be absorbed in the best-fit value for the advection velocity. Indeed, the same advection-dominated limit  in Eq.~(\ref{eq:appendix-f0-limits-approx}) with a given $u$ and $s_0$ can be obtained setting  $s_0=0$ and $u \rightarrow u(1+s_0)$, namely with a slightly larger value of the advection velocity.

\section{Analytic solution for stable nuclei}
\label{sec:appendix-stable-nuclei}
We start with Eq.~(\ref{eq:transp}), we neglect adiabatic losses (see Sect.~\ref{sec:appendix-adiabatic}) and let $\tau_r \rightarrow +\infty$ for stable nuclei.  The transport equation becomes
\begin{align}
\label{eq:appendix-transp-stable}
   -\frac{\partial}{\partial z} \left[D_{\alpha} 
 \frac{\partial I_{\alpha}}{\partial z} - u I_{\alpha}\right] & + \, 2\,h_d\,n_d v(E_k)\sigma_{\alpha}(E_k)\delta(z) I_{\alpha} = \\ \nonumber
 & h \delta(z)\left[\frac{\mathcal{Q}_{\alpha, \rm src}}{h}\, +\,
 2n_d\frac{h_d}{h} \mathcal{Q}_{\alpha, \rm spall}\right] 
\end{align}
where
\begin{equation}
    \begin{cases}
    \mathcal{Q}_{\alpha, \rm src} \equiv c A p^2 q_{0, \alpha}\\
    \mathcal{Q}_{\alpha, \rm spall} \equiv \sum_{\beta > \alpha} \, v(E_k) \sigma_{\beta\alpha} (E_k)I_{\beta}(E_k)
\end{cases}  
\end{equation}
are the injection from sources and the injection from spallation of heavier nuclei (per ISM atom),  respectively,  
and $H^*= h+H$ as described in Sect.~\ref{sec:prop-setup}.
We solve this equation with the assumptions of
Eqs.~(\ref{eq:D-def}, \ref{eq:u-def}, \ref{eq:free-esc}),  which we report here for completeness
\begin{equation}
    D = \begin{cases}
    D_h & \text{if $0 < z \leq h $ }\\
    D_H & \text{if $ h < z \leq H^* $ }
\end{cases}   
\end{equation}
\begin{equation}
    u = \begin{cases}
    0 & \text{if $0 < z \leq h $ }\\
    u & \text{if $ h < z \leq H^* $ }
\end{cases} 
\end{equation}
\begin{equation}
I_{\alpha}(H^{*}, E_k) = 0.
\end{equation}
The solution can be found similarly to the case presented in Sect.~\ref{sec:appendix-adiabatic}.
\subsection*{Solution $\mathbf{z\neq 0}$} 
The transport equation reduces to 
\begin{equation}
D \frac{\partial I_{\alpha}}{\partial z} - u I_{\alpha} = \mathcal{K}_{\alpha} = \, \rm const,
\end{equation}
and in particular to
\begin{equation}
\begin{cases}
    D_h \frac{\partial I_{\alpha}}{\partial z}  = \mathcal{K}_{\alpha} & \quad \text{if\; $0 < z \leq h $ \qquad\; (GD)} \\
    D_H \frac{\partial I_{\alpha}}{\partial z} - u I_{\alpha} = \mathcal{K}_{\alpha} & \quad\text{if\; $ h < z \leq H^* $\qquad  (GH) }.
\end{cases}
\end{equation}
Using the solution of Eq.~(\ref{eq:appendix-fz}) for $z>h$ we get
\begin{align}
    & I_{\alpha} = I_{\alpha h}\frac{1-e^{-\frac{u}{D_H}(H^*-z)}}{1-e^{-\frac{u\,H}{D_H}} }\\
    & \mathcal{K}_{\alpha} = -\frac{u\, I_{\alpha h}}{1-e^{-\frac{u\,H}{D_H}}},
    \label{eq:appendix-Ka-stable}
\end{align}
where $I_{\alpha h}$ is the particle flux at $z=h$. \\
Plugging the expression for $\mathcal{K}_{\alpha}$ into the equation for the GD we get
\begin{equation}
    D_h \frac{\partial I_{\alpha}}{\partial z}  = \mathcal{K}_{\alpha} =  -\frac{u\, I_{\alpha h}}{1-e^{-\frac{u\,H}{D_H}}},
\end{equation}
which can be integrated between $z=0^+$ and $z=h^{-}$, taking into account the continuity in $h$, namely $I_{\alpha, h^+} = I_{\alpha, h^-} = I_{\alpha h} $
\begin{equation}
    I_{\alpha h} - I_{\alpha 0} = - \frac{u\, h}{D_h}\frac{I_{\alpha h}}{1-e^{-\frac{u\,H}{D_H}}} 
\end{equation}
For the flux at $z=0$, $I_{\alpha 0}$, we obtain
\begin{equation}
    I_{\alpha 0}  = I_{\alpha h} \left[1+\frac{u\,h}{D_h} \frac{1}{1-e^{-\frac{u\,H}{D_H}}}\right].
\end{equation}
The latter expression can be rewritten in a more compact form by introducing 
\begin{equation}
    \begin{cases}
    \chi_h \equiv \frac{u\,h}{D_h}\\
    \chi_H \equiv \frac{u\,H}{D_H},
    \end{cases}
\end{equation}
namely
\begin{equation}\label{eq:appendix-Ia0-Iah}
    I_{\alpha 0}  = I_{\alpha h} \left[1+\frac{\chi_h}{1-e^{-\chi_H}}\right].
\end{equation}
\subsection*{Integration around $\mathbf{z= 0}$} 
Using the fact that 
$D_h \left. \frac{\partial I_{\alpha}}{\partial z}\right\vert_{0^+} = - D_h \left. \frac{\partial I_{\alpha}}{\partial z}\right\vert_{0^-} = \mathcal{K}_{\alpha}$
we get
\begin{equation}
-2\,\mathcal{K}_{\alpha} + 2\,h_d\,n_d v\sigma_{\alpha} I_{\alpha 0} = 
 2\,h \left[\frac{\mathcal{Q}_{\alpha, \rm src}}{2\,h}\, +\,n_d\frac{h_d}{h}\mathcal{Q}_{\alpha, \rm spall}\right].
\end{equation}
Using Eq.~(\ref{eq:appendix-Ka-stable}) and Eq.~(\ref{eq:appendix-Ia0-Iah}) we can write
\begin{equation}
    I_{\alpha 0}(E_k) = \frac{
\left(\frac{\mathcal{Q}_{\alpha, \rm src}}{2\,h}\, +\,n_d\frac{h_d}{h}\mathcal{Q}_{\alpha, \rm spall}\right)\,h
    \left[ 1-e^{-\chi_H} + \chi_h\right] }{u + h_d\,n_d\, v\, \sigma_{\alpha}\, (1-e^{-\chi_H} + \chi_h)}.
\end{equation}
To bring the latter equation in the form of Eq.~(\ref{eq:solution-stable-I0}) notice that
\begin{align}
    h \left[ 1-e^{-\chi_H} + \chi_h\right] & = h \left[ \frac{1-e^{-\chi_H}}{\chi_H} \chi_H + \chi_h\right] \\ \nonumber
    & = u\, \tau_{\alpha}^{h\,H},
\end{align}
where 
\begin{equation}
\tau_{\alpha}^{hH} =  \frac{h^2}{D_h} + \frac{h\,H}{D_H} \frac{1-\exp^{-\frac{uH}{D_H}}}{\frac{uH}{D_H}},
\end{equation}
as in Eq.~(\ref{eq:tau-hH-def}).\\ 
Notice that the term $\frac{1-\exp^{-\frac{uH}{D_H}}}{\frac{uH}{D_H}}$ was also found in Eq.~(\ref{eq:appendix-f0-tau-HH}) and can be interpreted as a modification of the diffusion coefficient in the GH, $D_H$, that takes into account the advection (see also discussion in Sect.~\ref{sec:appendix-adiabatic}).
Indeed, let us introduce the effective diffusion coefficient
\begin{equation}\label{eq:appendix-Deff-adv}
    D_H^{\rm eff} \equiv  D_H \frac{\frac{uH}{D_H}}{1-\exp^{-\frac{uH}{D_H}}}.
\end{equation}
The advection-dominated  and diffusion-dominated regimes read
\begin{equation}
D_H^{\rm eff} \longrightarrow
\begin{cases}
    u\,H  & \qquad\qquad\text{advective limit ($uH/D_H \gg 1$)}  \\
    D_H & \qquad\qquad \text{diffuive limit ($uH/D_H \ll 1$)},
\end{cases}
\end{equation}
as expected in the two cases.\\
Finally, the expression for $I_{\alpha 0}$ becomes
\begin{equation}
   I_{\alpha 0}(E_k)  = \frac{\tau_{\alpha}^{hH}}{1 + n_d\,\frac{h_d}{h}\, v(E_k)\sigma_{\alpha}\tau_{\alpha}^{hH}} \times
\left[\frac{\mathcal{Q}_{\alpha, \rm src}}{2h} + n_d\frac{h_d}{h} \mathcal{Q}_{\alpha, \rm spall} \right],
\end{equation}
as reported in Eq.~(\ref{eq:solution-stable-I0}).
%
%%%%%%%%%%%%%%%%%%%
\begin{figure*}
\begin{multicols}{2} \includegraphics[width=\linewidth]{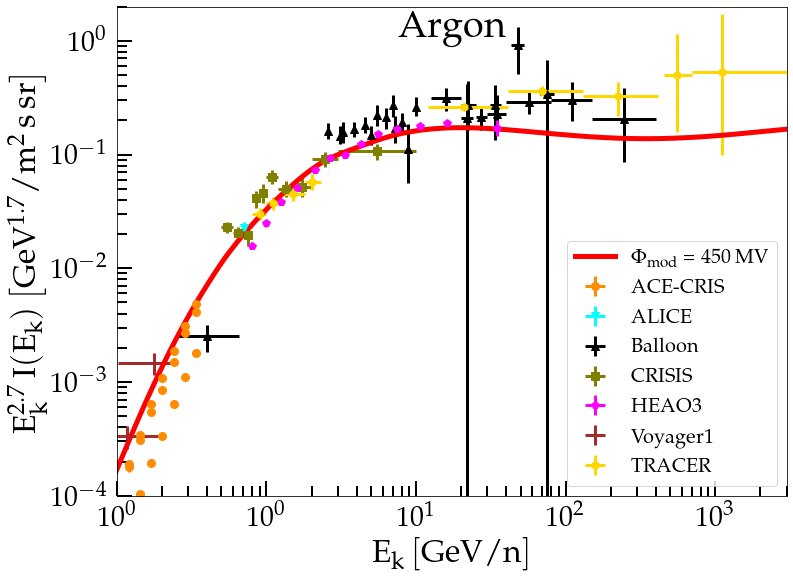}\par 
\includegraphics[width=\linewidth]{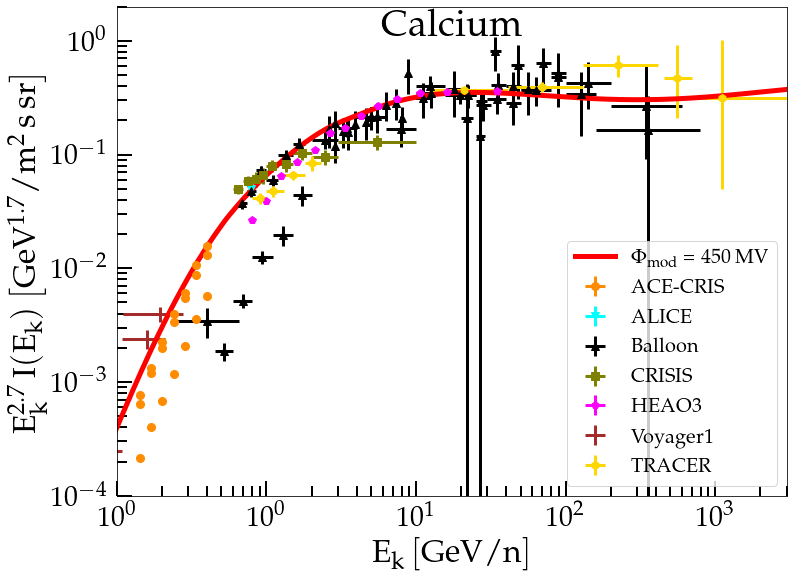}\par 
\end{multicols}
\begin{multicols}{2} \includegraphics[width=\linewidth]{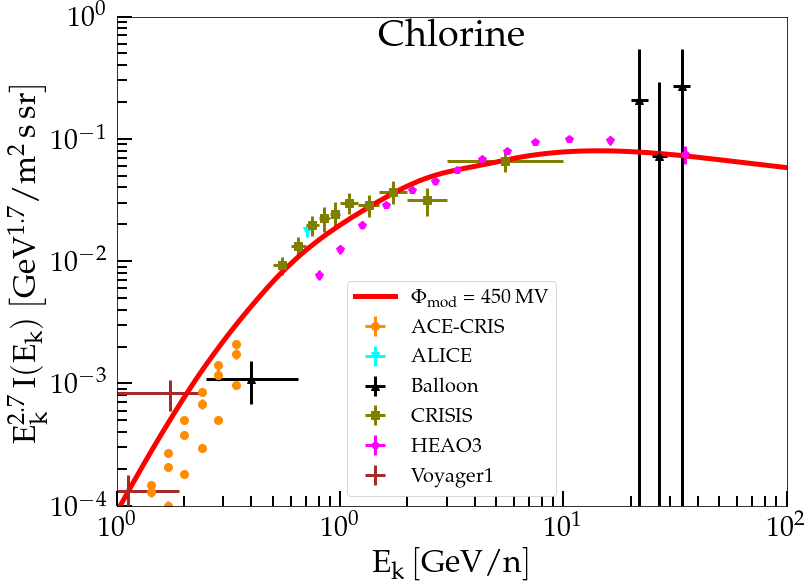}\par 
\includegraphics[width=\linewidth]{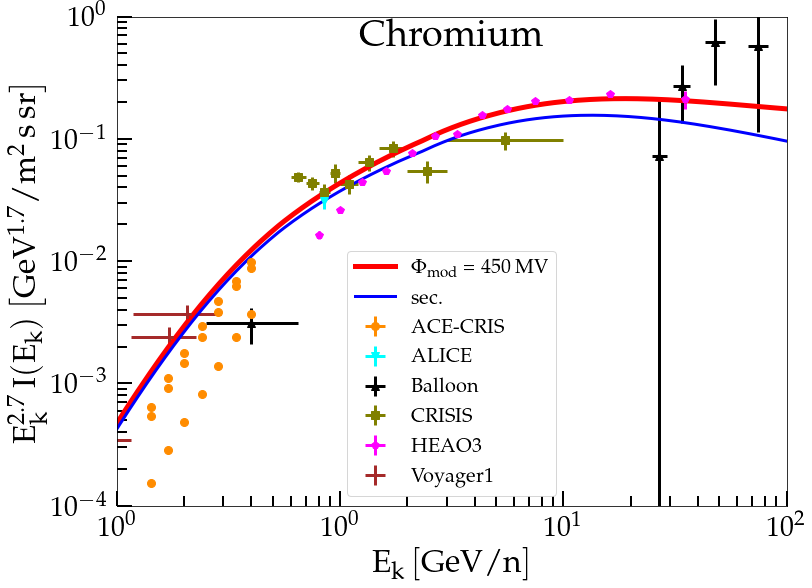}\par 
\end{multicols}
\includegraphics[width=0.5\linewidth]{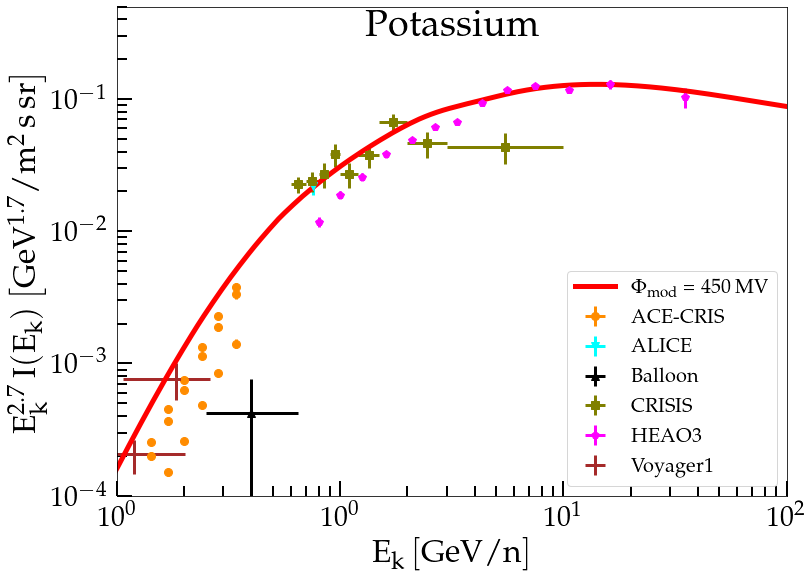}
\caption{Fluxes of Ar, Ca, Cl, Cr, K, compared to a variety of data (taken from \protect{\citealt{Maurin-2023-CRDB}}).
}
\label{fig:Ar-K}
\end{figure*}
%%%%%%%%%%%%%%%%%%%%%
\begin{figure*}
\begin{multicols}{2} \includegraphics[width=\linewidth]{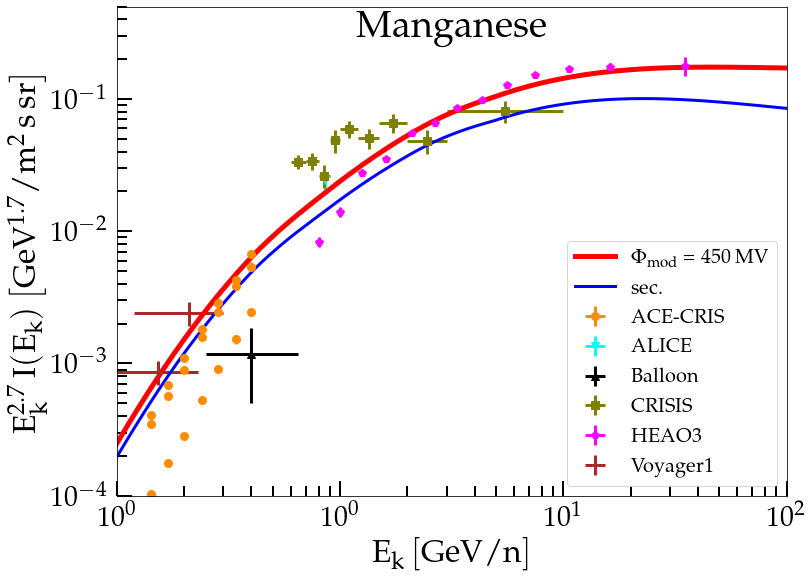}\par 
\includegraphics[width=\linewidth]{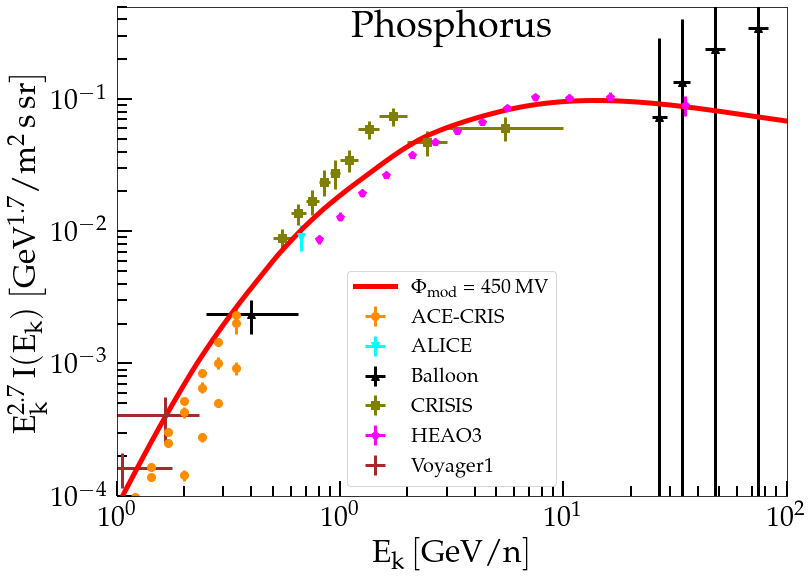}\par 
\end{multicols}
\begin{multicols}{2} \includegraphics[width=\linewidth]{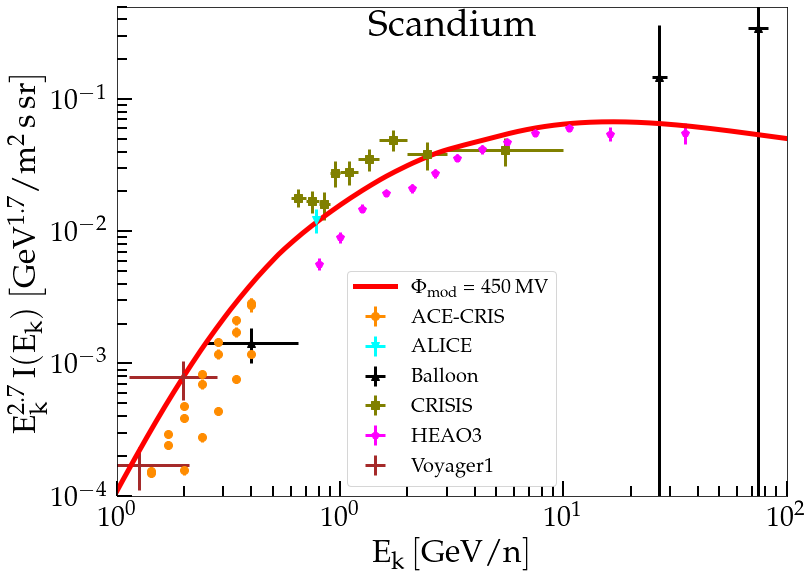}\par 
\includegraphics[width=\linewidth]{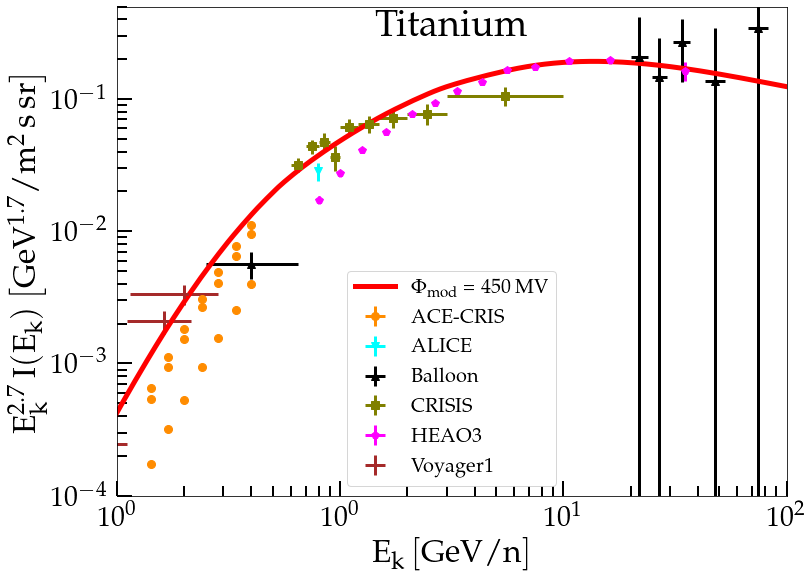}\par 
\end{multicols}
\includegraphics[width=0.5\linewidth]{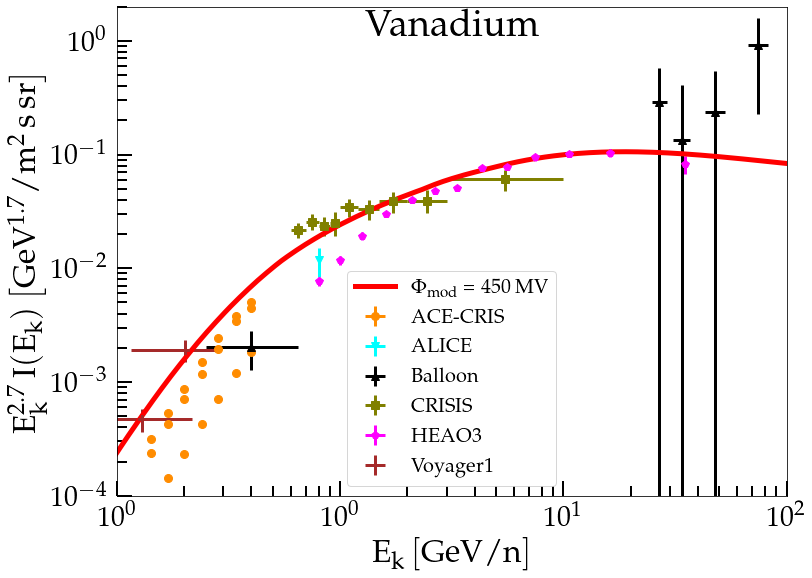}
\caption{Fluxes of Mn, P, Sc, Ti, V, compared to a variety of data (taken from \protect{\citealt{Maurin-2023-CRDB}}).
}
\label{fig:Mn-V}
\end{figure*}
%%%%%%%%%%%%%%%%
%%%%%%%%%%%%%%%%
\section{Analytic solution for unstable nuclei}
\label{sec:appendix-unstable-nuclei}
We follow the procedure used in Sect.~\ref{sec:appendix-stable-nuclei} and solve the 
transport equation 
\begin{align}
\label{eq:transp-unstable}
   -\frac{\partial}{\partial z}& \left[D_{\alpha} 
 \frac{\partial I_{\alpha}}{\partial z} \right]  +\nu_r I_{\alpha} = \\ \nonumber
 & - \, 2\,h_d\,n_d v(E_k)\sigma_{\alpha}(E_k)\delta(z) I_{\alpha} +
 2\,h\delta(z) \left[n_d\frac{h_d}{h} \mathcal{Q}_{\alpha, \rm spall}\right],
\end{align}
where we introduced the radioactive decay rate $\nu_r(E_k) = 1/\tau_r(E_k) = 1/\tau_{r0}\gamma(E_k)$ (where $\tau_0$ is the decay time at rest and $\gamma(E_k)$ the particle Lorentz factor) and removed the injection term from sources, since unstable nuclei are secondary. 

As for the advection in the GH, in order to keep the analytic solution simpler, we treat it with the effective diffusion coefficient $D_H^{\rm eff}$ defined in Eq.~(\ref{eq:appendix-Deff-adv}), namely
\begin{equation}
    D = \begin{cases}
    D_h & \text{if $0 < z \leq h $ }\\
    D_H^{\rm eff} & \text{if $ h < z \leq H^* $ }.
\end{cases}   
\end{equation}
We also impose the usual free-escape condition $I_{\alpha}(H^{*}, E_k) = 0$
\subsection*{Solution for $\mathbf{z\neq 0}$} 
Eq.~(\ref{eq:transp-unstable}) reduces to 
\begin{equation}
\begin{cases}
   D_h \frac{\partial^2 I_{\alpha}}{\partial z^2}  - \nu_r I_{\alpha} =0  & \text{if $0 < z \leq h $ \qquad (GD) } \\
   \\
    D_H^{\rm eff} \frac{\partial^2 I_{\alpha}}{\partial z^2}  - \nu_r I_{\alpha} =0 & \text{if $ h < z \leq H^* $\qquad  (GH) },
\end{cases}
\end{equation}
whose general solution reads
\begin{equation}
\begin{cases}
I_{\alpha}^{\rm GD} = a_h e^{\xi_h \frac{h -z}{h}} + b_h e^{-\xi_h \frac{h -z}{h}}  & \qquad \text{$ \xi_h \equiv \sqrt{\frac{\nu_r  h^2}{D_h}} $}  \\
   \\
I_{\alpha}^{\rm GH} = a_H e^{\xi_H \frac{H^* -z}{H}} + b_H e^{-\xi_H \frac{H^* -z}{H}}  & \qquad \text{$ \xi_H \equiv \sqrt{\frac{\nu_r H^2}{D_H^{\rm eff}}} $}.
\end{cases}
\end{equation}
The constants $a_h$, $b_h$ and $b_H$ as a function of $a_H$ are determined with the conditions
\begin{description}
    \item{\makebox[4.5cm][l]{$I_{\alpha}^{\rm GH}(H^*) = 0 $\hfill} (free escape)}
    \item{\makebox[4.5cm][l]{$I_{\alpha}^{\rm GH}(h^{+}) = I_{\alpha}^{\rm GD}(h^{-})$\hfill} (continuity in $h$)}
    \item{\makebox[4.5cm][l]{$D_h \left.\frac{\partial I_{\alpha}^{\rm GD}}{\partial z} \right\vert_{h^-} = D_H \left.\frac{\partial I_{\alpha}^{\rm GH}}{\partial z} \right\vert_{h^+} $\hfill} (integration around $h$)}
     \item{\makebox[4.5cm][l]{$I_{\alpha 0}
  \equiv I_{\alpha}^{\rm GD}(z=0),$\hfill}}
\end{description}
which gives
\begin{equation}\label{eq:appendix-Iz-unstable}
\begin{cases}
I_{\alpha}^{\rm GH} & = 2\,a_H \sinh\left(\xi_H \frac{H^* - z}{H}\right)\\
\\
I_{\alpha}^{\rm GD} & = 2\,a_H \sinh(\xi_H)\cosh\left(\xi_h \frac{h - z}{h}\right)\\
&\; + 2\,a_H \frac{H}{h}\frac{\xi_h}{\xi_H} \cosh(\xi_H)\sinh\left(\xi_h \frac{h - z}{h}\right)\\
\\
I_{\alpha 0} &  = 
2\,a_H \left[\sinh(\xi_H)\cosh(\xi_h) + \frac{H}{h}\frac{\xi_h}{\xi_H} \cosh(\xi_H)\sinh(\xi_h) \right].
\end{cases}
\end{equation}
\subsection*{Integration around $\mathbf{z= 0}$} 
We get
\begin{equation}
    -D_h \left.\frac{\partial I_{\alpha}^{\rm GD}}{\partial z} \right\vert_{0^+} + \,h_d\,n_d\, v\,\sigma_{\alpha} I_{\alpha 0} =  
    h \left[n_d\frac{h_d}{h} \mathcal{Q}_{\alpha, \rm spall}\right],
\end{equation}
where $\left.\frac{\partial I_{\alpha}^{\rm GD}}{\partial z} \right\vert_{0^+}$ can be derived from Eq.~(\ref{eq:appendix-Iz-unstable}).  With some cumbersome algebra we can finally derive an expression for $I_{\alpha 0}$, namely
\begin{equation}\label{eq:appendic-I0-unstable}
  I_{\alpha 0}(E_k) = \frac{n_d\frac{h_d}{h} \tau_{\alpha}^{hH}\mathcal{Q}_{\alpha,\rm spall}(E_k)}{1+ n_d\frac{h_d}{h} \tau_{\alpha}^{hH}\,v(E_k)\,\sigma_{\alpha}(E_k) },  
\end{equation}
where we defined the quantities
\begin{align}
   & \tau_{\alpha}^{hH} \equiv \xi_{\alpha}^{hH} \sqrt{\frac{h^2}{D_h} \tau_r}\\
   & \xi_{\alpha}^{hH} \equiv \frac{\sinh(\xi_H)\cosh(\xi_h) + \sqrt{\frac{D^{\rm eff}_H}{D_h}}\cosh(\xi_H)\sinh(\xi_h)}{\sinh(\xi_H)\sinh(\xi_h) + \sqrt{\frac{D^{\rm eff}_H}{D_h}} \cosh(\xi_H)\cosh(\xi_h)}.
\end{align}
Indeed, it can be shown that $\frac{H}{h}\frac{\xi_h}{\xi_H} = \sqrt{\frac{D^{\rm eff}_H}{D_h}}$.\\
Notice that $\tau_{\alpha}^{hH} $ is a timescale $\propto{\tau_h \tau_r}$, where $\tau_h \equiv h^2/D_h$ is the diffusion timescale in the GD.

Similarly to the case of stable nuclei, we can introduce the grammage as in Eq.~(\ref{eq:gramm-def}), and rewrite Eq.~(\ref{eq:appendic-I0-unstable}) as
\begin{equation}
  I_{\alpha 0}(E_k) = \frac{\frac{X_{\alpha}}{\mu v(E_k)}\mathcal{Q}_{\alpha,\rm spall}(E_k)}{1+ \frac{X_{\alpha}(E_k)}{X_{{\rm cr} \alpha}(E_k)}}.  
\end{equation}

\subsection*{Relevant limits}
We conclude by analyzing simpler case of diffusion in the GH only (namely $h \rightarrow 0$ and $D_h \rightarrow D_H$) and negligible spallation  $\sigma_{\alpha}\rightarrow 0$. Under such assumptions , the solution can be simplified through
\begin{description}
    \item{\makebox[4.5cm][l]{$h \rightarrow 0$\hfill} $\implies \xi_h \rightarrow 0$}\\
    \item{\makebox[4.5cm][l]{$D_h \rightarrow D_H$\hfill} $\implies \xi_{\alpha}^{hH} \rightarrow 1/\coth(\xi_H)$}
\end{description}
which gives (we also multiply and divide by $H$) 
\begin{equation}
    I_{\alpha 0} = \left(n_d\frac{h_d}{H}\right)\mathcal{Q}_{\alpha,\rm spall} \frac{\xi_H}{\nu_r\, \coth(\xi_H)}.
\end{equation}
Taking into account the diffusion timescale, $\tau_{\rm diff}(E_k) \equiv H^2/D^{\rm eff}_H(E_k) $, the following limits can be envisaged (see \citealt{Berezinkii-1990-bible}):
\begin{enumerate}
    \item \textbf{diffusion limit} (relevant at high $E_k$)\\
    $\tau_{\rm diff} \ll \tau_r$ ($\xi_H \rightarrow 0$;  $\coth(\xi_H) \rightarrow 1/\xi_H$) namely
    \begin{equation}
       \qquad I_{\alpha 0} \quad\longrightarrow\qquad n_d\,h_d\,\mathcal{Q}_{\alpha,\rm spall}\frac{H}{D^{\rm eff}_H} 
    \end{equation}
    which corresponds to the  solution for secondary stable nuclei, Eq.~(\ref{eq:solution-stable-I0}) when spallation is neglected and $h\rightarrow 0$;\\
     \item \textbf{decay limit} (relevant at low $E_k$)\\
     $\tau_{\rm diff} \gg \tau_r$ ($\xi_H \rightarrow 1$;  $\coth(\xi_H) \rightarrow 1$) namely
    \begin{equation}
    \qquad I_{\alpha 0} \quad\longrightarrow\quad \left(n_d\frac{h_d}{H}\right)\mathcal{Q}_{\alpha,\rm spall}\sqrt{\tau_{\rm diff}\tau_r} \propto \frac{\tau_r}{\sqrt{D_H^{\rm eff}\tau_r}},
    \end{equation} 
    which can be interpreted as follows.  
    In a time $\tau_r$ particles travel diffusively a distance $\Tilde{H} \equiv \sqrt{D^{\rm eff}_H\tau_r}$, and thus experience an average density $n_d h_d/\Tilde{H}$ instead of $n_d h_d/H$. Thus the equilibrium flux is given by 
    \begin{equation}
        \qquad I_{\alpha 0} =\left(n_d\frac{h_d}{\Tilde{H}}\right)\mathcal{Q}_{\alpha,\rm spall}\tau_r = \left(n_d\frac{h_d}{H}\right)\mathcal{Q}_{\alpha,\rm spall}\sqrt{\tau_{\rm diff}\tau_r},  
    \end{equation}
    where the latter equality has been obtained by multiplying and dividing by $H$.
\end{enumerate}

\end{appendix}  
%%%%%%%%%%%%%%%%%%%%%%%%%%%%%%%%%%%%%%%%%%%%%%%%%%

\end{document}